\documentclass[twocolumn]{aastex63}

\usepackage{comment}
\usepackage{hyperref}
\usepackage{amsmath}
\usepackage{xspace}

\makeatletter
\def\paragraph{\@startsection{paragraph}{4}{\z@}{-3.25ex plus
-1ex minus -.2ex}{1.5ex plus .2ex}{\small\it}}
\makeatother

\newcommand{\euclid}{\textit{Euclid}\xspace}
\newcommand{\wfirst}{\textit{Roman}\xspace}
\newcommand{\rosat}{\textit{ROSAT}\xspace}
\newcommand{\galex}{\textit{GALEX}\xspace}

\newcommand{\swift}{\textit{Swift}\xspace}

\newcommand{\hst}{\textit{HST}\xspace}

\newcommand{\ultrasat}{\textit{ULTRASAT}\xspace}
\newcommand{\uvex}{\textit{UVEX}\xspace}
\newcommand{\wise}{\textit{WISE}\xspace}
\newcommand{\lisa}{\textit{LISA}\xspace}

\newcommand{\jwst}{\textit{JWST}\xspace}

\newcommand{\lya}{Ly$\alpha$\ }

\newcommand{\msun}{M_{\odot}}
\newcommand{\mstar}{M_*}
\newcommand{\name}{\textit{UVEX}\xspace}

\newcommand{\LMLZ}{LMLZ\ }

\newcommand{\R}{\mbox{$\mathcal{R}$}\xspace}
\newcommand{\luv}{\mbox{$L_{\rm UV}$}\xspace}

\newcommand{\caltech}{Division of Physics, Mathematics, and Astronomy, California Institute of Technology, Pasadena, CA 91125, USA}
\newcommand{\berkeley}{Department of Astronomy, University of California, Berkeley, CA 94720-3411, USA}
\newcommand{\ipac}{IPAC, California Institute of Technology, Pasadena, CA 91125, USA}
\newcommand{\GRAPPA}{GRAPPA,  University of Amsterdam, Science Park 904, 1098 XH Amsterdam, Netherlands}
\newcommand{\carnegie}{The Observatories of the Carnegie Institution for Science, 813 Santa Barbara St., CA 91101, USA}
\newcommand{\liege}{Groupe d’Astrophysique des Hautes Energies, STAR, Universit\'{e} de Li\`{e}ge, Quartier Agora (B5c, Institut d’Astrophysique et de G\'{e}ophysique), All\'{e}e du 6 Ao\^{u}t 19c, B-4000 Sart Tilman, Li\`{e}ge, Belgium}

\newcommand\W{{$\lambda$}}
\newcommand\ii{{\sc ii}}
\newcommand\iii{{\sc iii}}

\shorttitle{Ultraviolet Explorer}
\shortauthors{Kulkarni et al.}

\graphicspath{{./}{Figures/}}

\begin{document}

\title{Science with the Ultraviolet Explorer (UVEX)}

\correspondingauthor{S.R. Kulkarni}
\email{srk@astro.caltech.edu}

\author[0000-0001-5390-8563]{S.\ R.\ Kulkarni}
\affiliation{\caltech}

\author[0000-0002-4226-8959]{Fiona A.\ Harrison}
\affiliation{\caltech}

\author[0000-0002-1984-2932]{Brian W.\ Grefenstette}
\affiliation{\caltech}

\author[0000-0001-5857-5622]{Hannah P.\ Earnshaw}
\affiliation{\caltech}

\author[0000-0003-3768-7515]{Igor Andreoni}
\affiliation{Joint Space-Science Institute, University of Maryland, College Park, MD 20742, USA}

\author[0000-0002-4153-053X]{Danielle A. Berg}
\affiliation{Department of Astronomy, The University of Texas at Austin, 2515 Speedway, Stop C1400, Austin, TX 78712, USA}

\author[0000-0002-7777-216X]{Joshua S.\ Bloom}
\affiliation{\berkeley}

\author[0000-0003-1673-970X]{S.\ Bradley Cenko}
\affiliation{Astrophysics Science Division, NASA Goddard Space Flight Center, 8800 Greenbelt Road, Greenbelt, MD 20771, USA}

\author[0000-0002-7706-5668]{Ryan Chornock}
\affiliation{\berkeley}

\author[0000-0002-8035-4778]{Jessie L. Christiansen}
\affiliation{\ipac}

\author[0000-0002-8262-2924]{Michael W.\ Coughlin}
\affiliation{School of Physics and Astronomy, University of Minnesota, Minneapolis, MN 55455, USA}

\author[0000-0002-9225-7756]{Alexander Wuollet Criswell}
\affiliation{University of Minnesota, Minneapolis, MN 55455, USA}

\author[0000-0003-4919-9017]{Behnam Darvish}
\affiliation{\caltech}

\author[0000-0001-8372-997X]{Kaustav K.\ Das}
\affiliation{\caltech}

\author[0000-0002-0786-7307]{Kishalay De}
\affiliation{MIT-Kavli Institute for Astrophysics and Space Research, 77 Massachusetts Ave., Cambridge, MA 02139, USA}

\author[0000-0003-0599-8407]{Luc Dessart}
\affiliation{Institut d'Astrophysique de Paris, CNRS-Sorbonne Universit\'e, 98 bis boulevard Arago, F-75014 Paris, France}

\author{Don Dixon}
\affiliation{Vanderbilt University, Nashville, TN 37235, USA}

\author{Bas Dorsman}
\affiliation{\GRAPPA}

\author[0000-0002-6871-1752]{Kareem El-Badry}
\affiliation{Center for Astrophysics $|$ Harvard \& Smithsonian, 60 Garden Street, Cambridge, MA 02138, USA}

\author{Christopher Evans}
\affiliation{UK Astronomy Technology Centre, Royal Observatory, Blackford Hill, Edinburgh, EH9 3HJ, UK}

\author[0000-0002-5956-851X]{K.\ E.\ Saavik Ford}
\affiliation{Department of Science, CUNY Borough of Manhattan Community College, 199 Chambers Street, New York, NY 10007, USA}

\author[0000-0002-4223-103X]{Christoffer Fremling}
\affiliation{\caltech}

\author[0000-0002-2761-3005]{Boris T.\ G\"ansicke}
\affiliation{Department of Physics, University of Warwick, Coventry, CV4 7AL, UK}

\author[0000-0003-3703-5154]{Suvi Gezari}
\affiliation{Space Telescope Science Institute, 3700 San Martin Drive, Baltimore, MD 21218, USA}

\author[0000-0002-6960-6911]{Y. G\"{o}tberg}
\affiliation{\carnegie}

\author[0000-0001-5417-2260]{Gregory M.\ Green}
\affiliation{Max Planck Institute for Astronomy, D-69117 Heidelberg, Germany}

\author[0000-0002-3168-0139]{Matthew J.\ Graham}
\affiliation{\caltech}

\author[0000-0002-1082-7496]{Marianne Heida}
\affiliation{ESO, Karl-Schwarzschild-Str 2, 85748 Garching b. München, Germany}

\author[0000-0002-9017-3567]{Anna Y.\ Q.\ Ho}
\affiliation{Miller Institute for Basic Research in Science, 468 Donner Lab, Berkeley, CA 94720, USA}

\author[0000-0002-3850-6651]{Amruta D.\ Jaodand}
\affiliation{\caltech}

\author[0000-0002-8828-6386]{Christopher M.\ Johns-Krull}
\affiliation{Department of Physics \& Astronomy, Rice University, 6100 Main St., Houston, TX, USA}

\author[0000-0002-5619-4938]{Mansi M.\ Kasliwal}
\affiliation{\caltech}

\author[0000-0003-3252-352X]{Margaret Lazzarini}
\affiliation{\caltech}

\author[0000-0002-1568-7461]{Wenbin Lu}
\affiliation{Department of Astrophysical Sciences, Princeton University, Princeton, NJ 08544, USA}

\author[0000-0003-4768-7586]{Raffaella Margutti}
\affiliation{\berkeley}

\author[0000-0001-5390-8563]{D.\ Christopher Martin}
\affiliation{\caltech}

\author[0000-0001-5382-6138]{Daniel Charles Masters}
\affiliation{\ipac}

\author[0000-0002-9726-0508]{Barry McKernan}
\affiliation{Department of Science, CUNY Borough of Manhattan Community College, 199 Chambers Street, New York, NY 10007, USA}

\author[0000-0003-4071-9346]{Ya\"{e}l Naz\'{e}}
\affiliation{\liege}

\author{Samaya M.\ Nissanke}
\affiliation{\GRAPPA} 

\author[0000-0002-3155-0385]{B.\ Parazin}
\affiliation{Northeastern University, Boston, MA 02115, USA}

\author[0000-0001-8472-1996]{Daniel A.\ Perley}
\affiliation{Astrophysics Research Institute, Liverpool John Moores
University, IC2, Liverpool Science Park, 
Liverpool L3
5RF, UK}

\author[0000-0002-9656-4032]{E.\ Sterl Phinney}
\affiliation{\caltech}

\author[0000-0001-6806-0673]{Anthony L.\ Piro}
\affiliation{\carnegie}

\author[0000-0002-9397-786X]{G.\ Raaijmakers}
\affiliation{\GRAPPA}

\author[0000-0003-4715-9871]{Gregor Rauw}
\affiliation{\liege}

\author[0000-0003-4189-9668]{Antonio C.\ Rodriguez}
\affiliation{\caltech}

\author[0000-0001-6656-4130]{Hugues Sana}
\affiliation{Institute of Astronomy, KU Leuven, Celestijnlaan 200D, 3001 Leuven, Belgium}

\author[0000-0002-9132-6561]{Peter Senchyna}
\affiliation{\carnegie}

\author[0000-0001-9898-5597]{Leo P.\ Singer}
\affiliation{Astroparticle Physics Laboratory, NASA Goddard Space Flight Center, Code 661, Greenbelt, MD 20771, USA}

\author[0000-0002-5547-3775]{Jessica J. Spake}
\affiliation{Division of Geological and Planetary Sciences, California Institute of Technology, Pasadena, CA 91125, USA}

\author[0000-0002-3481-9052]{Keivan G.\ Stassun}
\affiliation{Department of Physics and Astronomy, Vanderbilt University, Nashville, TN 37235, USA}

\author[0000-0003-2686-9241]{Daniel Stern}
\affiliation{Jet Propulsion Laboratory, California Institute of Technology, 4800 Oak Grove Drive, Pasadena, CA 91109, USA}

\author[0000-0002-7064-5424]{Harry I.\ Teplitz}
\affiliation{\ipac}

\author[0000-0002-6442-6030]{Daniel R.\ Weisz}
\affiliation{\berkeley}

\author[0000-0001-6747-8509]{Yuhan Yao}
\affiliation{\caltech}


\bigskip
\bigskip
\bigskip
\begin{abstract} 
\uvex\ is a proposed medium class Explorer mission designed to provide crucial missing capabilities that will address objectives central to a broad range of modern astrophysics.  The \uvex\ design has two co-aligned wide-field imagers operating in the FUV and NUV and a powerful broad band medium resolution spectrometer.  In its two-year baseline mission, \uvex\ will perform a multi-cadence synoptic all-sky survey 50/100 times deeper than \galex\ in the near/far ultraviolet, cadenced surveys of the Large and Small Magellanic Clouds, rapid target of opportunity follow-up, as well as spectroscopic followup of samples of stars and galaxies.   
The science program is built around three pillars.  First, \uvex\ will explore the low-mass, low-metallicity galaxy frontier through imaging and spectroscopic surveys that will probe key aspects of the evolution of galaxies by understanding how star formation and stellar evolution at low metallicities affect the growth and evolution of low-metallicity, low-mass galaxies in the local universe. Such galaxies contain half the mass in the local universe, and are analogs for the first galaxies, but observed at distances that make them accessible to detailed study. Second, \uvex\ will explore the dynamic universe through time-domain surveys and prompt spectroscopic followup capability will probe the environments, energetics, and emission processes in the early aftermaths of gravitational wave-discovered compact object mergers, discover hot, fast UV transients, and diagnose the early stages of stellar explosions. Finally, \uvex\ will become a key community resource by leaving a large all-sky legacy data set, enabling a wide range of scientific studies and filling a gap in the new generation of wide-field, sensitive optical and infrared surveys provided by the Rubin, \euclid, and \wfirst\ observatories.  This paper discusses the scientific potential of \uvex, and the broad scientific program.

\end{abstract}


\keywords{surveys --- ultraviolet: galaxies --- ultraviolet: stars --- ultraviolet: general --- instrumentation: photometers --- instrumentation: spectrographs}

\section{Introduction}

The Pathways to Discovery in Astronomy and Astrophysics for the 2020’s report (Astro2020, \citealt{NAP26141}) broadly considers the scientific landscape for the coming decades, identifying the areas of New Windows on the Dynamic Universe, and Unveiling the Drivers of Galaxy Growth as priorities motivating future investments. Astro2020 emphasizes the New Windows area because of the tremendous opportunities in multi-messenger astronomy enabled by the opening of the gravitational wave (GW) and particle windows, and the upcoming power of cadenced wide-field surveys with the Vera C. Rubin Observatory for finding explosive and time varying phenomena. Astro2020 also emphasizes the Galaxy Growth area, recognizing the power of the Atacama Large Millimeter/submillimeter Array (ALMA) and \textit{James Webb Space Telescope} (\jwst) to observe the seeds of galaxy growth in the early Universe, and of Rubin, the \textit{Nancy Grace Roman Space Telescope} and \euclid\ surveys for transforming our understanding of how galaxies and their contents grow and evolve over cosmic time.

Wide-field imaging and spectroscopy in the near- and far-ultraviolet (UV) bands are essential capabilities for addressing these priority science themes and for a rich and broad range of astrophysical studies. Wide-area surveys and followup spectroscopy are central for uncovering and understanding the predominantly low-mass galaxy population in the local universe, identifying local low-metallicity analogs of the seed galaxies in the early Universe that are being studied by \jwst. Finding the low-mass, low-metallicity galaxy population will significantly advance our understanding of star formation and evolution in such environments, providing templates for understanding measurements in the early Universe. Deep, wide-area UV surveys are an essential complement to Rubin, \euclid, and \wfirst\ for breaking fundamental degeneracies in photometric distance measurements, and for determining star formation rates for faint galaxies in the local universe. 

\begin{figure*}[hbt]
 \centering\includegraphics[width=16cm]{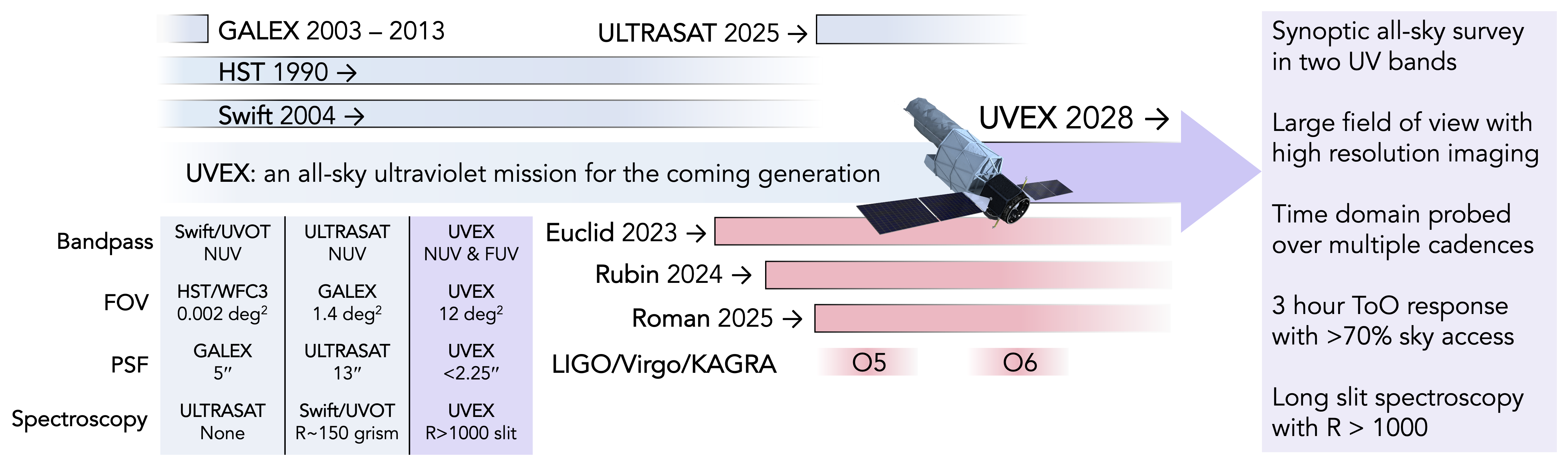}
  \caption{\small The timeline of \uvex\ in relation to other UV missions and the upcoming generation of multi-wavelength, multi-messenger facilities that \uvex\ will complement with its coverage of the dynamic UV sky.} 
 \label{fig:uvex_timeline}
\end{figure*}

In the time domain, the majority of explosive phenomena peak in the UV at early times, with this emission probing the hot initial phases of the expanding shocks and/or ejecta. Hot gas is rich in UV resonant line transitions, and early UV spectroscopy can measure elemental composition in the regions close to the progenitor. UV observations are therefore unique for both identifying transients in early phases through imaging, and diagnosing the aftermaths via photometry and spectroscopy. With a wealth of time-domain facilities operating in the coming decade, rapid UV followup of triggers from the collaboration of the Laser Interferometer Gravitational-Wave Observatory, Virgo, and Kamioka Gravitational Wave Detector (LIGO/Virgo/KAGRA), Rubin, the Deep Synoptic Array-110, the Square Kilometer Array (SKA), and more will open an entirely new view on the dynamic universe.  Considering the legacies of the \textit{Hubble Space Telescope} (\hst) and the \textit{Galaxy Evolution Explorer} (\galex), it is clear that UV observations are also broadly central to astrophysical studies of comets, planets, stars, galaxies, AGN, compact objects, as well as dust and gas in the Milky Way and beyond.

Figure~\ref{fig:uvex_timeline} shows a timeline of existing or approved missions in relation to \uvex\ and illustrates the key advances \uvex\ will make in UV capabilities, as well as the suite of multi-wavelength and multi-messenger facilities for which \uvex\ will provide necessary complementary UV data. No existing or upcoming mission\footnote{Here we include only missions that are adopted and/or are in development.} will have the deep, broadband (NUV and FUV) synoptic UV imaging, broadband spectroscopy, and rapid response target of opportunity (ToO) followup capabilities required to address many priority questions in astrophysics in the coming decades. While \hst\ continues to provide deep imaging and spectroscopy, it is over a narrow field of view (FOV) and with typical ToO turnaround times of two weeks, far too slow to study the relatively short-lived (few day-long) hot UV emission from early explosive stages.  Further, \hst\ is an aging observatory whose longevity is uncertain. The \textit{Neil Gehrels Swift Observatory} (\swift) UVOT instrument~\citep{Gehrels2004} has rapid turnaround capability, but covers only the NUV band, with limited FOV, sensitivity and spectral resolution. The upcoming \textit{Ultraviolet Transient Astronomical Satellite} (\ultrasat, \citealt{2014AJ....147...79S}) is designed for wide-field NUV imaging for transient identification, as well as followup of GW counterparts and other explosive phenomena, but lacks the depth, FUV coverage, spatial resolution, and spectroscopy to address many of the goals and objectives of the galaxy growth and dynamic universe themes.    

\begin{figure}[hbt]
 \centering\includegraphics[width=\columnwidth]{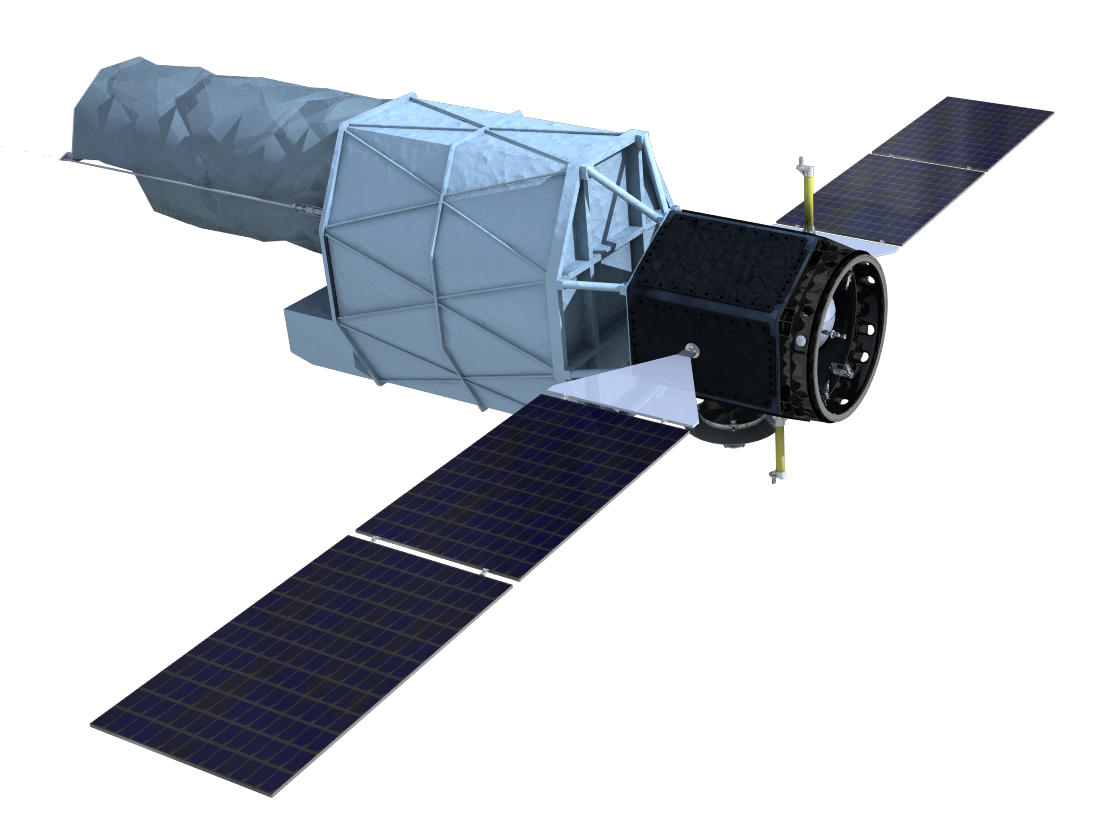}
  \caption{\small A render of the proposed \uvex\ telescope.}
 \label{fig:uvex_telescope}
\end{figure}

\begin{figure}[hbt]
 \centering\includegraphics[width=\columnwidth]{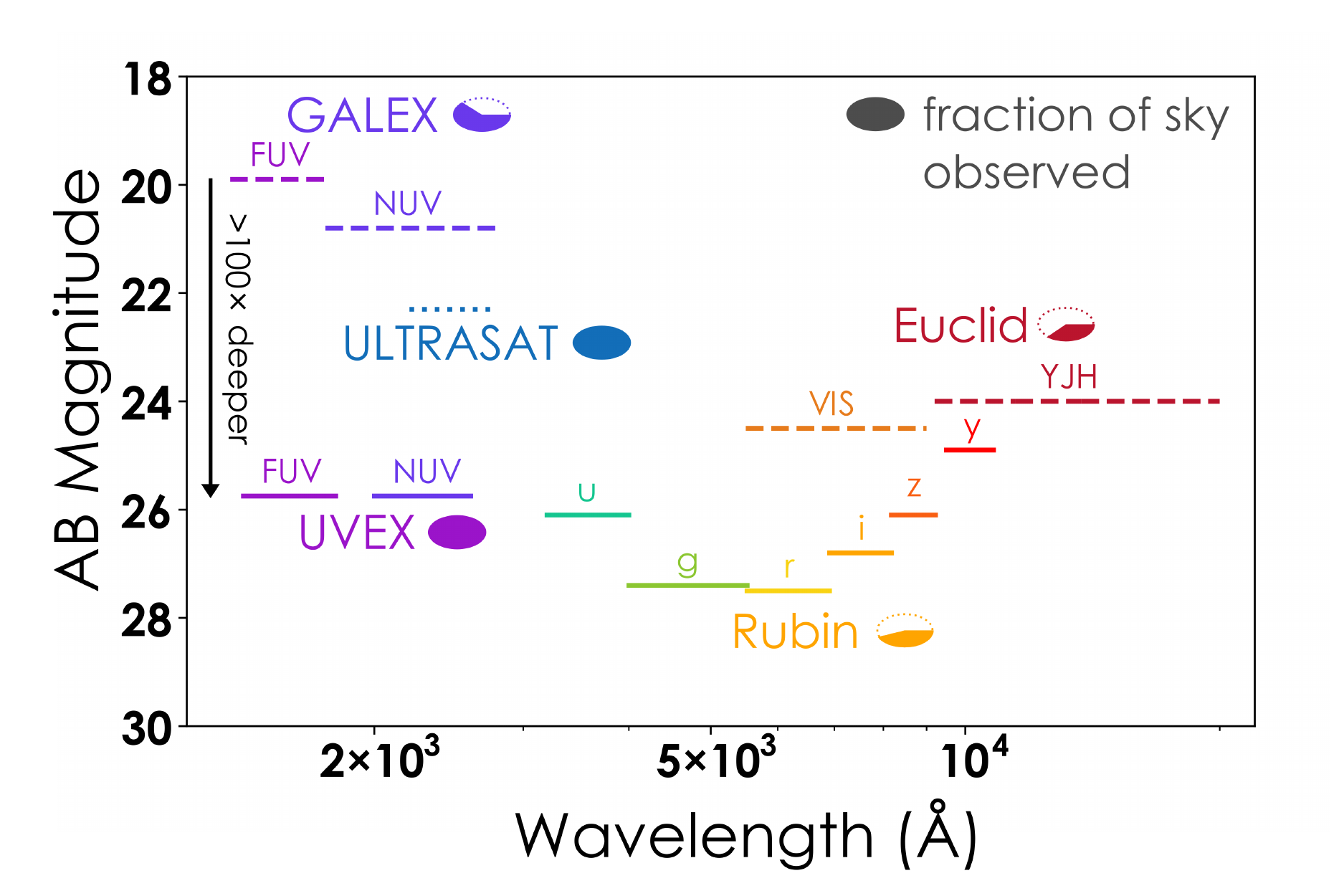}
  \caption{\small \uvex\ will provide deep, two-band UV data to
  complement planned deep, wide-field ($>$ 10,000 deg$^{2}$) optical and
  near-IR surveys by Rubin and \euclid.}
 \label{fig:survey_sensitivity}
\end{figure}

In this paper we describe the {\it Ultraviolet Explorer (UVEX)}, illustrated in Figure~\ref{fig:uvex_telescope}, a proposed Medium-Class Explorer (MIDEX) mission that is designed to provide crucial missing capabilities to address key elements of the Galaxy Growth and Dynamic Universe priority decadal science themes. \uvex\ will explore the low-mass and low-metallicity galaxy frontier, and will also provide a new view of the dynamic universe through cadenced wide-field NUV and FUV imaging for transient detection, as well as rapid photometric and spectroscopic followup of transients reported by other observatories. By performing a deep (50/100 times fainter than \galex\ in NUV/FUV) synoptic all-sky survey, \uvex\ will be an essential complement to the Rubin Observatory and \euclid\ mission (Figure~\ref{fig:survey_sensitivity}). \uvex\ will leave a legacy archive that will enable a rich range of community science investigations for decades to come.

We focus this paper on  describing the \uvex\ primary scientific goals and objectives; a followup paper will describe the details of the instrument and mission designs. Section~\ref{UVEX-description} provides a brief overview of the \uvex\ design, including the mission design, the optical telescope array, and the UV instrument module. Section~\ref{UVEX-pillars} describes the three primary science pillars: exploring the low-mass, low-metallicity galaxy frontier, providing a new window on the dynamic Universe, and contributing the missing piece to a legacy of deep, multi-wavelength, synoptic all-sky surveys. Finally, Section~\ref{UVEX-otherscience} describes a sample of the rich range of science beyond the primary mission that can be be pursued by the community using the \uvex\ archive.

\section{UVEX Design Overview}
\label{UVEX-description}

\uvex\ is designed for wide-field imaging simultaneously in FUV and NUV imaging bands, and for moderate resolution long-slit spectroscopy covering a broad FUV to NUV bandpass. Table~\ref{tab:missionparams} provides the observatory's top level design parameters. The FOV is large, and the design and orbit provide for a large field of regard (FoR) for high instantaneous sky accessibility. The instrument leverages modern high quantum efficiency CMOS detectors and coatings to achieve high sensitivity, eliminating the bright source constraints that precluded \galex\ from surveying the Galactic Plane and Magellanic clouds. Placed into a {\em TESS}-like lunar resonance orbit, \uvex\ achieves low and stable background and high observing efficiency. With frequent ground station contacts, \uvex\ has low data latency (data is transmitted every 6 hours) and rapid response to ToOs (average response time is three hours).

\begin{deluxetable}{lc}[htb]
\label{tab:missionparams}
\tablecaption{\uvex\ Mission Parameters}
 \tablewidth{0pt}
 \tablehead{\colhead{\uvex\ } & \colhead{Design} } 
 \startdata
 Imaging FOV & $3.5^{\circ} \times 3.5^{\circ}$ \\
 Image quality (HPD) & $\leq 2.25$\arcsec \\
 FUV imaging bandpass& 1390--1900\,\AA   \\
 NUV imaging bandpass & 2030--2700\,\AA \\
 Spectroscopy band/resolution & 1150--2650\,\AA, $R\geq1000$ \\
 Photometric sensitivity & $>$24.5 AB (SNR 5, 900 s) \tablenotemark{a} \\
 Sky survey depth &  $>$25.8 FUV and NUV \\
 Instantaneous sky accessibility  & $>70$\% \\
 Average ToO response & $<$3 hrs \\
 Data latency & $<$6.5 hrs \\   
 Orbit & elliptical  $17 R_e \times 15 R_e$ \\
 Orbital period  & 13.7 days\\
 Sun exclusion angle & $45^{\circ}$ \\
 Baseline mission duration & 2 years \\
 Target launch date &   Fall 2028  \\
 \enddata
 \tablenotetext{a}{for a point source in an extragalactic field with average background}
\end{deluxetable}

\subsection{\uvex\ Instrument}

Detector and coating technologies have advanced significantly over the last two decades. The development of backside-illuminated silicon solid state detectors (\citealt{nikzad1994, hoenk2009}) enables dramatic improvements in quantum efficiency over the microchannel plate detectors employed by \hst, \galex, and \swift. Solid state detectors also do not suffer from damage due to illumination by bright sources, enabling observations of the Galactic Plane and Magellanic Clouds, which have a high density of bright stars. By employing new detector and coating technologies, \uvex\ achieves significantly improved sensitivity compared to \galex with a relatively modest aperture consistent with a MIDEX-scale mission.

\uvex\ has a single instrument consisting of an UV-optimized optical telescope array (OTA) and the UV Instrument Module (UVIM). The OTA employs a standard all-reflective three-mirror anastigmat design, with an effective aperture of 75\,cm. At the UVIM entrance, a dichroic splits the light into FUV and NUV channels that are simultaneously imaged by two focal planes, each of which is composed of a $3 \times 3$ array of 4k $\times$ 4k CMOS detectors. Light for the long-slit spectrograph avoids the dichroic, passes through the slit, and is then dispersed by a grating onto a single CMOS detector. All the detectors and their modular readout electronics are identical except for individualized coatings for out-of-band light suppression. All imaging and spectroscopic data are compressed onboard and then sent to the ground once every six hours for scientific analysis. 

The OTA provides a field-averaged point spread function (PSF) with half-power diameter (HPD) \textless2.25\arcsec\ across the $3.5^{\circ}\times3.5^{\circ}$ FOV. This is Nyquist sampled by the 1\arcsec\ pixels. A standard 900-s dwell consists of 3 $\times$ 300-s imaging exposures, with each exposure read out in a high dynamic range (HDR) mode to avoid saturation on bright sources. The use of CMOS devices enables a shutterless rolling readout and a exposure duty cycle of 99\%.

The dichroic beamsplitter enables simultaneous imaging in both the FUV (1390--1900\,\AA) and NUV (2030--2700\,\AA) bands. Bandpass filters (\citealt{hennessy2021}) suppress the out-of-band background associated with zodiacal light and geocoronal \lya\ emission (e.g., \citealt{colina1996, leinert1998, murthy2014}). The aperture for the long-slit spectrograph is offset from the imaging field so that light bypasses the dichroic and is transmitted through a 1$^{\circ}$ long slit. The width of the fixed slit varies along its length, with apertures ranging from 2\arcsec\ to 16\arcsec. A grating disperses the light, with resulting spectral resolution ranging from $R \sim 1600$ at 1150\,\AA\ to $R \sim 3500$ at 2750\,\AA\ (for the portion of the slit with 2\arcsec\ width).

\begin{figure}
 \centering\includegraphics[width=\columnwidth]{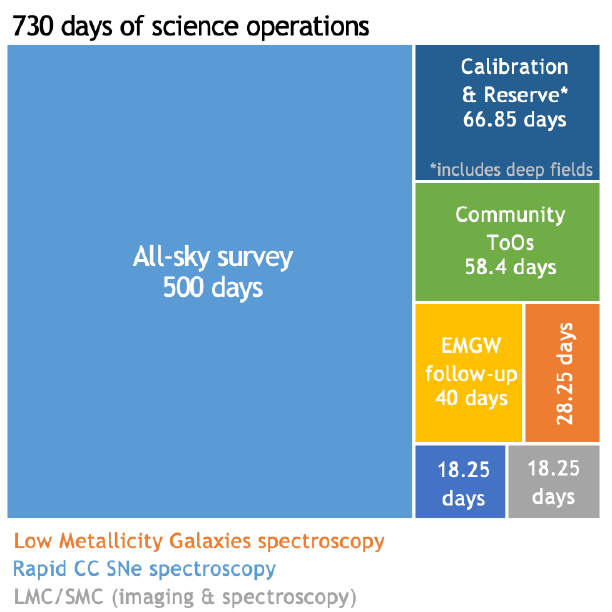}
  \caption{The anticipated distribution of \uvex observing time during the two-year baseline mission.} 
 \label{fig:mission}
\end{figure}

\section{The UVEX Baseline Science Program}
\label{UVEX-pillars}

Three scientific `pillars' provide the primary scientific motivation for \uvex, and define the primary requirements for the baseline mission design and observing program: (I) Exploring the low-mass, low-metallicity galaxy frontier; (II) Providing new views on the dynamic universe; and (III) Leaving a broad legacy of modern, deep synoptic surveys adding to the panchromatic richness of 21st century astrophysics.

During its two-year baseline mission, \uvex\ will undertake a deep, synoptic all-sky survey, as well as targeted, cadenced observations of the Large and Small Magellanic Clouds, spectroscopic followup of selected samples of stars and low-mass, low-metallicity galaxies, and ToO followup of GW events, supernovae, and transient events both discovered by \uvex and triggered by the community, many of which will be discovered by other facilities (Figure \ref{fig:mission}). These observations serve to fulfill the \uvex\ top level mission requirements, and will also provide a rich data set that will be promptly made available to the community for a wide variety of investigations.

In the sections below we describe the primary scientific pillars, as well as examples of the broad range of science that can be undertaken by the community using archival observations from the baseline mission. In Section~\ref{UVEX-otherscience} we describe additional scientific observations that could be pursued in an extended mission phase through Guest Observer (GO) observations.

\subsection{Pillar 1: The Low-Mass, Low-Metallicity Galaxy Frontier}
\label{UVEX-pillar1}

Our knowledge of galaxies and galaxy halos is based largely on
studies of those with masses comparable to or larger than the Milky
Way ($M\sim 5 \times 10^{10} \msun$). However, the properties of
these galaxies (e.g., Solar metallicity, dusty) are not well-matched
to the low-mass ($M\sim 10^{5}-10^{9}\msun$), low-metallicity (1--50\%
solar)  systems that dominate the hot, metal-poor early universe,
are thought to power cosmic reionization ($6 \lesssim z \lesssim 20$),
and are believed to be the majority of galaxies in the local universe.
Although low-mass, low-metallicity (LMLZ) systems are central to a
broad range of astrophysics, they are among the least explored
galaxy frontier, because only a small fraction of the large expected
\LMLZ galaxy population is known at any redshift and the birth,
evolution, and death of stars at low metallicities is poorly
understood.

\begin{figure}
 \centering\includegraphics[width=\columnwidth]{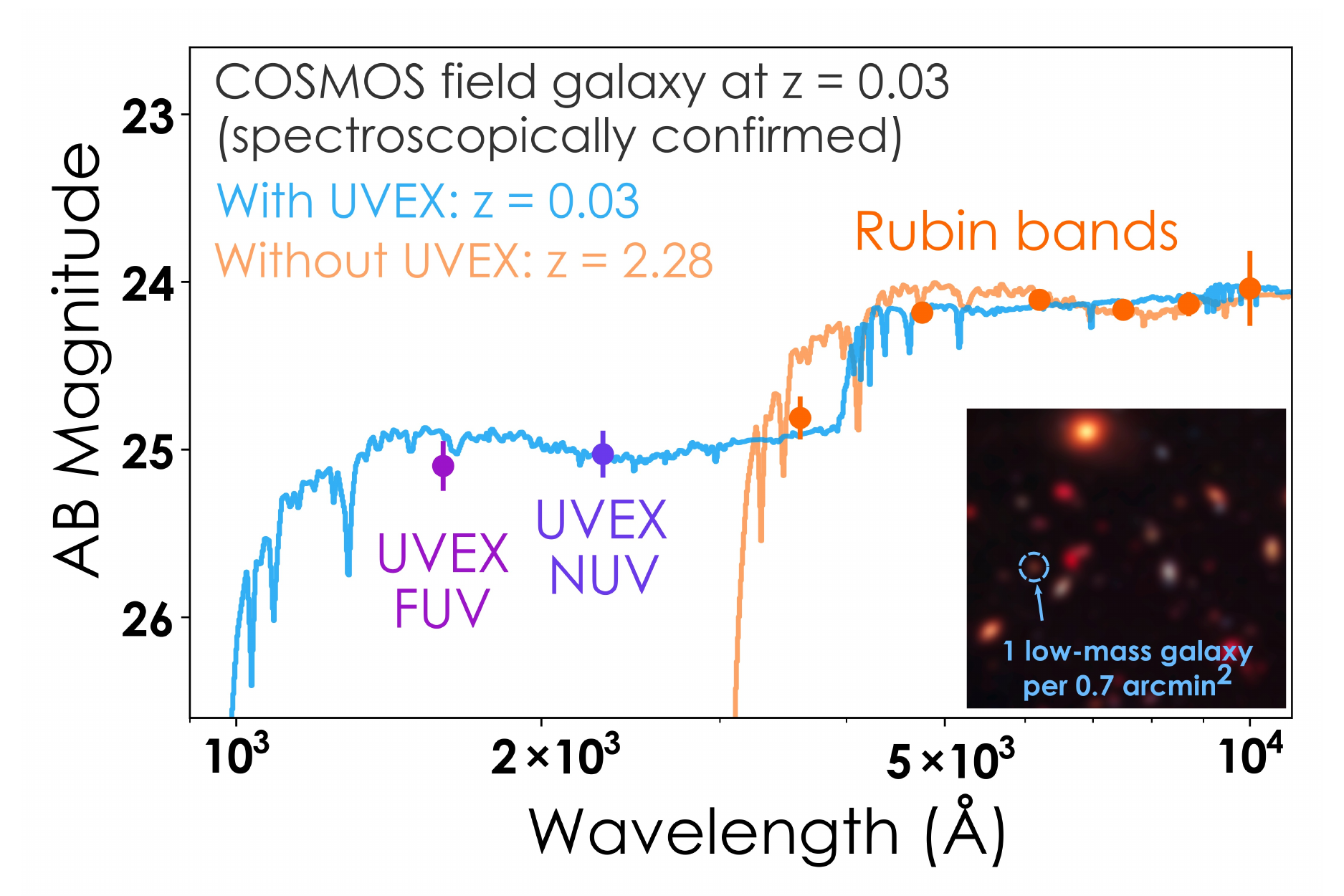}
  \caption{\small \uvex\ imaging picks out low-mass, $z<0.3$ galaxies
  by providing the crucial UV photometry needed to differentiate
  the Balmer break for a low-redshift system (blue) from the Lyman break
  in far more numerous high-redshift galaxies (orange).}
 \label{fig:galsed}
\end{figure}

A definitive study of local \LMLZ galaxies is key to understanding
the processes of galaxy formation, stellar evolution and death, and
the formation of compact objects in metal-poor environments. UV
heating and radiation pressure from massive metal-poor stars and
the explosive deaths of metal-poor single and binary stars regulate
star formation in ways that are different from the Milky Way
(\citealt{dessart17}) and may be responsible for initial mass function (IMF) variations and
bursty, chaotic star formation in many known \LMLZ systems (\citealt{uvex12}; \citealt{uvex13}, \citealt{uvex14}). With
reduced opacities, low-metallicity stars can grow to and maintain
larger masses and sizes, and many more interact in binary systems
\citep{uvex15,uvex11}. In close binaries, lower-opacity
winds reduce mass loss rates (e.g., \citealt{smith2014}) which leads
to significant, but poorly understood changes to stellar evolution
such as fewer red supergiants (\citealt{uvex14}), enhanced production
of single and binary black holes (\citealt{uvex15}), and a broader
diversity of supernova (SN) types.

Deep \name all-sky imaging, undertaken during the baseline mission, along with upcoming O/IR
surveys (e.g., Rubin, \textit{Euclid}), will help uncover millions of nearby
($D\leq$ 100 Mpc; $z\le 0.03$) \LMLZ galaxies that are predicted
to exist and measure their basic properties (e.g., mass, age, star
formation rate, dust).  Targeted \name spectroscopy of the youngest,
strongest star-forming \LMLZ systems will provide crucial rest-frame
UV nebular emission templates needed to interpret observations of
the first galaxies in the early Universe.  The unique and powerful
capabilities of \name will define the \LMLZ frontier for decades to
come.

\subsubsection{Finding the Low-Mass Galaxy Population in the Local Universe}
\label{sec:finding_lmlz}

With an all-sky survey $\geq50\times$ deeper than \galex, \name
will find the missing local population of LMLZ
galaxies. Key questions that will be addressed include: 
Where are the local LMLZ galaxies located? 
What are their properties? How do these properties vary with environment?

Our current census of the nearby \LMLZ galaxies is highly incomplete.
Within 100 Mpc ($z\le0.03$), theoretical matching of stellar and
dark matter halo masses predicts the existence of $\sim$10–200
million \LMLZ galaxies.  Yet only $\sim$20,000 \LMLZ galaxies are
known in this volume (\citealt{uvex10}; \citealt{uvex27}, \citealt{uvex28}), 
far fewer than even the most conservative theoretical estimates.

Finding local \LMLZ galaxies is challenging. They are intrinsically
faint and spread across the sky, tracing the local cosmic web from
low-density filaments and voids to high-density groups and clusters.
Their properties appear to vary with environment: star-forming and
gas-rich galaxies dominate the field, while groups and clusters host more
diverse populations. But these conclusions are based on small,
incomplete samples.

Mapping the nearby \LMLZ population requires a wide-area, sensitive
UV imaging survey. Degeneracies in age/metallicity/dust/redshift
mean that optical/infrared (O/IR) colors (e.g., from Rubin, 
\textit{Euclid}) alone cannot
distinguish local \LMLZ galaxies from more massive higher-redshift
($z>2$) interlopers (Figure \ref{fig:galsed}). Due to redshift and
intergalactic gas absorption, massive background galaxies have
little flux in rest-frame UV bands, whereas local \LMLZ field
galaxies have little dust and are predominantly star-forming
(e.g., \citealt{Geha2012}, \citealt{uvex13}), making them UV bright. 
Moreover, joint analysis of \textit{GALEX}+SDSS data shows that the UV is 
essential for measuring basic properties (e.g., mass, age, star 
formation rate) of nearby \LMLZ galaxies (\citealt{salim2016}).

\uvex\ will identify millions of \LMLZ galaxies within 100~Mpc down to $\mstar \sim 10^6 \msun$ when combined with optical imaging from Rubin and Northern-hemisphere counterparts (e.g., UNIONS, DESI Legacy Survey).  Due to the lack of deep, wide-field observations, we base the expected numbers that \uvex\ will find on theoretical models, which come with large uncertainties (e.g., \citealt{garrison-kimmel2017}). We estimate the total number of low-mass galaxies ($\mstar \sim 10^6-10^9 \msun$) within 100 Mpc using two different stellar halo mass relationships (SHMs; e.g., \citealt{behroozi2013}, \citealt{conroy2015}) that bracket the accepted range of low-mass galaxy formation models and current empirical constraints ($M_{\star} \propto\ M_{\rm halo}^{\alpha}$, where $\alpha \sim 1.6 - 3.1$; e.g.,  \citealt{garrison-kimmel2017}). These relations predict between $\sim 1.5 \times 10^7$ and $\sim 3\times10^8$ luminous $\mstar >10^6 \msun$ galaxies within 100 Mpc. The lowest-mass galaxies dominate the population by number due to the steepness of the SHM relationship. 

The number of dark matter halos predicted by SHMs agrees well with large-volume N-body simulations at slightly higher masses (\citealt{eliahi2018}), which unfortunately cannot be used directly for our estimates because no simulations include large enough volumes and low enough halo mass ranges. Local galaxies with $\mstar< 10^6~\msun$ are often quenched in groups due to environmental effects or reionization in the early Universe. However, most local galaxies with $\mstar~ > 10^6 ~\msun$ are known to be star-forming. For slightly higher mass galaxies ($\mstar~>10^7 \msun$), \citet{Geha2012} estimates that only 0.06\% of field galaxies and 24\% of galaxies associated with groups lack recent star formation. Smaller, less complete samples of low-mass galaxies within 10 Mpc yield similar estimates (e.g., \citealt{uvex31}). Conservatively, we assume that $\sim$1/3 of galaxies with $\mstar~>10^6~\msun$ are quenched and thus undetectable in the UV due to a lack of massive stars. In total, our estimate is that there are 10–200 million star-forming galaxies with $\mstar>10^6\msun$ within 100 Mpc, in good agreement with estimates from other approaches aimed at finding low-mass galaxies (e.g., via transients; \citealt{conroy2015}). \uvex, through its all-sky survey, is designed to detect all such galaxies in the extragalactic sky.

A large sample is essential for anchoring the SHM relation, providing 
the 3D maps of the low-mass, low-density Universe (probing poorly 
surveyed filaments and voids), enabling the first large-scale study 
of how the lowest-mass halos evolve as a function of environment, 
and finding the most extreme examples (lowest metallicity and 
youngest) for followup (e.g., \citealt{uvex29}).
For example, if low mass galaxies continue to form, the probability
of finding one within 10 Myr of formation is $\sim$10 Myr/10 Gyr,
or 1/1000. Thus we require a very large sample to isolate the rarest
and most interesting forming galaxies.

Adequately sampling rare \LMLZ galaxies,
and studying large-scale cosmic structures requires surveying the
entire extragalactic sky (i.e., 20,000 deg$^2$). A
typical $10^6\,\msun$ star-forming galaxy has $M(UV)=-10$\,mag (AB).  
Detecting such galaxies to 100 Mpc, and constraining star formation rates
requires SNR $\ge5$ to a depth of $m_{UV}=25$ in both FUV and NUV.  
We estimated this minimum luminosity by drawing on 
\galex\ studies of star-forming dwarfs in the Local Volume (e.g., \citealt{uvex_35}). 
Pairing UV luminosities from \citet{uvex_35} with UV+optical stellar masses from \citet{uvex14}, we find that an actievly star-forming galaxy with $\mstar \sim 10^6 \msun$  has a typical UV luminosity of M$_{FUV}$ of –13 mag, which includes Milky Way foreground and internal extinction corrections (\citealt{uvex_35, uvex14}). Then, as shown in \citet{uvex14}, due to long duty cycles and the bursty nature of star formation, galaxies with $\mstar < 10^7 \msun$ spend $\approx$ 80–90\% of their time in post-burst states (i.e., with little or no H$\alpha$ and reduced UV luminosities) that can make star-forming galaxies with $\mstar = 10^6 \msun$ as faint as $M_{FUV} = –10 $ mag. Thus, in order to capture a large, unbiased sample of low-mass, star-forming galaxies (i.e., not just the brightest, most active systems), \uvex must reach this sensitivity limit over the extragalactic sky.

We estimate the metallicity range (1–50\% $Z_{\odot}$) for our LMLZ sample based on the well-established relationship between gas-phase metallicity and stellar mass (see \citealt{berg16} for an example), the latter of which is estimated as described above.

Figure \ref{fig:survey_sensitivity} shows the estimated depth of the \name
all-sky UV survey (in a typical high-latitude extragalactic field),
highlighting both its generational improvement enabling it to explore 
the low-mass nearby galaxy population, and its complementarity
to modern wide-area O/IR surveys. \name reaches 4–5 mag fainter
than the \galex wide area surveys over the entire extragalactic sky.
\galex reached $m_{UV}\sim25$ over only 80 deg$^2$, far too small
an area to obtain a global census of low-mass systems (\citealt{uvex29}).
The \hst FOV is far too small to undertake the needed survey. Many
\LMLZ galaxies are too faint ($m>24$) for optical spectroscopy,
while because of the bursty star formation histories narrow-band
imaging (e.g., H$\alpha$) is a less reliable tracer of star-formation
in many \LMLZ galaxies compared to the UV.

There is, and will continue to be, rich complementary data to \uvex\ for characterizing and studying the large local LMLZ population. As previously mentioned, wide area O/IR surveys (e.g., Rubin, UNIONS) will immediately help \uvex\ with the first steps in making fundamental progress on discovery and characterization (e.g., masses, star formation rates [SFRs], distances). Beyond the optical and IR regime, there is a wealth of multi-wavelength data that will further enhance the science delivered by \uvex. For example, \uvex\ imaging when paired with H~I and H$\alpha$ will provide new insight into the process by which gas is turned into stars at low metallicities. Pathfinder examples of such science have already been shown using \galex, but the number of systems are small (e.g., \citealt{donovan15}). When combined with H~I (e.g., from ALFALFA, SKA and its pathfinders), \uvex\ SFRs will provide new insight into the star formation process in regimes in which H~I dominates over H~II, regimes in which standard relations (e.g., Kennicutt-Schmidt) are not applicable (e.g. \citealt{uvex11}). The process of converting gas into stars at such low metallicities, and how it may be affected by surrounding environments (e.g., groups vs voids; \citealt{uvex29}), is poorly understood observationally and theoretically. Similarly, the millions of LMLZ galaxies discovered by \uvex\ will enable follow-up observations of rare objects (as in \citealt{salzer20}). When combined with existing and planned datasets in the optical, near-IR and radio, \uvex\ will enable fundamental and transformative progress in a variety of areas, such as measuring the stellar mass function and reducing the 2 dex uncertainty in the SHM (e.g., \citealt{garrison-kimmel2017}), exploring claims of systematic variations in the high-mass IMF (e.g. \citealt{uvex11,uvex13,uvex14}), quantifying the bursty nature of star formation and its duty cycle (e.g. \citealt{uvex12}), determining galaxy-wide dust properties at low metallicities (e.g. \citealt{salim18}), identifying rare classes of LMLZ systems (e.g., various analogs to higher-redshift systems such as green peas, extreme bursting systems, rare massive star populations; \citealt{uvex46,uvex40,uvex50}), and much more as highlighted in Section~\ref{sec:GalaxyFormation}.  

\subsubsection{Nebular Emission in the Lowest Mass, Lowest Metallicity Systems}

\name will diagnose LMLZ galaxies dominated by
radiation from hot stars, and polluted by early generations of SN.
This is essential to the quest to understand the first galaxies at
high redshift. Key questions that will be addressed include: What is the radiation environment created by the first generation of stars?  What are the feedback processes in high-z star forming regions at very low metallicity?

What we learn from ALMA, \jwst, and the ELTs about the first galaxies
and their stars will come from integrated nebular emission lines.
Emission lines are powerful diagnostics of the baryonic processes
that shape galaxy evolution (e.g., star-formation history, SN
feedback, ionizing radiation field). However, interpreting emission
lines from the first galaxies is challenging due to a lack of local
anchors.  The extremely low metallicities, strong radiation fields,
and high star formation rates of the first galaxies are not captured
in typical nearby calibration samples of resolved stars (e.g., Milky
Way, SMC) or integrated light observations of most nearby low-metallicity
dwarf galaxies.

To understand the extreme environments of high-redshift galaxies we need 
integrated spectral templates of extremely low-metallicity, strongly
star-forming, young, nearby galaxies. These spectra will anchor the stellar 
population synthesis models used to interpret observations of primordial 
galaxies. Rest-frame UV spectra are particularly urgent, as they contain 
several key diagnostic nebular lines, such as \ion{He}{2} $\lambda1640$, 
\ion{O}{3}] $\lambda\lambda1661,1666$, \ion{Si}{3}] $\lambda\lambda1883,1892$,
and \ion{C}{3}] $\lambda\lambda1907,1909$, all of which are being detected 
in high-redshift observations. The most powerful constraints include the C/O 
ratio (sensitive to the star formation history, SN feedback, 
gas-phase metallicity, and age of the current bust of star formation) and
the \ion{C}{3}]/\ion{O}{3}] vs. \ion{C}{4}/\ion{He}{2} diagnostic (sensitive 
to the shape of the ionizing spectrum).

\begin{figure}
 \centering\includegraphics[width=\columnwidth]{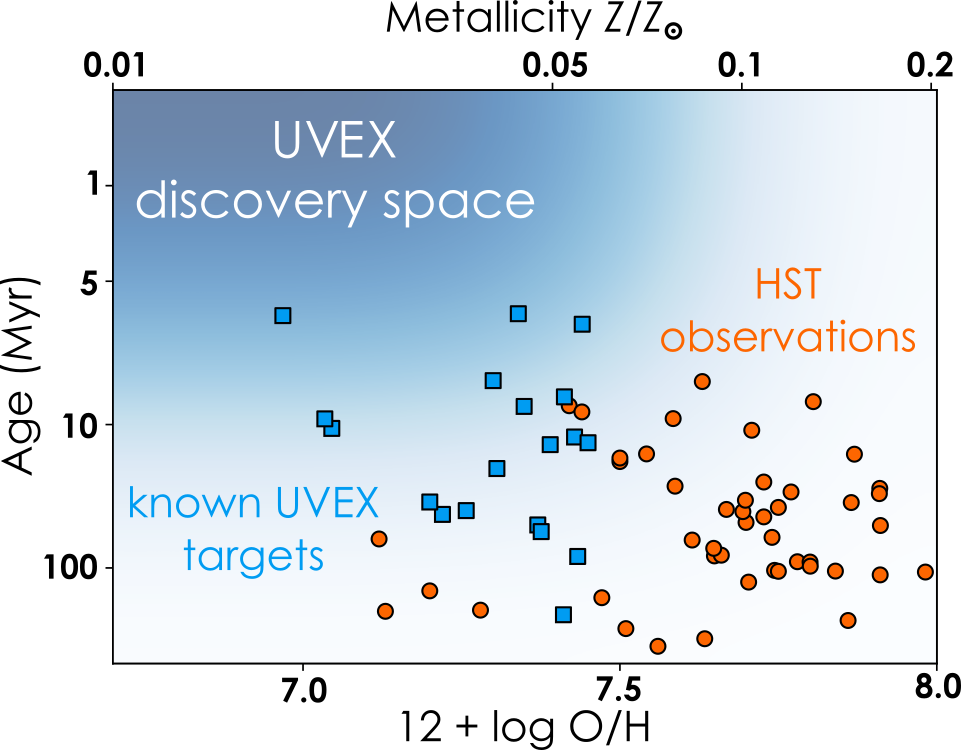}
  \caption{\small \uvex\ will obtain spectra of the lowest-metallicity
  galaxies in the local universe. Orange dots are measurements from
  \hst, blue squares indicate the known sample selected for \uvex\
  followup. \hst\ can still make some progress in the lighter shaded
  blue regions but  probing the darker blue region requires \uvex.}
 \label{fig:age_metallicity}
\end{figure}

\hst has opened this field by obtaining UV spectra of several dozen
local, strongly star-forming, modestly low-metallicity systems.
However, progress with \hst in understanding the lowest-metallicity and 
youngest galaxies is fundamentally limited by several factors. For one, 
only a very small number of systems at metallicities below $\sim 5\%~Z_\odot$
($12+\log\mathrm{O/H}<7.35$: in the regime of 
extremely metal-poor galaxies; XMPs)
with light-weighted effective ages $<10$~Myr are known (Figure
\ref{fig:age_metallicity}). This is a product of severe observational biases:
most XMPs known were discovered in SDSS spectroscopy, which was broadly
restricted to only the brightest galaxies in continuum magnitudes. And second,
\hst UV spectroscopic throughput is limiting: the available detector/grating
combinations on \hst/STIS and COS have throughputs that fall precipitously 
redwards of $\sim 1950$~\AA, limiting studies to a handful of the brightest and
lowest-redshift LMLZ systems. In contrast, \uvex\ is optimized for
sensitivity and resolution across the range from 1500–2000\,\AA,
making it ideal for acquiring the sorely lacking UV spectra in
low-redshift \LMLZ galaxies.

\uvex\ will revolutionize this area: the all-sky survey will reveal substantial numbers of new metal-poor and young LMLZ galaxies. We estimate the total number of galaxies with $10^6 \msun < M_\star < 10^7 \msun$ and low metallicities 12+log(O/H) $< 7.35$ (e.g., \citealt{berg16}) within 100 Mpc to be $\sim9\times10^6$ to $240\times10^6$, using the same SHM relations as in Section~\ref{sec:finding_lmlz}. 
Of these, $\sim0.4$\% are expected to be dominated by young stars that are just entering a burst phase (i.e., ensuring the light-weighted ages will be $<10$ Myr; \citealt{stark16}, \citealt{williams18}, \citealt{meurer09}, \citealt{tweed18}). 
This suggests that \uvex\ will discover $2.8\times10^4$ to $5.2\times10^5$ galaxies in the largely unexplored region of parameter space illustrated in Figure \ref{fig:age_metallicity}.
There are $\sim$10 such systems currently known from the SDSS in this area of parameter space, which is likely a lower limit due to various selection effects in detecting and characterizing SDSS galaxies and the known high degree of incompleteness in SDSS for galaxies with $M_\star < 10^7 \msun$ (e.g., \citealt{Geha2012}). 

As part of the baseline mission, \uvex\ will take spectra of 100 strongly star-forming low-mass galaxies over a range of metallicity and age. The improved sensitivity of the spectrograph onboard \uvex\ has the potential to increase the nearby LMLZ spectroscopic sample by orders-of-magnitude, providing a platform for fundamental progress in the study of such young systems that are our best  analogs to the chemically-young systems of the early Universe (e.g., \citealt{stark16}, \citealt{williams18}, \citealt{endsley21}). Twenty very low-metallicity galaxies too faint for \hst but accessible to \uvex have already been identified (\citealt{uvex44, uvex50, uvex51, senchyna19}; see Figure~\ref{fig:age_metallicity}). New systems will be selected from the first year of \uvex\ imaging based on UV-optical spectral energy distributions (SEDs), good proxies for age, metallicity, and dust \citep{uvex41}, and confirmed with follow-up optical spectroscopy. \uvex\ exposure time calculator (ETC) simulations indicate 2–80 ksec exposures for the existing sample of 20. Exposures for the \uvex sample will be at the lower end of this range as bright systems will be prioritized.  We provide more details (e.g., extinction, targeted excitation states) on the sample and calculations for the UV line measurements in Appendix~\ref{sec:co_ratios}.

\begin{figure}
 \centering\includegraphics[width=\columnwidth]{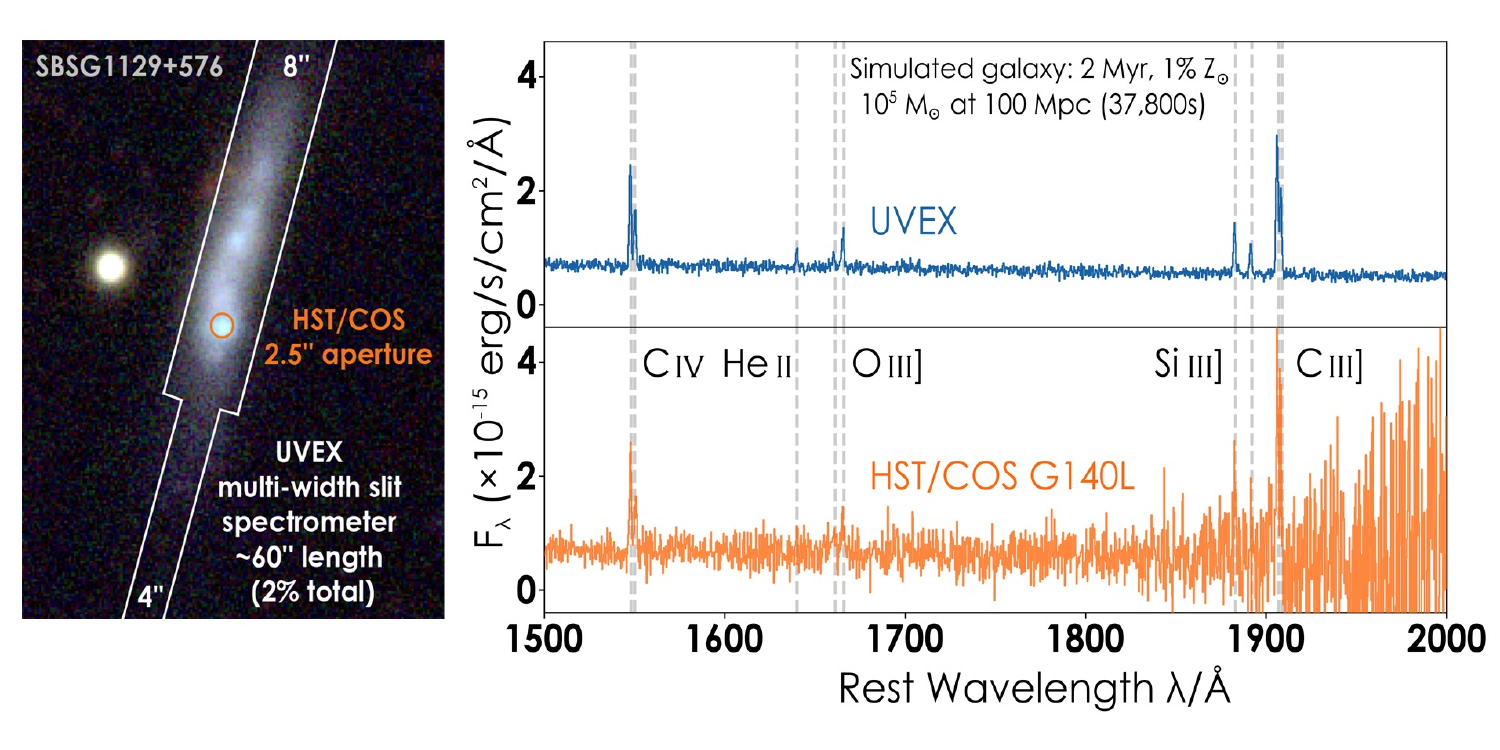}
  \caption{\small The \uvex\ spectrograph is optimized for observing
  nebular emission lines over the crucial wavelength range of
  1500--2000\AA. (Left)
  A \textit{gri} image of a local extremely metal-poor galaxy showing
  \hst/COS and \uvex\ spectroscopic apertures. (Right) A simulated
  \uvex\ spectrum of a $\sim$1\% Z$_{\odot}$ low-mass galaxy at 100 Mpc
  compared to \hst/COS for similar integration times.}
 \label{fig:uvex_spectrum_sim}
\end{figure}

\subsubsection{The Magellanic Clouds: A Laboratory for Low-Metallicity Stars}
\label{sec:massivestars}
\begin{figure*}[htbp]
 \centering
 \includegraphics[width=\textwidth]{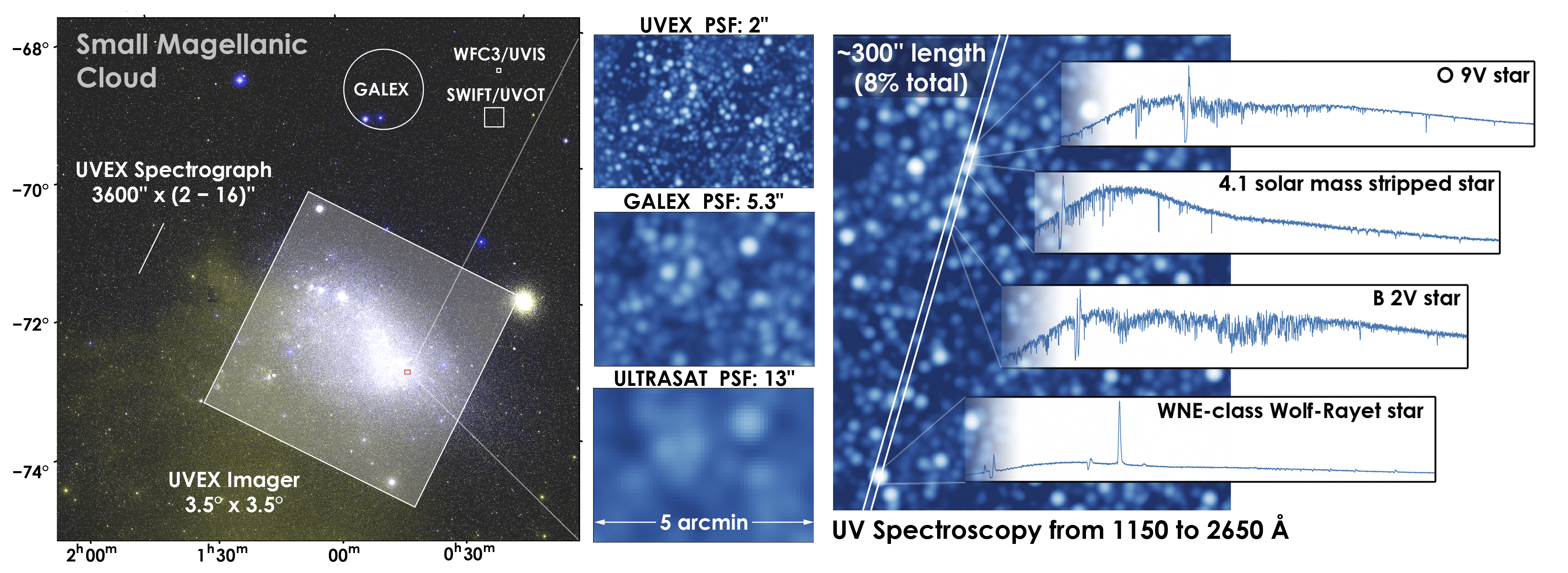}
  \caption{\small The \uvex\ survey of the Magellanic Clouds
  illustrated.  (Left) The \uvex\ field of view overlaid on an optical
  image of SMC. Select other UV facility footprints are shown for
  comparison.  At 12 deg$^2$, \uvex\ will be able to image the main
  bodies of the LMC and SMC in just seven pointings. \uvex\ will
  visit the LMC and SMC weekly over its prime mission to obtain
  deep and cadenced imaging as well as spectroscopy for $>1000$ hot
  stellar systems.  (Middle) A simulated image in a central portion
  of the SMC illustrating the spatial resolving power of \uvex.
  The PSF of \uvex\ ensures it will be able to resolve stars in all
  but the most crowded regions; a dramatic improvement over other
  UV facilities.  (Right) Simulated \uvex\ spectroscopy for a handful
  of objects that \uvex\ will observe during its prime mission.
  The SNR and spectral resolution of \uvex\ are comparable to
  \hst/COS G140L.  }
 \label{fig:uvex_smc}
\end{figure*}

Massive star evolution is key for understanding galaxy evolution, and mass loss and multiplicity are key for understanding massive stars. Mass loss from stellar winds in massive single and binary stars are driven by high energy radiation pressure which is uniquely observed through absorption line spectroscopy of resonance lines in the UV -- only stars with the very highest mass loss rates exhibit wind features in the optical \citep{hillier2020}.
Metallicity is a key parameter in wind-driven mass loss due to the strong metallicity dependence of wind opacity \citep{vinkMetallicitydependentWindParameter2021}. However, at fixed metallicity and spectral type, there is significant variation in wind properties. These variations are currently not understood and may reflect variation in the past history and current evolutionary stage of the objects, including possible consequences of the presence of a nearby companion and/or prior interaction. Indeed, the binary fraction increases with stellar mass while interactions between close binaries are more common at lower metallicities as stars can grow to larger sizes \citep{uvex16}.

Binary interactions have dramatic evolutionary consequences. A third of all massive stars are expected to lose their hydrogen-rich envelopes via mass transfer or common envelope ejection \citep{uvex15,demink14,uvex16}, leaving hot and compact helium cores exposed (i.e., a stripped star). As prolific sources of ionizing radiation, low-metallicity stripped stars helped power cosmic reionization \citep{2016MNRAS.456..485S,2019A&A...629A.134G,2020ApJ...904...56G,2020ApJ...901...72S}. Binary stripped stars are thought to be precursors to many merging compact objects (e.g., merging neutron stars [NSs]; \citealt{2017ApJ...846..170T,2020PASA...37...38V,2020ApJ...888L..10Y}); while extremely compact stripped binaries may themselves be sources of GWs detectable with the \textit{Laser Interferometer Space Antenna} (\textit{LISA}) \citep{2020ApJ...904...56G,2004MNRAS.349..181N,2020A&A...634A.126W}. However, our knowledge of stripped stellar systems is essentially unconstrained by data. Only a handful candidate stripped stars have been identified so far, most of them in the form of an sub-dwarf O star (sdO) paired with a rapidly rotating Be star \citep{thaller95,peters08,peters13,wang17,wang21}. Some objects have been caught soon after the mass transfer before the stripped star has reached thermal equilibrium and is still bloated and cooler than its sdO counterpart \citep{shenar20,bodensteiner20,frost22}. In many cases, these systems cannot be sufficiently characterized from the optical to constrain evolutionary models, while stripped stars descending from truly massive stars have remain elusive.

Understanding how massive single and binary stars, and their descendants, shape galaxy evolution requires us to understand how their wind properties change with metallicity. Current models are insufficiently constrained so that the way forward relies on observations of large samples at lower metallicities. Fortunately, nature has provided us with nearby, sub-solar metallicity laboratories that are close enough to be spatially resolved: the LMC ([Fe/H] = –0.5) and SMC ([Fe/H] = –1.0) \citep{mcconnachie2012}. 

\begin{figure}[t!]
 \centering\includegraphics[width=\columnwidth]{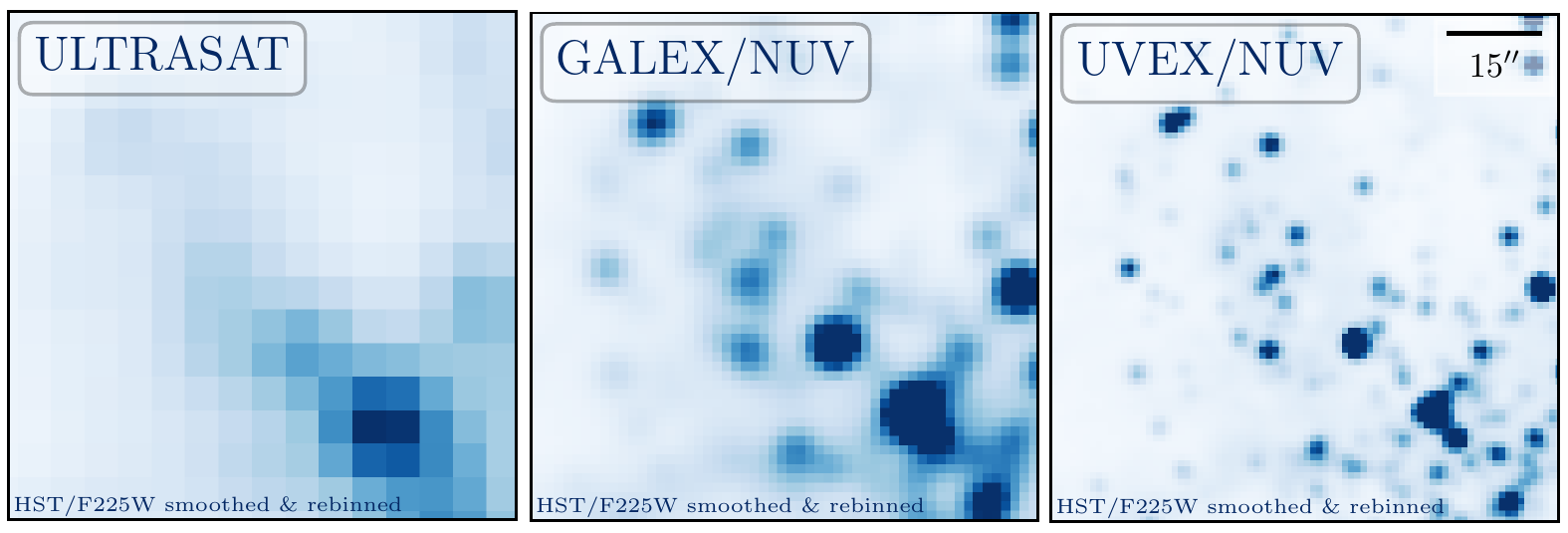}
  \caption{\small The spatial resolution of \uvex\ compared to select UV facilities. This \hst\ UV image on the outskirts of 30~Doradus (from program GO-11360) has been convolved with the PSF of each facility. 
 This stellar field is unresolved by  \ultrasat, marginally resolved by \galex, and resolved by \uvex. As other regions in the LMC and SMC are less crowded, \uvex\ will reveal the detailed distribution of stars without being significantly affected by crowding throughout most of the LMC and SMC. }
 \label{fig:uvex_lmc_image}
\end{figure}

By exploiting these nearest low-metallicity laboratories where individual stars can be resolved (see Figure~\ref{fig:uvex_smc} and Figure~\ref{fig:uvex_lmc_image}), \uvex\ will probe the mass-loss driven evolution of hot and massive stars as a function of metallicity and binarity across the Hertzsprung-Russell (H-R) diagram. As part of the baseline mission, \uvex\ will (i) perform deep, cadenced imaging of the LMC and SMC, making a near-complete census of hot and stripped stars and (ii) obtain UV spectra of 100 \uvex-identified stripped and 1,000 hot (O and B) single and binary stars, measuring wind velocities using the blue edge of P-Cygni profiles \citep{crowther2016}. Binarity will be determined from \uvex\ cadenced imaging and/or ground-based optical surveys \citep{uvex67, cioni2019}. Key questions \uvex\ will address: How does mass loss from hot, massive stars depend on their properties? What are the demographics of stripped stars, and what influences their formation?

\begin{figure*}
 \centering
  \includegraphics[width=\textwidth]{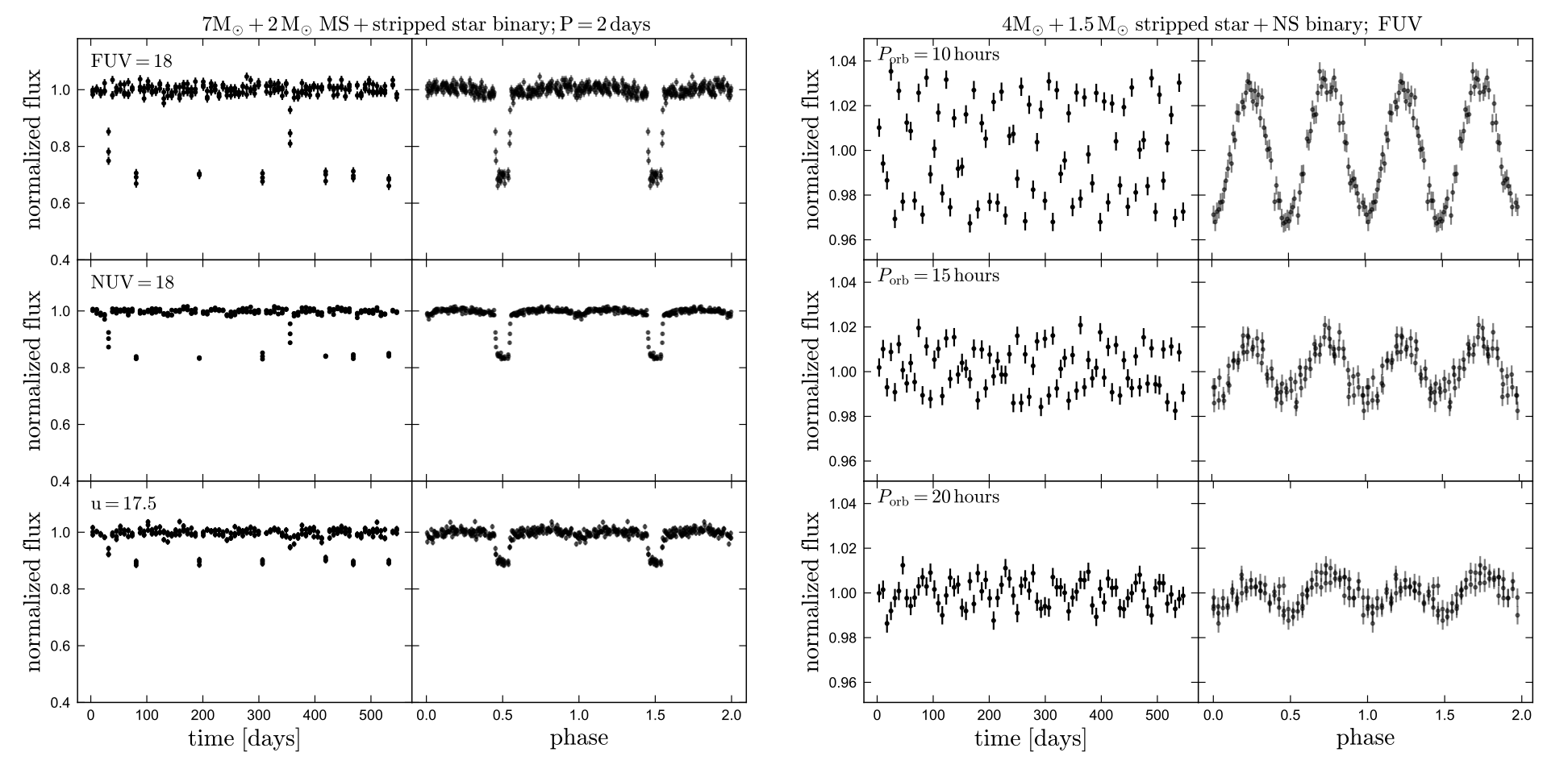}
   \caption{\small Predicted light curves for two types of binaries
   containing stripped stars detectable with \uvex time-series
   photometry in the Magellanic Clouds. Left two panels show an
   eclipsing binary containing a $7 \msun$ main-sequence star
   and a $2 \msun$ stripped star. Deep eclipses are apparent
   in the UV, where the stripped star contributes almost half of
   the total light. Shallower eclipses are apparent in the optical
   (e.g. Rubin/LSST $u$-band), but these would be misinterpreted
   as being due to a normal main-sequence companion without the UV
   data. Right panels show a $4 \msun$ stripped star with a
   neutron star companion at different orbital periods in the FUV;
   the predicted variability is due to a combination of ellipsoidal
   variability and Doppler beaming. Such a system will evolve to
   become a binary neutron star.}
 \label{fig:example_light_curves}
\end{figure*}

\begin{figure}[t!]
 \centering\includegraphics[width=\columnwidth]{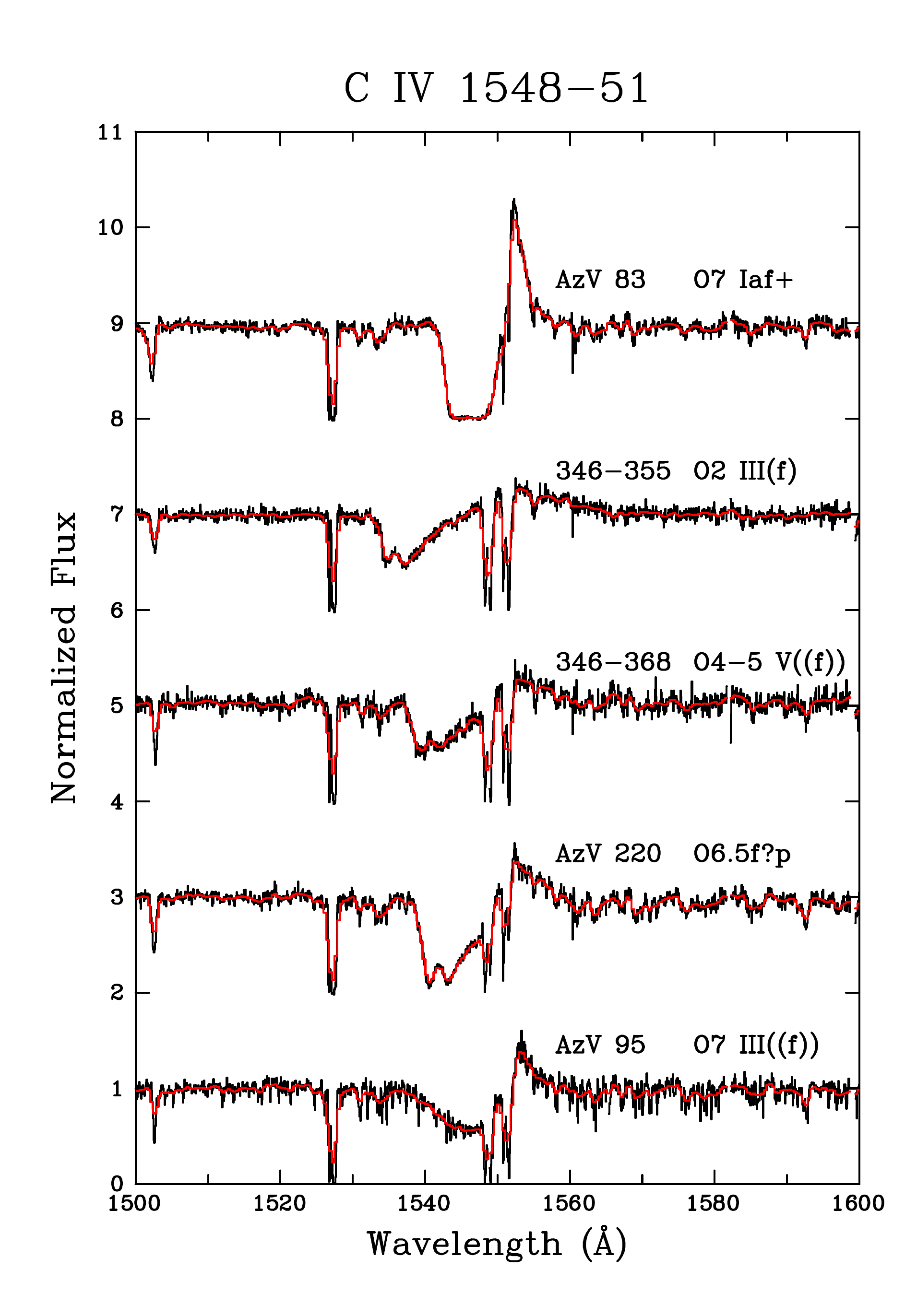}
  \caption{\small Example spectra showing the C~IV doublet in 
O-type stars in the SMC. STIS/E140M observations (\citealt{walborn2000})
are shown in black. The same spectra smoothed to \uvex\ resolution are shown in red. Because key features of the UV resonance lines (e.g., terminal velocities, peak intensities) are so prominent, they are preserved at the resolution of \uvex.  The low metallicity of the SMC means these wind features are among the weakest and least pronounced.  The UV line fidelity is equally well-preserved for stronger winds typically found in the higher metallicity LMC (\citealt{crowther2016}).}
 \label{fig:CIV_smc}
\end{figure}

\begin{figure}[t!]
 \centering\includegraphics[width=\columnwidth]{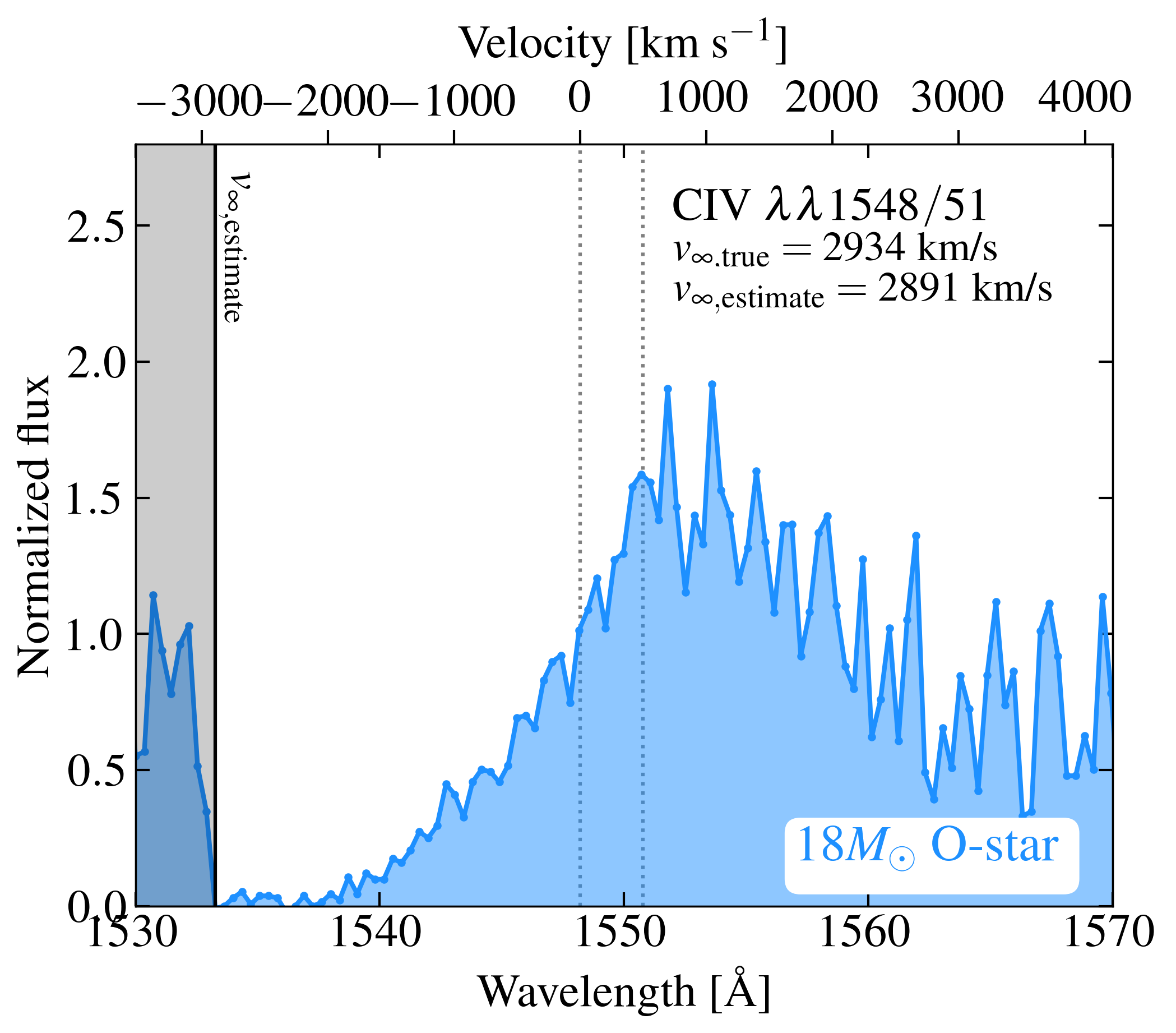}
  \caption{\small The simulated \uvex\ spectrum (\textit{SNR} = 10 at the C~IV doublet) of a typical LMC metallicity O-star spectrum with an average wind velocity, zoomed in on the C~IV resonance line.  By fitting for the blue edge of the C~IV P-Cygni profile, we recover the wind velocity to within a few percent. \uvex\ will enable the accurate determination of massive star wind speeds for 1000 hot, massive single and binary stars in the LMC and SMC.}
 \label{fig:ostar_wind}
\end{figure}

\begin{figure}[h]
 \centering\includegraphics[width=\columnwidth]{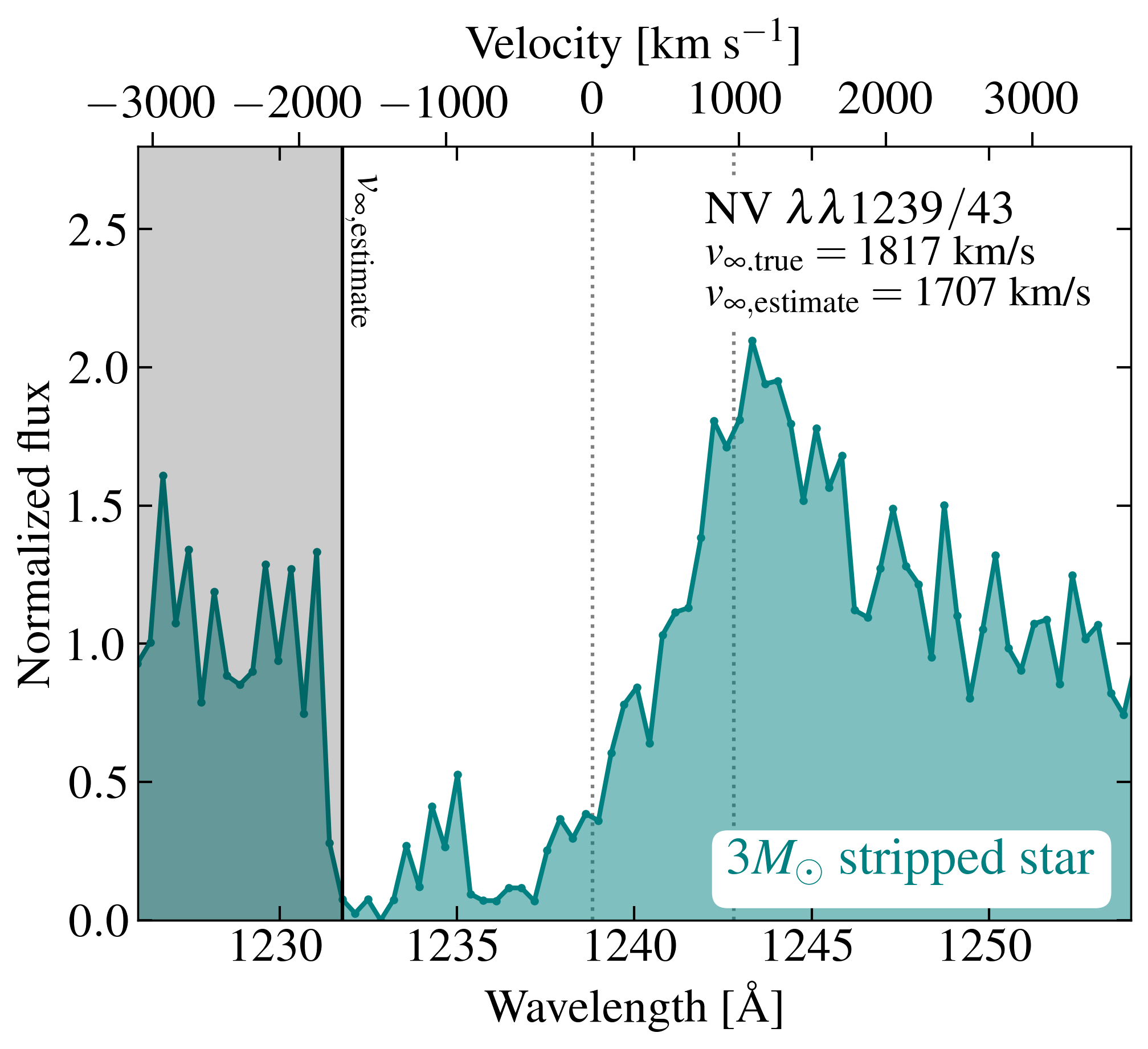}
  \caption{\small The simulated \uvex\ spectrum (\textit{SNR} = 5 at the N~V doublet) of a $3 \msun$ stripped star with an expected typical wind velocity zoomed in on the N~V resonance line.  By fitting for the blue edge of the N~V P-Cygni profile, the wind velocity can be recovered to within $\sim10$\%; 20\% for the weakest wind cases. \uvex\ will enable the accurate determination of massive star wind speeds for $\sim100$ stripped stars in the LMC and SMC.}
 \label{fig:stripped_star_wind}
\end{figure}

A \uvex\ survey of the LMC/SMC will transform our understanding of hot single and binary stars. \uvex\ imaging will discover and characterize thousands of hot stars, from single stars on the main sequence (MS) to rare binary stars in various post-MS evolutionary stages (e.g., stripped stars), efficiently identified via  photometric techniques \citep[e.g., UV excess,][]{2018A&A...615A..78G}. Cadenced UVEX imaging will identify and characterize eclipsing hot, massive systems, and systems containing a stripped star (Figure~\ref{fig:example_light_curves}). \uvex\ absorption line spectroscopy (e.g., CIV, NV, HeII) will measure terminal wind velocities and mass loss rates for single, binary, and stripped stars across the H-R diagram (Figures~\ref{fig:CIV_smc},\ref{fig:ostar_wind},\ref{fig:stripped_star_wind}).

Only \uvex\ has the required sensitivity, resolution and FOV to identify populations of hot and stripped stars.  With its $\sim2$\arcsec\ PSF, \uvex\ will resolve individual stars in all but the most crowded star formation regions (e.g., 30 Doradus) in the LMC and SMC (Figure~\ref{fig:uvex_lmc_image}).  Second, its exquisite sensitivity enables very deep imaging.  Separating the lowest mass stripped stars (0.37 $\msun$) from the main sequence via the color excess method requires a color precision of $NUV-g \le 0.1$ (\citealt{2018A&A...615A..78G}) at $m_{\rm NUV} = 25$.  Moderately deep optical imaging with DECam (\citealt{nidever2017}) will enable the identification of intermediate and massive stripped stars.  When combined with deep Rubin imaging, \uvex\ will uncover stripped stars of all possible masses in the LMC and SMC.  Though extinction in the LMC and SMC is modest (e.g., \citealt{zaritsky2004}), the addition of the FUV band will help to mitigate mild degeneracies between temperature and extinction. 
By comparison, the \galex\ and \swift\ imaging surveys are too shallow; \textit{hst}’s FOV is too small. In the LMC/SMC the \hst\ FOV will contain 0.3 stripped stars, while a single \uvex\ field at similar depth will contain 1400 \citep{2019A&A...629A.134G}. By design, the \hst\ Ultraviolet Legacy Library of Young Stars as Essential Standards (ULLYSES) spectroscopic survey only includes a handful of close binaries in the LMC and SMC, covering a small range of the H-R diagram \citep{roman-duval_2020}. Further context for the uniqueness of \uvex\ imaging and spectroscopic measurements in the Magellanic clouds, and the relationship between ULLYSES and other surveys is provided in Appendix~\ref{appendix:MC_context}, while details of our expected spectroscopic exposure times are presented in Appendix~\ref{sec:lmc_exp_time}.

\subsection{Pillar 2: New Views of the Dynamic Universe}

The coming decade will be a golden era in time-domain and multi-messenger astronomy. Rubin in the O/IR and the Square Kilometer Array (SKA) in the radio will join an existing, vibrant suite of panchromatic wide-field observatories that will identify hundreds of thousands of variable and transient events, opening tremendous discovery space. \uvex\ will be a unique and powerful tool in this exciting new era: it will explore the GW window opened by major upgrades to the LIGO/Virgo sensitivity that will be online by 2028; it will perform the first rapid UV spectroscopic observations of infant SNe; and it will, through a community-driven ToO program, provide the first deep rapid UV spectroscopic followup capability, opening a new window on the dynamic universe.

\subsubsection{The Gravitational Wave Frontier}
\label{sec:emgw}

\uvex\ will probe the physics of compact object mergers, as well as the nature and energetics of the material ejected in NS mergers. Key questions that \uvex\ will address include: What mechanism powers the early UV emission from NS-NS mergers? What are the properties (mass, com-position, velocity) of the early ejecta? How rapidly does the remnant collapse to a black hole?

When two NSs, or a NS and black hole (BH) merge, the NS will be tidally distorted, and ejected material can produce an electromagnetic (EM) counterpart. Indeed, on August 17, 2017, LIGO detected its first binary NS (BNS) merger, GW170817 \citep{abbott2016}, at a surprisingly close distance of 40 Mpc. EM radiation was subsequently observed across the spectrum, from gamma-rays to radio waves.
An astounding number of scientific results came from the single BNS GW170817 event: detection of a short gamma-ray burst (GRB) 1.7s after the merger \citep{uvex75} confirmed the connection between short GRBs and NS mergers \citep{uvex76,uvex77}; the optical localization \citep{uvex78} and host redshift measurement yielded an independent determination of the Hubble constant \citep{uvex79}; the fading UV/O/IR light indicated the presence of a moon-mass worth of neutron-rich ejecta that generated heavy elements, indicating NS mergers are a long-sought source of r-process material \citep{uvex80,Drout2017,uvex82,uvex83,Kasliwal17}; and the late-rising X-ray and radio light indicated that a relativistic jet was produced but viewed $\sim 20^{\circ}$ off axis \citep{uvex85,uvex86,uvex87,uvex88,uvex89,uvex90}.

A striking feature of the EM emission from GW170817 was the prominent UV “blue bump” detected by \swift\ UVOT follow-up beginning 0.6 days after the event \citep{Evans2017}. This emission was bright ($\sim 10^{42}$ erg~s$^{-1}$), blue (peaking at 10,000 K), and had a high velocity ($\geq$ 0.2c) \citep{uvex92}. The origin of this emission is hotly debated \citep{Arcavi2018}, and understanding it holds the key to understanding fundamental issues, including the amount, composition, and velocity distribution of the first ejecta, the NS mass ratio, and the question of how quickly the merger remnant collapsed to form a BH, which can be used to constrain the equation of state (EOS) of exotic, ultra-dense matter \citep{uvex94}. One idea is that the blue emission was radioactively powered, referred to as a “blue kilonova” (BKN) \citep{Metzger10}, with the blue colors implying an ejecta composition dominated by relatively light r-process elements (e.g., Se, Br, Kr). The source of the ejecta could be neutrino-driven winds from an accretion disk \citep{Kasen+2015,uvex97} or material squeezed from the interface of the colliding stars. However, challenges exist with both models \citep{uvex98,uvex99}. The composition and mass of the ejecta constrain the NS mass ratio and EOS.

\begin{figure}[h]
 \centering\includegraphics[width=\columnwidth]{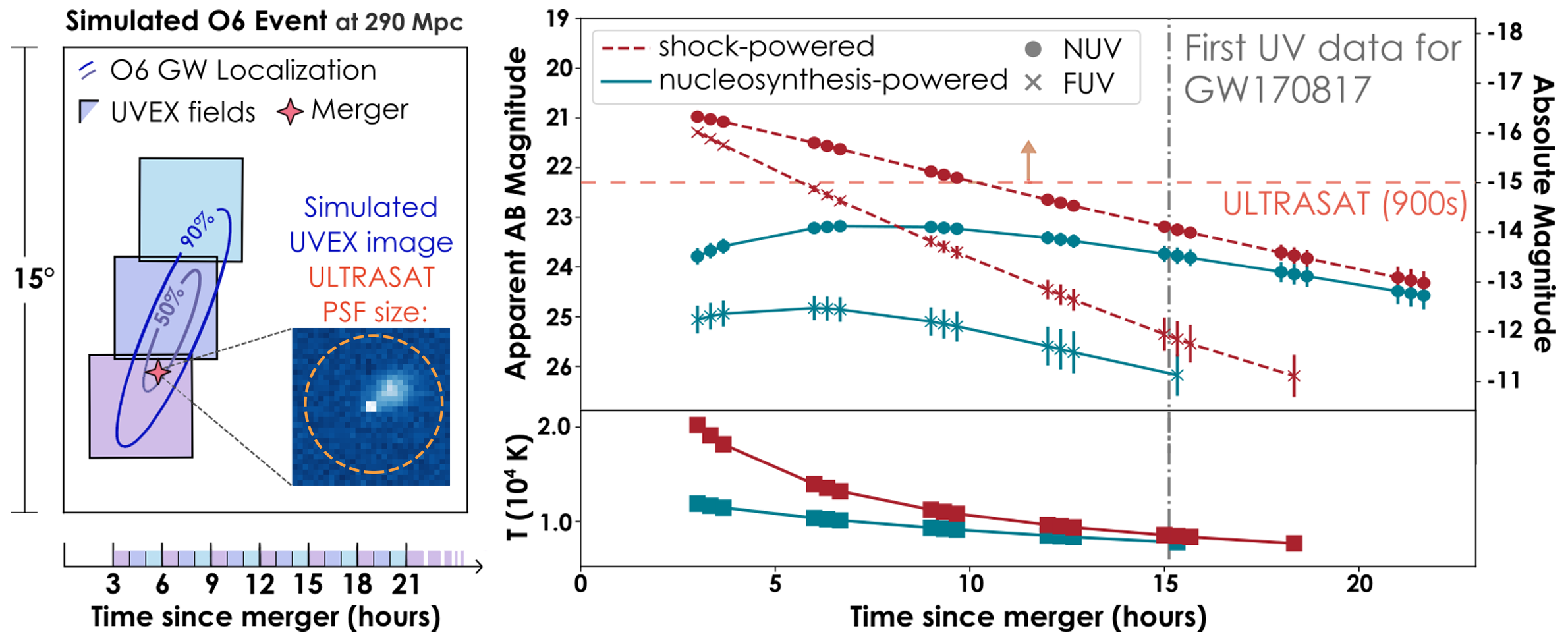}
  \caption{\small (Left) \uvex\ promptly localizes BNS mergers, locating counterparts within their host galaxy. (Right) Two UV bands can measure the evolution speed and peak luminosity (top) and constrain temperature (bottom), discriminating between models.}
 \label{fig:proposal+d-6}
\end{figure}

The merger of a NS and BH offers another potential, but less certain, opportunity to probe extreme physics, BH formation, and relativistic phenomena. LIGO reported detections of the first NS-BH mergers in its O3 observing run \citep{uvex102}. However, either the mass ratio was too large for EM emission, or poor GW spatial localizations precluded meaningful EM counterpart constraints. If there is a large amount of debris \citep{uvex104}, UV observations combined with the GW signal will afford better constraints on the BH mass and spin distribution \citep{uvex105,uvex106}. Many of the same processes inferred in GW170817 may be present in a NS-BH merger, and early-time UV detections or strong limits on the EM flux are the best way to probe these processes \citep{Metzger15,uvex108}.

The future for probing EM emission from GW events is bright over the next decade (Figure~\ref{fig:uvex_timeline}). The LIGO/Virgo/KAGRA GW interferometers are currently executing a major upgrade (termed A+) that will come online in the fifth observing run (O5) in 2025–26. The sixth observing run (O6) is planned to last 18–24 months beginning in 2028 with LIGO India added, dramatically improving the number of well-localized BNS events. In O4 to O5 to O6, the expected rate of BNS mergers localized to $<$100\,deg$^2$ increases from 7 to 47 to 117 per year \citep{uvex79,uvex109}. \uvex\ will follow up a sample of $\geq$20 NS mergers with localizations $<$100\,deg$^2$ within hours of the event, which requires it to detect events as far away as 250 Mpc (for details on the simulations that give rise to these numbers, see Section~\ref{sec:MMA}). Given the \uvex sensitivity, it will reach events to 850 Mpc in NUV and 450 Mpc in FUV for either BKN or shock models. For these events, \uvex\ will obtain high-quality two-band light curves (Figure~\ref{fig:proposal+d-6}).

For achieving the scientific goals enumerated above, only \uvex\ provides the envisioned capabilities. Two-band measurements extending to the FUV are crucial to understanding the early explosive stages, when the emission peaks in the \uvex\ FUV band. Quantitative constraints on emission models require a significant sample of events to probe viewing angle differences and spectral information to remove degeneracies in models. While \hst\ has the sensitivity, its FOV is too small and it does not respond fast enough. While \swift\ has rapid response, its FOV and sensitivity are insufficient to find the UV counterparts at early times (GW170817 was first detected in the optical band). \ultrasat\ will have a larger FOV and rapid response in NUV but lacks the needed two-band coverage extending into the FUV. \ultrasat\ is also much less sensitive than \uvex, corresponding to an event rate $\sim$7 times less than \uvex, without considering sensitivity losses from host Galaxy contamination due to \textit{ULTRASAT}’s coarser (13\arcsec) PSF.

\subsubsection{A New Window on Core Collapse Supernovae}
\label{sec:ccsne}

\uvex\ will open a new window on stellar death and galactic chemical enrichment by massive stars by acquiring the first UV slit spectroscopy of core collapse explosions from hours to days after stellar demise. \uvex\ will address key outstanding questions, including: How do massive stars evolve in the very final stages of their lives? What is the chemical composition, ionization state, and kinematics of late-stage eruptive mass-loss events?

\begin{figure*} 
 \centering
  \includegraphics[width=0.75\textwidth]{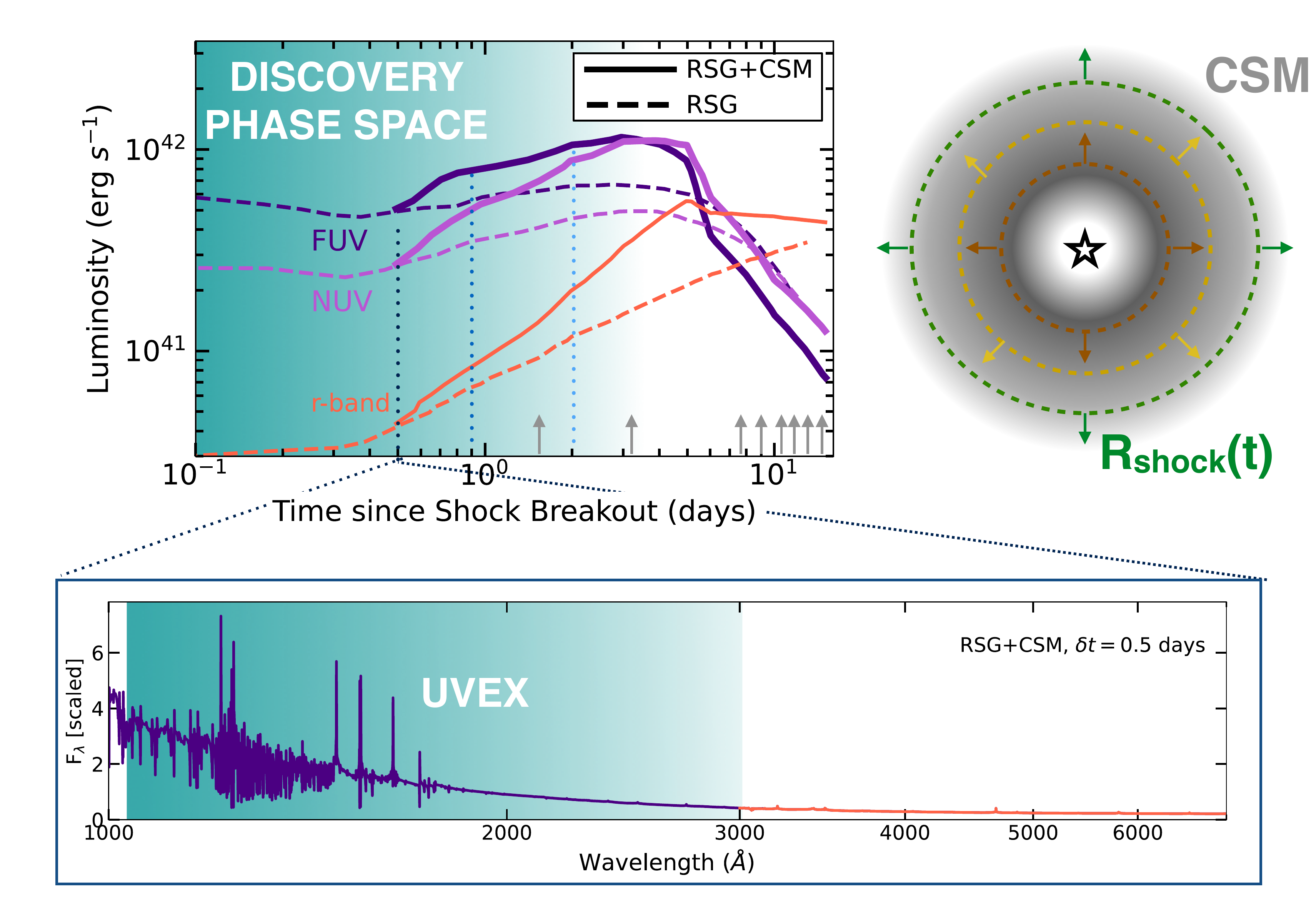}
   \caption{\small 
   The explosion of a red supergiant (RSG) star surrounded by a dense shell of CSM (grey shell in upper right) is a bright source of FUV (dark purple) and NUV emission (light purple). As the explosion’s shock propagates into the CSM (dashed lines in upper right), it ionizes material that then recombines, producing a rich UV spectrum (lower panel) that carries direct information on the unknown chemical composition of this material.
   At early times the optical emission is significantly fainter (see $r$-band) and has less prominent spectral features, as shown in the lower panel (red line). Dashed lines in the upper-left panel: FUV, NUV and $r$-band light curves of the same RSG explosion without a thick CSM. Grey vertical arrows: epochs of acquired UV spectra of SNe, including \textit{IUE} observations of SN\,1987A and the earliest \hst\ spectra of SN\,2020fqv. \uvex will thus explore a completely pristine part of the parameter space and will provide the characterization of stellar explosions of all types in their earliest, hottest phases. Simulated spectra from \citet{dessart17}.}
 \label{Fig:PhaseSpace}
\end{figure*}

\uvex\ spectroscopic observations will provide an unprecedented view of the final years in the life of stars, the chemical composition of their winds, and ejected mass that enriches galaxies in the elements. These are among the least understood aspects of stellar evolution \citep{uvex22}. The explosion of a massive star launches a shock into the circumstellar medium (CSM), heating the material up to $10^5$~K, which subsequently radiates copious UV continuum and line emission (Figure~\ref{Fig:PhaseSpace}). The gas rapidly expands and cools, shifting the peak of the emission towards longer wavelengths. Combined with the higher line blanketing at lower temperatures, this suppresses the UV flux within a few days. Because of its much larger velocity compared to any pre-explosion mass ejections, the SN shock acts as a time machine, and a UV spectroscopic sequence during the earliest ($<$2 days post-explosion) stages offers a unique opportunity to probe the mass-loss history and chemical composition of the exploding star in the last years of its evolution.

Recent observations have revealed an unexpected diversity of massive star behavior in the decades before core collapse, including enhanced and eruptive mass loss. A leading model suggests that this results from nuclear burning instabilities in the final stages of stellar evolution \citep{Quataert12,uvex114}. Our current knowledge of the composition, ionization stage, and kinematics of the ejected material is limited to the information from optical spectroscopy. But for young hot objects, the optical is relatively poor in bright emission lines \citep{Groh14}. As a result, fundamental properties of the progenitor star and physical conditions that lead to enhanced mass loss are unconstrained. Even for the extremely well-monitored SN2013cu, with exquisite optical spectroscopy starting as early as 15.5hr after collapse \citep{Gal-Yam14}, state-of-the art modeling cannot determine the progenitor type \citep{Groh14}.

Early UV spectroscopy will be transformative. Compared to the optical, UV accesses a significantly larger number of spectral transitions, adding crucial constraints to an otherwise under-constrained problem (Figure~\ref{Fig:PhaseSpace}). Virtually all young stellar explosions have SEDs peaking in the UV, where can be found resonance lines such as CIV, HeII, and NIV that have high optical depths, enabling detection of much lower wind densities than the optical. UV also probes highly ionized Fe lines at $\lambda$ 1200–1450 which can be used to directly measure the close CSM metallicity (at these high temperatures no optical Fe transition is available).
  
No current or planned observatory can obtain rapid-response broadband UV spectroscopic sequences. To date, the best-observed core collapse event is SN~1987A, and the best constraints come from the \textit{International Ultraviolet Explorer (IUE)} \citep{Pun95}. In subsequent decades, no comparable UV spectral sequence has ever been acquired. The earliest \hst\ UV spectrum of a SN was 3.3 days after explosion \citep{Tinyanont21}. The \swift\ UVOT slitless grism has rapid-response NUV capability, but limited sensitivity and low resolution, and \swift\ does not extend to the feature-rich FUV. \uvex\ will uniquely fill this observational gap.

\subsubsection{A Community Resource for Exploring the Dynamic Sky}

\uvex\ will open a new window on the dynamic universe by providing the first rapid ($<$1 day) sensitive UV spectroscopic follow-up of variable and explosive events. New Windows on the Dynamic Universe is one of three priority science areas in the Astro2020 decadal survey because of the enormous opportunities opened by facilities such as LIGO/Virgo, Rubin, the IceCube Neutrino Observatory, SKA, and many more. A wide variety of explosive phenomena have spectral energy distributions (SEDs) peaking in the UV at early times, and in many cases, spectroscopy holds the key for addressing fundamental questions. \uvex\ will play a unique and central role in this theme by providing the first sensitive, rapid response UV spectroscopic follow-up capability.

\subsection{Pillar 3: A Legacy of Deep Synoptic All-Sky Surveys}

The coming decade will see modern synoptic and deep wide-area sky surveys with Rubin, \wfirst, and \euclid, for which \uvex\ will provide the crucial, deep, complementary two-band UV data. Each point on the sky will be visited a minimum of 10 times during the prime \uvex\ mission, with cadences ranging from 12 hrs to 6 months. The LMC/SMC survey will be performed through observations taken on a weekly cadence, and the deep extragalactic fields required for validating the extragalactic dwarf galaxy survey will be cadenced to provide regular instrument calibrations. The combination will provide static images and time-domain information with enormous legacy value, enabling a broad range of science limited only by the ingenuity of the community. In addition, the spectral images from \textit{UVEX}’s long (1$^\circ$), offset, varying width slit will be taken with every pointing. These will be archived, affording significant discovery space in the spectral domain. Finally, unlike \galex, \uvex\ will cover the entire sky in FUV and NUV since its modern detectors need not avoid bright objects, providing the first ever deep exploration of the Milky Way in UV.

Combined, all this data will be a rich legacy for the astronomical community to address many areas of science beyond our core mission pillars. The following Section is dedicated to exploring in more depth a selection of these important scientific questions.

\section{A Community Resource: Archival and Extended Mission Science}
\label{UVEX-otherscience}

In addition to the primary scientific objectives that we address with the first two pillars of our baseline mission (Section~\ref{UVEX-pillars}), \uvex\ has the capacity to address many other important areas of investigation requiring its broad UV capabilities. The extensive data provided by the legacy all-sky survey and community-driven ToO follow-up observations, as well as a GO program in an extended mission, will become the basis for transformational advancement in a wide variety of astronomical fields. In this Section, we explore a selection of areas which will benefit from the data provided by \uvex, including stellar astronomy (Section~\ref{sec:StellarAstronomy}), galactic archaeology in the Milky Way and Magellanic Clouds (Section~\ref{sec:GalacticArcheology}), galaxy formation (Section~\ref{sec:GalaxyFormation}), cosmic explosions (Section~\ref{sec:CosmicExplosions}, active galactic nuclei (AGN; Section~\ref{sec:AGN}), tidal disruption events (TDEs; Section~\ref{sec:TidalDisruptionEvents}), multi-messenger astronomy (Section~\ref{sec:MMA}), and exoplanets (Section~\ref{sec:Exoplanets}).

\subsection{Stellar Astronomy}
 \label{sec:StellarAstronomy}

\begin{figure*}[!htbp]
 \centering
  \includegraphics[width=\linewidth]{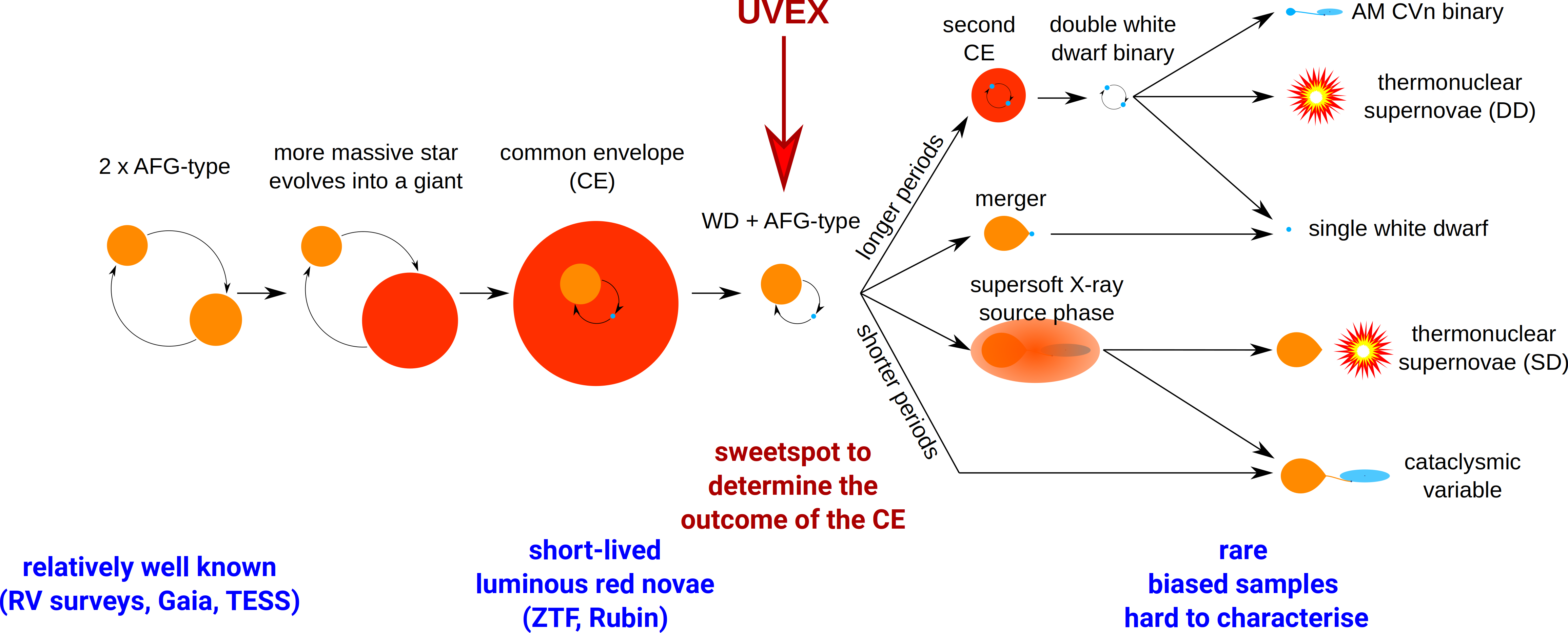}
   \caption{\small Schematic overview of the evolution of two
   AFG-type stars into a variety of compact binaries, which include
   low-frequency GW sources that \lisa\ will detect
   in large numbers, and the progenitors of all types of thermonuclear
   SNe, including SN Ia. The stellar masses and the orbital
   separation of the white dwarf + AFG-type star binaries emerging from the
   common envelope determine the future evolution. \uvex will be able to
   deliver FUV spectroscopy of many post-common envelope white dwarf + FGK 
   binaries, i.e., systems within a critical phase
   that determines their future evolution.}
    \label{fig:wd_binaries}
\end{figure*}

In the previous century, astronomers developed basic understanding of (single) star formation through stellar death and separately were able to explain how, as a result of stellar nuclear synthesis and stellar outflows from winds and explosions, the periodic table was populated with ``metals" (elements other than hydrogen and helium).
With the foundations thus established, in this century, astronomers are working to address the next levels of complexity due to (1) binarity and (2) metallicity. 
It turns out that  50--70\% of stars in the Universe are in binary systems, with this fraction reaching unity for the most massive stars.
For some fraction of these systems, as each of the stars evolve, mass can be lost from one star and gained by the other star. The result, in most cases, is a ``common envelope event" in which one star plunges into the envelope of the other star and forms a tight binary consisting of the core of that star and the companion.
The phase space  determined by the masses of the two stars and the orbital parameters (separation, eccentricity, mutual inclination) is exceedingly large, and poorly understood processes such as common envelope evolution and the impact of metallicity add further complexity.
In Figure~\ref{fig:wd_binaries}, we display some of the key evolutionary pathways for systems with at least one white dwarf -- the most common type of evolved binaries, and for which \uvex\ will provide critical observations.

Even for single stars, the evolution of a star depends on not just the mass but also its metallicity and rotation. 
Mass loss from stars, particularly that from massive stars, is a critical part in the evolution of a galaxy, and it is clear that mass loss is directly tied to metallicity. However, our present understanding of this important physical process is poor. Observations are needed to make further progress.

In terms of metallicity, the Sun is an average star in the Milky Way and we live in an average neighborhood. The frontier now lies in understanding star formation at low metallicities, which in turn provides insight into star formation in the young Universe. Locally, the Magellanic Clouds have metallicities of 10--20\% and 40--50\% relative to the Sun\footnote{We list the generally accepted ranges. The exact value depends on the adopted Solar metallicity scale and abundance pattern, type of metallicity tracer (e.g., stars, nebular emission), radial variations in metallicity, etc.} for the SMC and LMC, respectively (e.g., \citealt{mcconnachie2012}). These offer convenient nearby laboratories for the study of star formation and stellar evolution that are representative -- in at least one very important aspect -- of the early Universe (see Section~\ref{sec:massivestars}).

In the following subsections, we describe further specific scientific areas for which \uvex\ will be enable significant progress in the field of stellar astronomy. First, we discuss the opportunity to probe the physics of accretion in both the context of star formation (Section~\ref{sec:star_form}) and in the context of outflows from accreting compact objects and classical novae (Section~\ref{sec:ns_bs}), and the most common type of evolved binaries: ones hosting at least one white dwarf (Section~\ref{sec:wd}). We also discuss the ability of \uvex\ to contribute to the study of angular momentum evolution of evolved stars (Section~\ref{sec:additional}), and low-metallicity stars in the Local Group (Section~\ref{sec:dwarfs}).

\subsubsection{Star Formation and Accretion}
 \label{sec:star_form} 

Stars first become visible in the optical and UV during the later stages of their formation when the newly formed protostar is present, surrounded by an active accretion disk in which planets form and accrete their initial atmospheres. The stars themselves are very magnetically active, showing kilogauss level average surface magnetic fields \citep{Johns-Krull2007}.
The UV through X-ray emissions produced by the magnetic activity drives the chemistry in the atmospheres of the young planets. While the disk is still present, these high energy emissions help determine the ionization structure and chemistry in the disk, which is important in the action of both the magneto-rotational instability (MRI; e.g., \citealt{Balbus1998}) and the formation of magneto-centrifugally driven disk winds (e.g. \citealt{Gressel2015}), the two main candidates for producing the viscosity that leads to disk accretion onto the star. For accreting young stars, it is this accretion of disk material onto the stellar surface that is the dominant contributor to the UV and blue optical emission observed.
The accretion of disk material represents the last stage in the mass assembly of newly formed stars, and generally occurs within the first 10 Myr after the star is formed (e.g., \citealt{Wyatt2008}) while it is evolving along the pre-main sequence (PMS) evolutionary tracks.

\begin{figure}[htpb]
 \centering
  \includegraphics[width=\columnwidth,trim=0 10 0 0,clip]{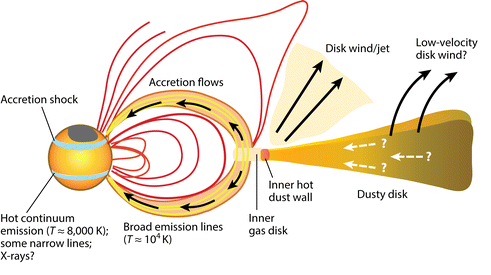}
  \caption{\small The close circumstellar environment of a young
  star (taken from \citealt{Hartmann2016}).  Magnetospheric accretion occurs
  as disk material is captured by the stellar magnetic field, flowing toward
  the star and accreting near the stellar poles.  At the base of
  the magnetospheric flow, the material falling at near free-fall
  velocities impacts the surface, creating a strong shock which
  produces substantial UV emission.}
 \label{fig:ctts_cartoon} 
\end{figure}

The current paradigm for accretion onto young, PMS stars is
magnetospheric accretion (for a recent review, see \citealt{Hartmann2016}),
which posits that the strong magnetic fields on
the surface of young stars truncate the disk near the co-rotation
radius, forcing the accreting disk material to flow along the field
lines such that the material impacts the stellar surface at near
free-fall velocities. Theoretical work (e.g., \citealt{Konigl1991},
\citealt{Shu1994}, \citealt{Long2005}, \citealt{Zanni2013},
\citealt{Romanova2018}) suggests that the coupling between the young
star's magnetosphere and its circumstellar disk is sufficient to
regulate the stellar angular velocity for the lifetime of the disk.
In this so-called ``disk-locking" picture, angular momentum is
magnetically transferred from the star to the disk and eventually
ejected in a mass outflow. 
This picture of magnetospheric accretion and disk locking has been 
underpinned primarily by observations from nearby star-forming
regions such as Taurus, rho Ophiuchus, the Upper Scorpius region
and a handful of others. However, these regions are generally considered
low density, relatively low mass star-forming regions, and have very few 
of the highest mass,
hottest stars (O-type stars) which produce copious amounts of ionizing
radiation and strong stellar winds that have important feedback
effects in the star formation process \citep{Rosen2020}. Studies
of the Orion Nebula Cluster (ONC) have also been significant in the
development of these ideas. While higher in overall mass and
containing the high-mass stars of the Trapezium, this region is still
not representative of the highest mass star-forming regions
\citep{Portegies2010} in which the majority of stars in our Galaxy
and probably most galaxies form (see review by \citealt{Portegies2010}),
and in which O-type stars and their feedback are expected to have
a substantial impact on other stars forming there.

\begin{figure*}[htbp]
 \centering\includegraphics[width=\textwidth]{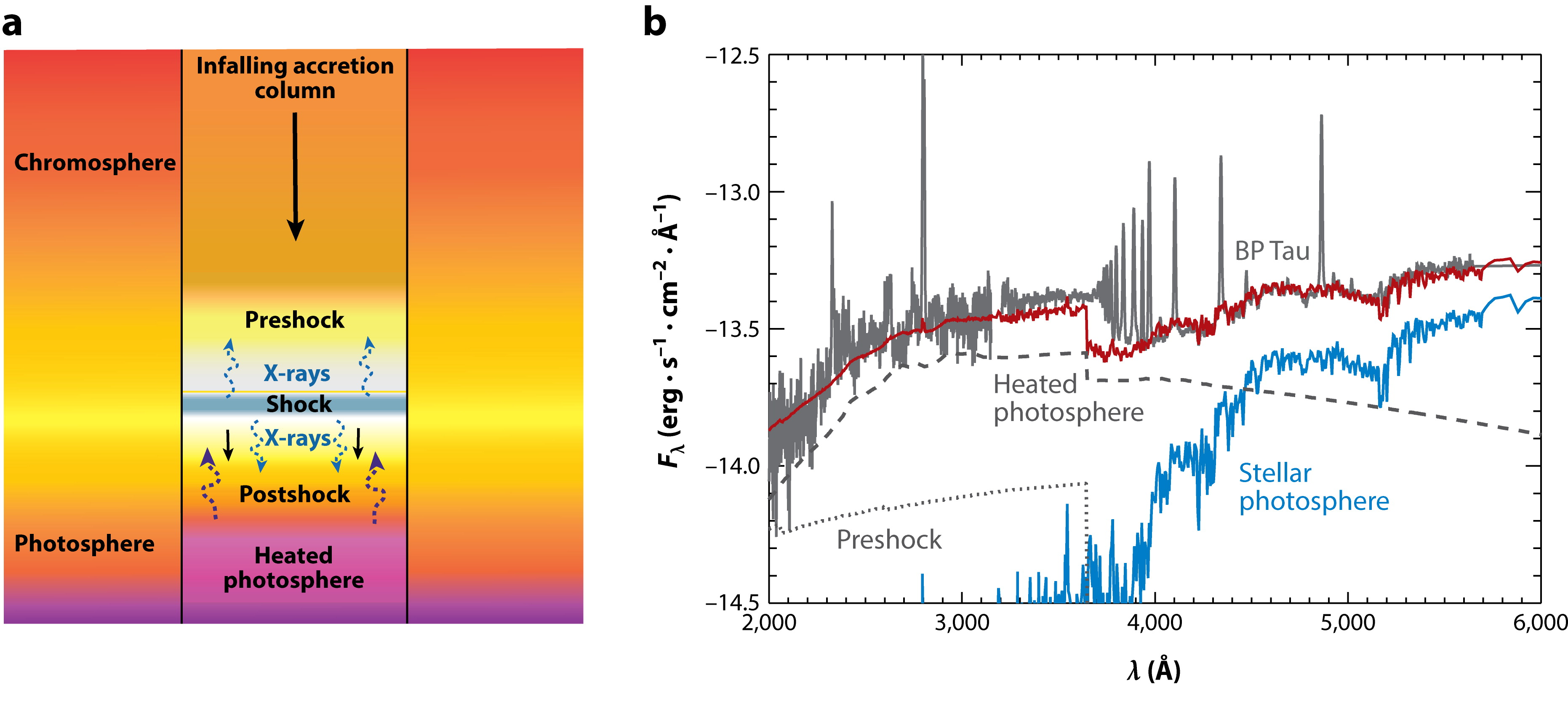}
  \caption{\small From \citet{Hartmann2016}: (a) Schematic diagram of
  accretion shock structure showing the precursor or preshock region,
  the postshock or cooling region, and the heated photosphere below
  the shock. (b) Spectral energy distribution of the classical T
  Tauri star BP Tau (gray solid line), stellar photosphere (blue
  line), and accretion shock model (red line) showing contributions
  from the preshock (gray dotted line) and heated photosphere postshock
  (gray dashed line) regions.}
 \label{fig:acc_shock}
\end{figure*}

Taking extinction and distance into consideration, perhaps the most
favorable region of massive star formation for the study of feedback
processes on low-mass star formation and early stellar evolution
is the Carina star forming complex. The Carina nebula complex (CNC)
is located in the Carina spiral arm (e.g., \citealt{Vallee2014}),
and is one of the most active massive-star-forming regions
in the Milky Way. Over 140 massive OB-stars \citep{Alexander2016}
and more than 1,400 young stellar objects \citep{Povich2011,Feigelson2011} 
have so far been identified in the CNC, which is suggested to contain 
10 times the young stellar content of the ONC \citep{Townsley2011}. 
The distance to the CNC, $\sim
2.3$ kpc, has been measured accurately using near-IR spectroscopy
(e.g., \citealt{Allen1993, Smith2006}). The number of O-stars in
the CNC is comparable to that in other massive-star-forming regions
in the Galaxy such as W43 and W51 (e.g., \citealt{Blum1999,
Okumura2000}), but the CNC is two or three times closer than those
regions. Therefore, the CNC offers an excellent opportunity to study
the physics of accretion onto newly formed low-mass stars in a
region that is more representative of the regions in which most
stars form. Study of the low-mass stellar population and its
accretion activity will provide a unique way to uncover the role
of massive star feedback on the formation and early evolution of
young stars.

As shown in Figure~\ref{fig:ctts_cartoon}, when material accreting
along the stellar magnetic fields reaches the star, an accretion shock
is expected to form. The material impacts the star at near free-fall
velocities, which is on the order of $\sim 300$ km s$^{-1}$. As a
result, a strong shock forms, initially heating the material up to
a temperature of $\sim 10^6$ K.
As the material cools from $10^6$ through $10^5$~K, it emits strong lines 
in the X-ray and UV which in
turn heat the photosphere below and immediately around the accretion
shock as shown in Figure~\ref{fig:acc_shock}a. Many authors have studied the
emission that results from this process 
(\citealt{Valenti1993}, \citealt{Calvet1998}, \citealt{Gullbring1998},
\citealt{Ingleby2013}), and the accretion-related
emission dominates the stellar flux in the UV (Figure~\ref{fig:acc_shock}b),
making short wavelength observations the most sensitive to accretion
onto young stars. For the most strongly accreting stars, the
accretion luminosity is also detected in the blue optical, 
but as the accretion rate falls, NUV and FUV observations are
required to make reliable accretion rate estimates for young stars.

Extinction will play a role in our ability to measure mass accretion rates; however, there are at least two well-established ways to estimate the extinction from broadband colors for cool stars. In the $J$-$H$ versus $H$-$K$ color-color diagram, reddening and IR excess from circumstellar emission moves stars in almost orthogonal directions (e.g., \citealt{meyer97}). In addition, over the spectral type range of cool stars, these objects define a nearly horizontal line in this two-color diagram. As a result, the extinction can be reliably estimated for stellar types K-M. In addition, narrower band optical photometry can also be used to estimate the effective temperature and extinction of low mass stars (see Figure 5 of \citealt{dario12} and accompanying discussion). Averaging results from these two methods will provide sufficiently accurate $A_v$ values for the statistical nature of accretion studies that \uvex\ photometry will allow. This would require obtaining supporting ground-based observations, albeit from existing observatories.

\begin{figure}[htbp]
 \centering\includegraphics[width=\columnwidth,trim=0 10 10 0,clip]{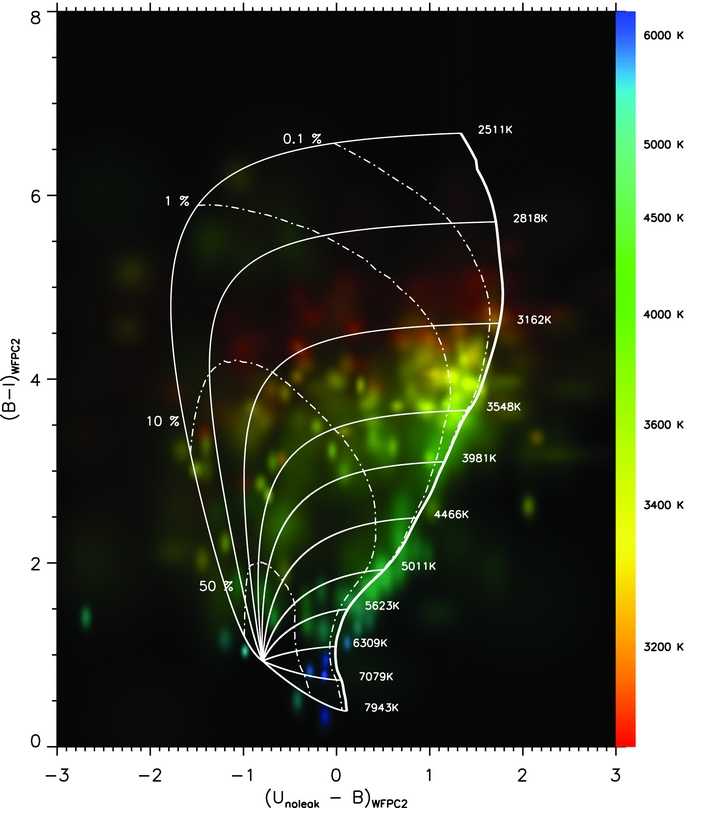}
  \caption{\small From \citet{Manara2012}:  Two-color diagram of
  stars in the ONC. Each source has been corrected for extinction
  and represented as a normalized 2D Gaussian, corresponding to the
  photometric errors.  Sources are color-coded according to
  their $T_{\rm eff}$ (scale at right). The thick line represents the calibrated isochrone for no
  accretion; thin lines represent the simulated displacements
  (for different $T_{\rm eff}$) from the photospheric colors, obtained
  by adding an increasing amount of a model of the accretion
  luminosity analogous to that shown in Figure~\ref{fig:acc_shock}.}
 \label{fig:onc_2cd}
\end{figure}

\uvex operating with supporting ground-based facilities (e.g. Rubin)
will provide the observations needed to measure accretion onto
thousands of young stars in the CNC. An example of how these
accretion rate determinations will be performed is shown in 
Figure~\ref{fig:onc_2cd} taken from \citet{Manara2012}. This figure shows
a two-color diagram (using colors from $U$ to $I$) of young stars in
the ONC. Red optical and near-IR colors are used to estimate
spectral types and reddening for each source. Similar data for the
CNC will be available from LSST (e.g., \citealt{Bonito2018}) and
other existing or planned ground-based surveys. The nearly vertical,
thick solid line shows the locus where stars without accretion fall.
The thinner white lines moving to the left and down from this locus
shows where stars with differing amounts of accretion land.
As shown in Figure~\ref{fig:onc_2cd}, many of the stars lie very close to the
non-accreting locus. These stars are either not accreting at all,
or accreting at levels too low to be measured using colors confined
to the optical bands. Adding in FUV and NUV photometry from \uvex
will effectively stretch this diagram out in both directions, but
most importantly will stretch it in the horizontal direction, making
it easier to accurately measure accretion rates and to distinguish
the accretion emission from low accretion rate objects.

Fortunately, the entire CNC can fit into a single pointing of \uvex.  
A \uvex CNC survey could detect accreting young stars down to 
$\sim 0.3 \msun$, providing the first comprehensive study of 
accretion onto low-mass stars in a canonical high-mass star formation region.  

\galex did not survey this region and did not have the
sensitivity to reach the required depth. \hst is capable of making
similar observations with WFC3; however, $\sim$4400 pointings
would be required to cover a single \uvex pointing. As a result, \uvex
is the best instrument to make the requisite observations to explore
the role of high mass star feedback on the accretion physics of low-mass young stars.

\subsubsection{Outflows from Accreting Compact Objects}\label{sec:ns_bs} 

Accretion occurs on a wide range of physical scales, from super-massive 
nuclear black holes to protostars. Here, we address accretion onto 
compact stellar objects. Galactic
accreting white dwarfs (i.e., cataclysmic variables; CVs) and 
stellar-mass black holes and neutron stars (i.e., X-ray binaries; XRBs) are
excellent laboratories to study accretion processes and the
accretion-ejection coupling mechanisms in great detail, across
time scales accessible to human beings \citep{fender16}. Moreover,
accretion in these systems spans a broad range in accretion rates, from
$10^{-5}-10^{2}$ times the Eddington rate.

Most accreting stellar remnants spend the majority of their time in a
quiescent state, punctuated by outbursts lasting a few days to
months. Massive outflows are launched
during those outbursts, which play an important role in the
evolution of these systems. UV observations provide a unique
window on these outflows. In this section, we focus on two distinct
types of outbursts, each with their own phenomenology and open
questions: outbursts caused by the disk instability mechanism, and 
classical novae.

The disk instability mechanism (DIM; for a
review see \citealt{hameury2020}) occurs in a subset of CVs called 
dwarf novae (DNe) and in most XRBs that accrete material through
Roche lobe overflow -- mainly low-mass X-ray binaries (LMXBs). 
In these systems, mass is transferred
at a relatively constant rate from a non-degenerate donor star
through Roche lobe overflow onto an accretion disk around the compact
object. In quiescence, the mass transfer rate through the disk onto
the compact object is low, the disk itself is cold and not ionized,
and the system is faint. In this state, matter builds up in the
disk and the temperature rises, until a critical point is reached
and an outburst starts. In outburst, the mass transfer rate through
the disk is greatly increased and the disk itself is extremely
bright.

Although the DIM works well in explaining the basic properties of
these outbursts, it is clear that more ingredients are necessary
to match the observations -- most importantly, irradiation of the disk
and outflows. 
Powerful disk winds are launched in the high state of these outbursts 
\citep{Fender:2004} and these winds carry away a significant amount of
mass and angular momentum. Indeed, most of the transferred mass 
from the star never gets accreted onto the compact object. In
this way, winds can fundamentally change the evolution of
these systems. 

The wind launching mechanisms for both XRBs and DNe are still very poorly understood. Line-driven winds are one of the candidates \citep{proga02}, especially for disk winds in XRBs with high mass accretion rates. Thermal winds \citep{Begelman:1983} and magnetic winds \citep{Ferreira:1995,Petrucci:2008,Begelman:2015} likely also play a role. More observations are needed to determine which mechanism dominates in different regimes. 

Outflows from accretion disks of DNe are best studied in the UV, specifically UV spectroscopy.  {\it HST} has provided some UV spectra of LMXB outbursts and DNe, but generally only one per outburst \citep[e.g.,][]{Sion:2004,Merritt:2007}.
In fact, due to the fact that {\it HST} cannot perform very fast ToOs,
the DNe outbursts were only caught by accident. \uvex, which will
be able to repoint on $\sim$hour timescales, will be far better able to catch
these outbursts. By obtaining multiple UV spectra over
the course of DN and LMXB outbursts, coordinated with multi-wavelength 
follow-up, we will be able to study the launching mechanism of the winds and their
relation to the jet.

The second type of outburst we discuss is classical novae, outbursts driven
by runaway thermonuclear burning of hydrogen accreted onto the
surface of a white dwarf from a binary companion (\citealt{Bode2008}, \citealt{DellaValle2020}, \citealt{Chomiuk2020}). 
Although known for centuries, our understanding of these explosions
has undergone a renaissance in the last decade -- beginning with
the discovery of $\gamma$-ray emission (by {\it Fermi}, \citealt{Abdo2010_novae})
and correlated optical--$\gamma$-ray variability (\citealt{Li2017},
\citealt{Aydi2020}), bright radio synchrotron emission (\citealt{Weston2016a},
\citealt{Weston2016b}), and hard X-ray emission (\citealt{Nelson2019},
\citealt{Sokolovsky2020}). Together, these observations
provide evidence of internal shocks between multiple outflows that
can power a substantial fraction of the optical luminosity of novae.

Although the basic picture of shocks between multiple outflows 
explains several multi-wavelength aspects of nova observations, 
a consistent picture remains elusive, particularly as some do not
exhibit correlated optical-$\gamma$-ray behavior \citep{Li2020}.
However, the shock interaction region between fast polar and 
slower equatorial outflows may be an ideal environment for the
formation of dust \citep{Derdzinski2017}.
A powerful testable consequence
of this scenario is the predicted i) viewing angle dependence of
multi-wavelength evolution and the formation of dust in novae and
ii) variations in the shock and dust formation properties as a
function of the underlying white dwarf mass, which drives the amount
of mass ejected and the photometric evolution of the nova
(\citealt{DellaValle1992}, \citealt{Yaron2005}).

\begin{figure}[htbp]
 \centering
  \includegraphics[width=\columnwidth]{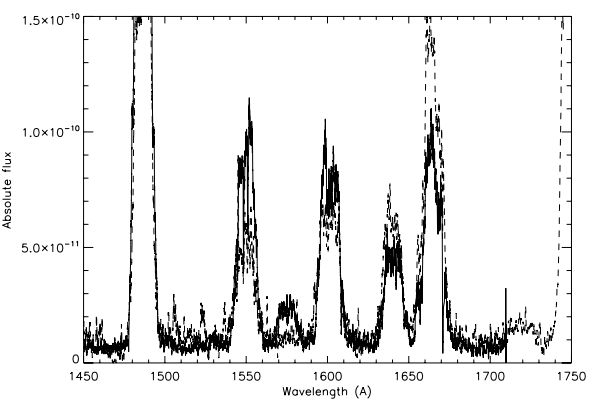}
   \caption{\small Comparison of the UV spectra of Nova~Mon~2012 (solid)
   and Nova V1974 Cyg (dashed), taken from \citet{Shore2013}.} 
 \label{fig:novamon}
\end{figure}

UV spectroscopy is powerful in that the strongest lines of C, N,
Ne, Mg that are produced in the ejecta are in that wavelength range
\citep{Shore2012}. These lines can be used to trace the ejecta
density profiles, clumping, and filling factor to derive accurate
ejection masses. 
Due to limited ToO capabilities, {\it HST} and {\it IUE} have
provided a limited number of UV spectra of only the brightest and
most nearby novae. Figure~\ref{fig:novamon} presents two
{\it HST} spectra of novae, showing a diversity of line profiles. The
line profiles can further be used to estimate the ejecta geometry
and density profiles (e.g., \citealt{Shore2013}). Combining the
density and structure diagnostics to map out the variation in the
multi-wavelength emission (radio, $\gamma$-rays) together with the
ejecta dynamics/geometry and white dwarf mass of the nova. will
allow us to develop a holistic picture of mass ejection in nova
outbursts. Based on current estimates of the nova rate in the Milky
Way \citep{De2021} and the Magellanic Clouds \citep{Mroz2016}, a
total of $\sim$10 novae are expected to be accessible for UV
spectroscopy during the \uvex 2-year baseline mission (accounting 
for Galactic extinction).

\subsubsection{White Dwarf Companions}
 \label{sec:wd} 

Historically, identifying stellar binaries and obtaining the time series
imaging and spectroscopy necessary to fully characterize these
systems has been observationally taxing. However, modern all-sky imaging 
surveys (e.g., {\it Gaia}, {\it Kepler}, {\it TESS}) that provide 
high precision photometry
and astrometry as a function of time have revolutionized our
understanding of `vanilla’ low-mass binaries (e.g., two comparably
low-mass main sequence stars), as well as facilitated the 
discovery of a small number of more exotic systems (e.g.,
main sequence-white dwarf binaries). Similarly, ground-based optical 
spectroscopy has helped to characterize
a small set of these systems in the Milky Way, but numbers remain small.
Concerted efforts for large-scale optical spectroscopic identification
of such binaries is only getting underway (SDSS-V, DESI, WEAVE,
4MOST).

Despite progress with optical binaries in the Milky Way, there remains
very little exploration of binary systems with a hot component
(e.g., systems with OB stars, stripped stars, white dwarfs). In many cases
the hot companion is virtually impossible to discern with optical
observations. Single hot companions may not affect the optical
spectra or perhaps only leave indirect trace signatures (e.g., odd
optical emission line combinations) in the otherwise normal spectrum
of the optically dominant star.

In other cases, even when a binary system is identified (e.g., from
optical light curves or spectra), key aspects of the system (e.g.,
mass loss, wind speeds) or basic characteristics of the hot component
(e.g., mass, temperature) are only accessible in the UV, wavelengths
at which few observations exist. Thus, though we now believe some
of the most influential astrophysical phenomena originate from hot
star binary systems, our census and physical understanding of these
systems is lacking.

In addition, binaries born as two AFG-type stars
will go through a phase where the more massive star has already
evolved into a white dwarf, whereas its companion is still on the
main sequence. At optical wavelengths, the white dwarf is totally
swamped by the companion, but it dominates the UV emission of the
system.

The \uvex all-sky survey will identify $\sim$10,000 FGK-type stars
that exhibit a UV excess indicating the presence of a white dwarf
companion that is undetectable at optical wavelengths. Some pilot
studies have been carried out with \galex \citep{Parsons:2016},
but many of the nearby FGK-type stars were too bright for 
\galex. The \uvex detectors do not have a bright source limit.
Moreover, \uvex will increase the FUV footprint with respect to \galex
by 50\%. This will allow an unbiased statistical study of the
population of these binaries within the Milky Way. 
This will both enable a full all-sky search for white dwarf + FGK
binaries, as well as allow the identification of the closest systems,
hence those most suitable to detailed follow-up studies.

The key parameters for the future outcome of white dwarf + FGK-type binaries
are the masses of both stellar components as well as their orbital
period. Whereas the mass of the main sequence star and the orbital
period can be obtained from optical photometry and spectroscopy,
measuring the white dwarf mass can only be done in the FUV.
{\it HST} has observed $\sim$25 white dwarf + FGK binary candidates, 
confirming a white dwarf component in most of them. 
However, a population study
that is sufficiently large to sample the full parameter space of
stellar mass and orbital period requires a dedicated survey that
is beyond the limited resources that are left to {\it HST}. \uvex
will enable a population study that spans the full parameter range
in stellar mass and orbital period, which is essential to establish
tight observational constraints on the branching ratio of these
systems with respect to their future evolution
(Figure~\ref{fig:wd_binaries}).

Low-resolution ($R\sim1000$) spectroscopy covering 1150--1800\AA\
is essential, as the Stark-broadened photospheric Ly$\alpha$ is
sensitive to both the temperature and the surface gravity of the
white dwarf, and, in conjunction with a mass-radius relation,
provides a measurement of the mass. Detection of emission lines,
primarily \ion{C}{4} 1550\AA, will be a signature of ongoing mass
transfer \citep[e.g.,][]{Parsons:2015}. 
The main-sequence masses and orbital periods can be established from
optical data, and hence those two components of the parameter space
can be mapped out in advance of the \uvex survey. However, the
properties of the white dwarf will be unknown prior to \uvex
spectroscopy. 
A \uvex spectroscopic sample
can then be used to extrapolate to the full, unbiased all-sky
population established from the \uvex imaging, which in turn will
put tight constraints on the low-frequency gravitational background
of double-degenerates descending from white dwarf + FGK binaries, as well as
on rates of the various sub-types of thermonuclear supernovae.

\subsubsection{Evolution of Angular Momentum of 
Stars Across the HR Diagram}
 \label{sec:additional}

The typical story of a star’s life begins with the Jeans collapse
of a molecular cloud, before eventually contracting onto the Zero
Age Main-Sequence (ZAMS). Conservation of angular momentum suggests 
that ZAMS stars have measurable rotation rates stemming
from the initial angular momentum of their respective parent molecular
clouds. However, observations show that there is a
strong separation between fast and slow rotating stars at a temperature
of $\sim 6200\ K$. This separation, commonly referred to as the Kraft
break, represents a rough boundary between stars with radiative
envelopes and stars with convective envelopes \citep{Kraft1967}.

The Kraft break highlights the significance of convective envelopes
to angular momentum evolution in stars as they are essential to
powering dynamo action, which generates the self-sustaining magnetic
field that steals angular momentum via interaction with stellar
winds \citep{Weber1967}. This connection ties a star’s rotation to
both its magnetic field and its age, both of which are notoriously
difficult to characterize. In general, magnetic activity and age
are quantified in terms of angular momentum using rotation-activity
relations and gyro-chronology, respectively.

Empirical rotation-activity relations come in several forms, one
of which is a comparison of UV emission to rotation period.
\citet{Stelzer2016} attempted to derive a UV rotation-activity for
M dwarfs using \galex photometry and {\it Kepler} rotation periods
to better understand fully convective dynamos, but was limited by
a lack of observations and commented that follow-up UV observations would
be a powerful constraint for a M dwarf rotation-activity relation.
Additionally, \citet{Dixon2020} found rotation-activity relations
for giants using \galex and APOGEE, but also suffered from a lack
of fast rotating stars.

With significantly improved angular resolution and greater sensitivity 
than \galex, \uvex presents an opportunity to derive
well constrained UV rotation-activity relations. This is especially
true as an increasing number of stellar rotation periods are becoming available from large-scale time-series surveys.  ESA's approved {\it PLAnetary Transits and Oscillations of stars}
({\it PLATO}) mission, with a targeted launch date in 2026, will support \uvex in this
science by delivering ultra-precise optical light-curves for more than 2 million FGK dwarfs and subgiants and nearly 300,000 M~dwarfs 
\citep{Montalto2021}.

Finally, it should be possible to associate large numbers of these stars with space-based rotation and activity measures to regions of the Galaxy with well estimated ages, including the recently identified stellar ``strings" that represent coherent stellar populations spanning large regions of the Galaxy \citep[e.g.,][]{Kounkel:2020}. This will allow the development of robust rotation-activity-age relations for stars across the HR diagram, and in turn allow the mapping of stellar ages across large swathes of the Milky Way. 

\subsubsection{Lower-metallicity Massive Stars in the Local Group} 
 \label{sec:dwarfs}

Despite their importance to wide-ranging next generation astrophysics (e.g., as discussed in Section~\ref{sec:massivestars} and illustrated in Figure~\ref{fig:CIV_smc}), theoretical models for massive stars remain essentially untested below the metallicity of the SMC. The requirement of deep spectroscopy for individual resolved massive stars severely limits the environments to which this stellar calibration work can be applied. Early results from Wolf–Lundmark–Melotte and IC~1613 identified these galaxies as likely SMC-like in stellar abundances, stymieing the first attempts to measure mass-loss rates at lower metallicity (e.g., \citealt{2015MNRAS.449.1545B}). While challenging at $\gtrsim 1$~Mpc, the dwarf irregulars Leo~A (e.g., \citealt{2007ApJ...659L..17C}), the Sagittarius Dwarf Irregular Galaxy (SagDIG; \citealt{2018MNRAS.474L..66G}), and Sextans~A (\citealt{2016A&A...585A..82C}, \citealt{2019MNRAS.484..422G}) harbor populations of massive stars that are likely our best hope for calibrating stellar models at metallicities below the SMC. As for stars in the SMC and LMC, UV is crucial to characterizing these stars and their winds (Section~\ref{sec:massivestars}). But while \hst\ has begun the work of collecting spectra for these foundational targets, the resolution, sensitivity, and wavelength coverage are of varying utility for characterizing winds and photospheric abundances, and many other potential targets await a dedicated UV survey.

\begin{figure}[htbp]
 \centering
  \includegraphics[width=0.3\textwidth]{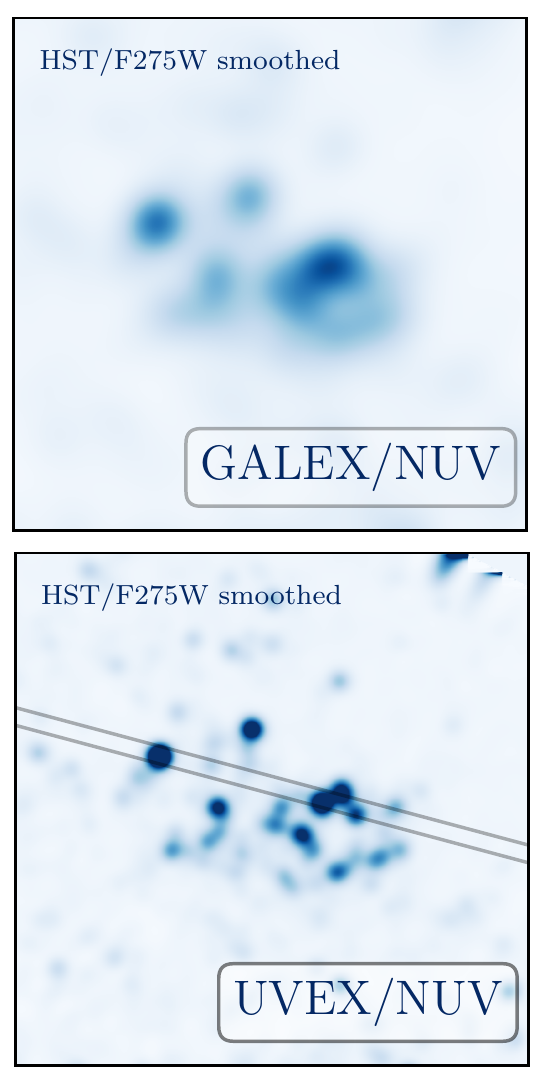}
    \caption{\small \uvex\ will provide access to individual massive stars in the distant, low-metallicity dwarf irregular galaxies such
    as Sextans~A ($D\sim1.2$~Mpc). Both full-field
    imaging from the all-sky survey and deep spectroscopic followup
    of the most UV luminous stars will yield unique constraints on
    young stellar populations below the metallicity of the SMC.}
 \label{fig:lowzdwarfs_longslit}
\end{figure}

\uvex\ has the capabilities to fulfill the promise of these metal-poor dwarf irregular galaxies for massive star model constraints. First, all-sky photometry will immediately provide the most complete picture of unobscured sub-SMC metallicity massive stars in the Local Group, where \galex\ is severely limited by crowding and \hst\ NUV coverage is incomplete (Figure~\ref{fig:lowzdwarfs_longslit}). Compared to previous work restricted to optical selection, deep \uvex\ imaging will substantially improve the census of luminous blue stars in these galaxies, especially very hot metal-poor stripped binary products.  

\uvex\ spectroscopy will provide the capabilities for the first systematic investigation of stellar wind strengths at sub-SMC metallicity. Exposures of order 3--30~ks will suffice to provide detailed constraints on the resonant wind complexes and photospheric indices discussed in Section~\ref{sec:massivestars}, enabling unique measurement of wind terminal velocities and mass loss rates unavailable from the optical, as well as quantities such as clumping filling factor, temperatures, and abundances. Approximately ten targets without UV data have already been identified with extant published optical spectroscopy (seven O to early-B giants in Sextans~A from \citealt{2016A&A...585A..82C} and \citealt{2019MNRAS.484..422G}; 2--3 OB stars in SagDIG from \citealt{2018MNRAS.474L..66G}). \hst\ will deliver UV spectra over part of the needed wavelength range for a total of 12 OB stars below SMC metallicity (eight in Sextans~A, three in Leo~A, and one star in Leo~P). \uvex\ will improve upon the resolution, coverage, and SNR of these spectra where they may be insufficient to measure robust wind properties (e.g., \citealt{2019arXiv190804687G}). Though not part of the prime mission, a spectroscopic survey with \uvex\ would be able to, at a minimum, more than double the sample of sub-SMC massive stars with UV wind constraints.

\subsection{Galactic Archaeology}
 \label{sec:GalacticArcheology}
 
\uvex\ provides a facility for mapping the Milky Way and nearby galaxies in the UV, providing insight into the distribution and physical properties of dust and precise measurements of stellar metallicity from photometry.
 
\subsubsection{Milky Way Dust Maps}
 \label{sec:mwdust}

Many areas of extragalactic science require high-precision maps of dust extinction and reddening at high Galactic latitudes. Maps based on far-IR (FIR) dust thermal emission are widely used (e.g., \citealt{SFD98}, \citealt{Planck2014-XI-Dust}), but suffer from a number of systematics.  More concretely, these maps only trace dust extinction indirectly, and systematic errors can be introduced by incorrect modeling of dust temperature, spectral index, or column density, or by variations in the ratio of dust extinction at optical wavelengths to thermal emission in the FIR. In addition, dust maps based on FIR emission are contaminated by large-scale structure (through dust emission in distant galaxies; see \citealt{ChiangMenard2019}), which is of particular concern for cosmology.  

Dust maps based on optical and near-IR stellar photometry (e.g., \citealt{Marshall2006-3D-Dust}, \citealt{Green2015-3D-Dust}, \citealt{JuvelaMontillaud2016-Dust}) more directly measure dust extinction and reddening, and are thus less affected by these systematics. However, dust maps based on stellar photometry typically achieve lower signal-to-noise ratios than FIR emission-based maps at high Galactic latitudes, where the sky density of stars is lower and per-star extinction is lower.

At high Galactic latitudes, it is therefore critical to observe stellar photometry in the UV. The UV colors of stars are extremely sensitive to small amounts of dust, and can thus boost the SNR of per-star reddening measurements. Recent work using \galex\ photometry in combination with LAMOST spectroscopy has mapped extinction at high Galactic latitudes (\citealt{Sun2021-UV-Dust-GALEX-LAMOST}). \uvex\ will expand the number of stars with UV photometry by a factor of $\sim$5 relative to \galex, enabling the creation of dust extinction maps with higher resolution and higher signal-to-noise.

\begin{figure}[htbp]
 \centering
  \includegraphics[width=\columnwidth]{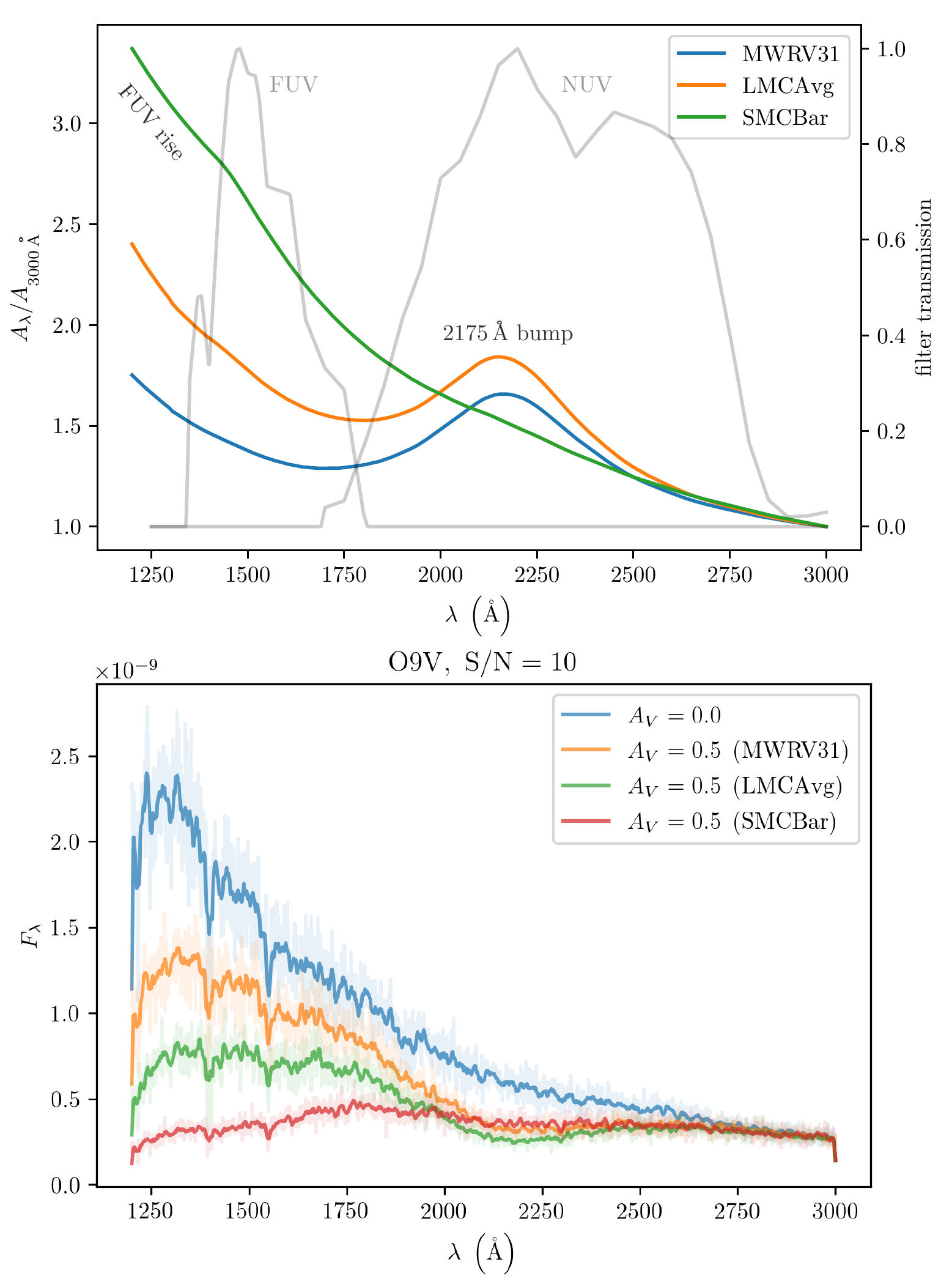}
   \caption{\small (Top) Three representative dust wavelength-extinction relations,
   from the Milky Way ($R\left(V\right)=3.1$), the LMC, and the SMC
   bar. The \uvex\ NUV/FUV-band transmission curves are overplotted.
   Extinction in the NUV band is primarily sensitive to the strength
   of the 2175\,\AA\ bump, while extinction in the FUV
   band is primarily sensitive to the slope of the FUV rise. 
   (Bottom) The effect of different dust extinction laws on
   an O9V-star spectrum, observed at a native signal-to-noise ratio
   of 10 (shaded curves), and then smoothed to \uvex\ resolution to show that coarse features are clearly discernible. The lack of a 2175\,\AA\ bump and strong FUV rise in the SMC bar extinction law are readily apparent.}
   \label{fig:dust_figure}
\end{figure}

\paragraph{LMC and SMC} \label{sec:lmcdust}

Dust extinction laws exhibit striking variations depending on dust chemistry and the interstellar environment (Figure~\ref{fig:dust_figure}). A well-known example is variation in dust extinction curves along different sight lines in the SMC (\citealt{Gordon2003-Ext-Curves}). 

The origin and extent of variations in dust extinction is poorly understood. This is particularly true at UV wavelengths. While the optical extinction law is well-characterized by a single parameter, $R\left(V\right)$, the UV extinction law exhibits more variability, which is only weakly correlated with $R\left(V\right)$ (\citealt{PeekSchiminovich2013}). In the UV, the NUV-band extinction is sensitive to the strength of the 2175\,\AA\ bump, thought to be carried by polycyclic aromatic hydrocarbons (PAHs) or graphite grains (\citealt{Draine2003-Dust-Review}), while the FUV-band extinction is sensitive to the slope of the far-UV rise, driven by very small grains (\citealt{MishraLi2015-Carbon-UV-Ext}).  
\begin{figure*}
 \centering
   \includegraphics[width=\textwidth]{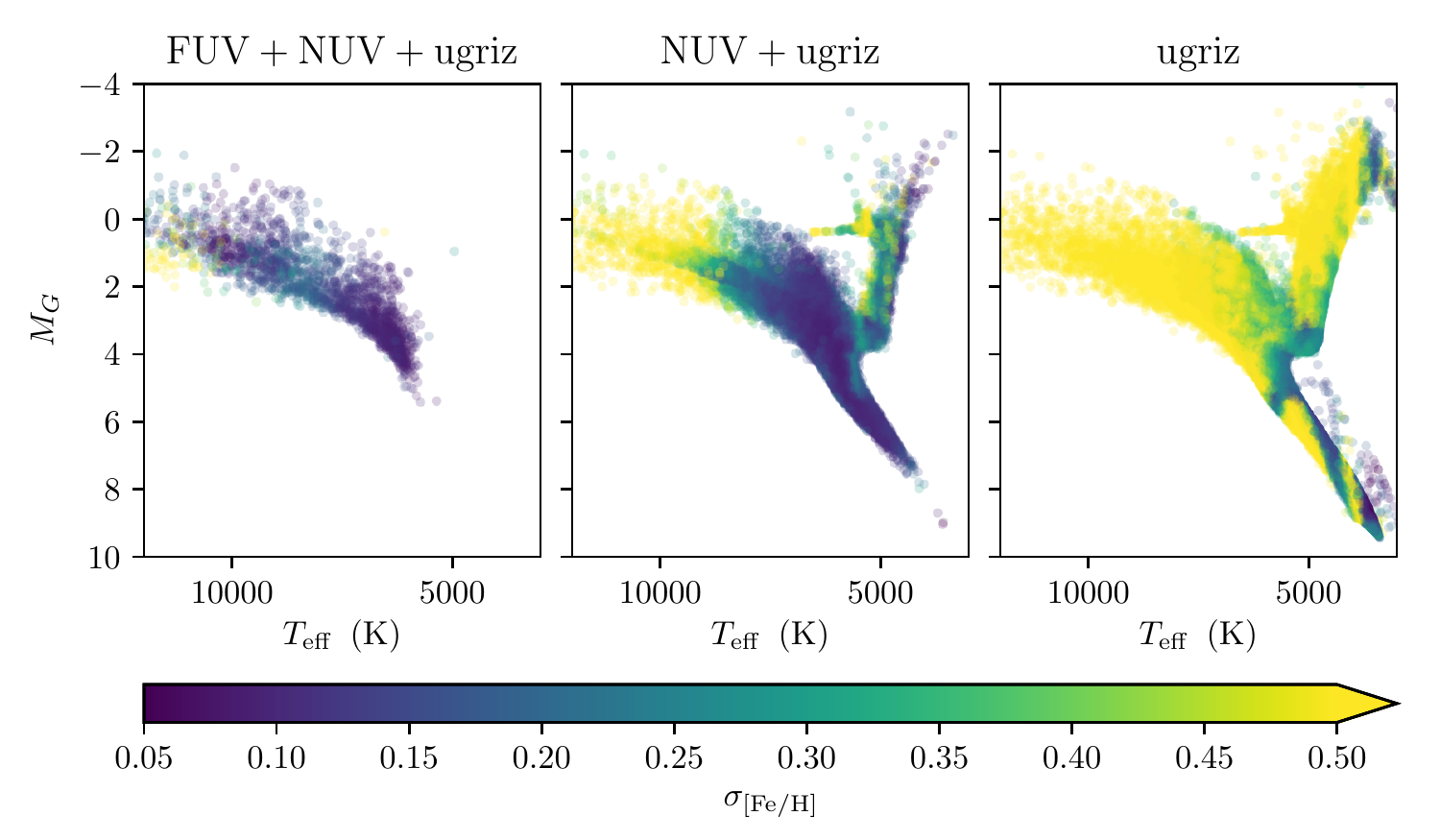}
    \caption{\small Cram\'{e}r-Rao bounds on the uncertainties 
    (i.e., the theoretical precision) in
    $\left[\mathrm{Fe/H}\right]$ obtained using different
    combinations of \uvex\ and Rubin band photometry, assuming photometric
    uncertainties of 0.02~mag. Though \uvex\ detects fewer low-temperature
    stars than Rubin, it will achieve lower uncertainties
    in $\left[\mathrm{Fe/H}\right]$, particularly for hot (and
    therefore preferentially young) stars.}
 \label{fig:feh_cramer_rao}
\end{figure*}

The gold standard for measuring dust properties is UV spectroscopy. Because shorter wavelength light is more easily scattered by dust, the amount of extinction as a function of wavelength, along with other variation (e.g., the 2175\,\AA\ bump) are the empirical anchors for our knowledge of the UV extinction laws. Within the Galaxy, the UV extinction law, and its relation to dust physics, has been measured from UV spectroscopy from samples of a few hundred stars (e.g., \citealt{2007ApJ...663..320F}, \citealt{2019ApJ...886..108F}, \citealt{2020ApJ...891...67M}). Far less is known about UV extinction laws at sub-Solar metallicities. The commonly used LMC and SMC dust curves are determined from averaging over a small number of sight lines, which exhibit substantial variance. 

\uvex\ spectroscopy of 1000 hot OB type stars in the LMC and SMC will provide qualitatively new insights into the UV extinction curve at sub-Solar metallicities. For example, the average SMC UV extinction curve exhibits no 2175\,\AA\ bump. However, the bump is present in a handful of sight lines with UV spectroscopy in the wing of the SMC (\citealt{Cartledge2005-Magellanic-Sightlines}, \citealt{LiMisseltWang2006-SMC-Dust}). The precise carrier of the 2175\,\AA\ bump has not yet been determined, though PAHs are a candidate (\citealt{Li2020-SpitzerPAHs}). A census of the precise properties (amplitude, central wavelength and width) of this feature in a variety of interstellar environments (e.g., different metallicity and interstellar radiation field) will limit the range of possible physical models of interstellar dust at low metallicity (e.g., \citealt{Draine2003-Dust-Review}, \citealt{Hensley2021-PostPlanck}). By piggy-backing off of \uvex\ spectroscopic surveys of O-stars in the LMC and SMC (see Section~\ref{sec:massivestars}), we can expand this sample by $\sim$1000 stars. Because dust affects the broad UV spectrum, even fairly low-SNR ($\approx10$) and moderate-resolution ($R\sim1300$) spectra obtained by \uvex\ are well-suited for measuring variations in the UV dust extinction curve in the SMC and LMC.

\begin{figure*}[htbp]
 \centering
  \includegraphics[width=\textwidth]{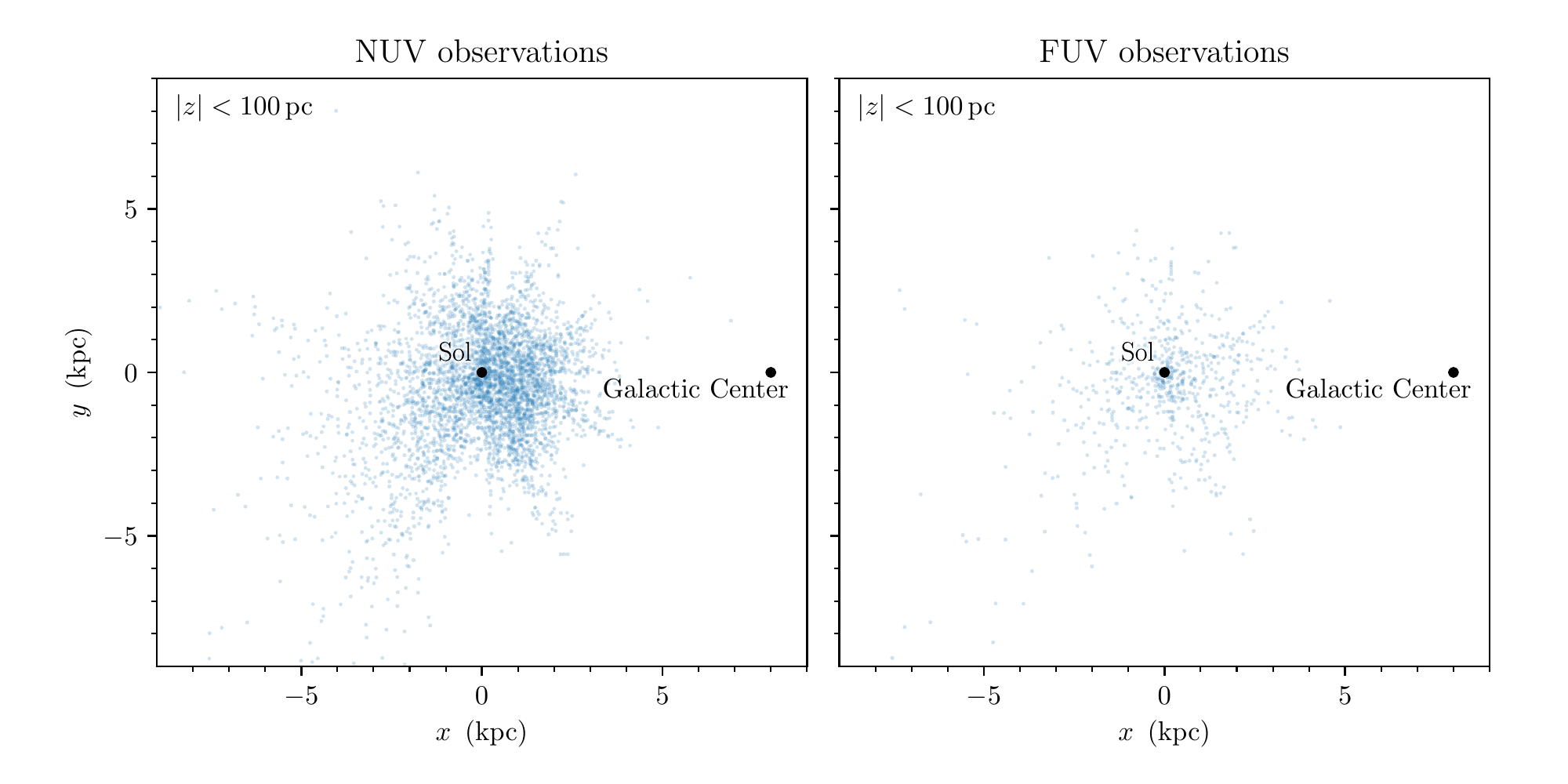}
   \includegraphics[width=\textwidth]{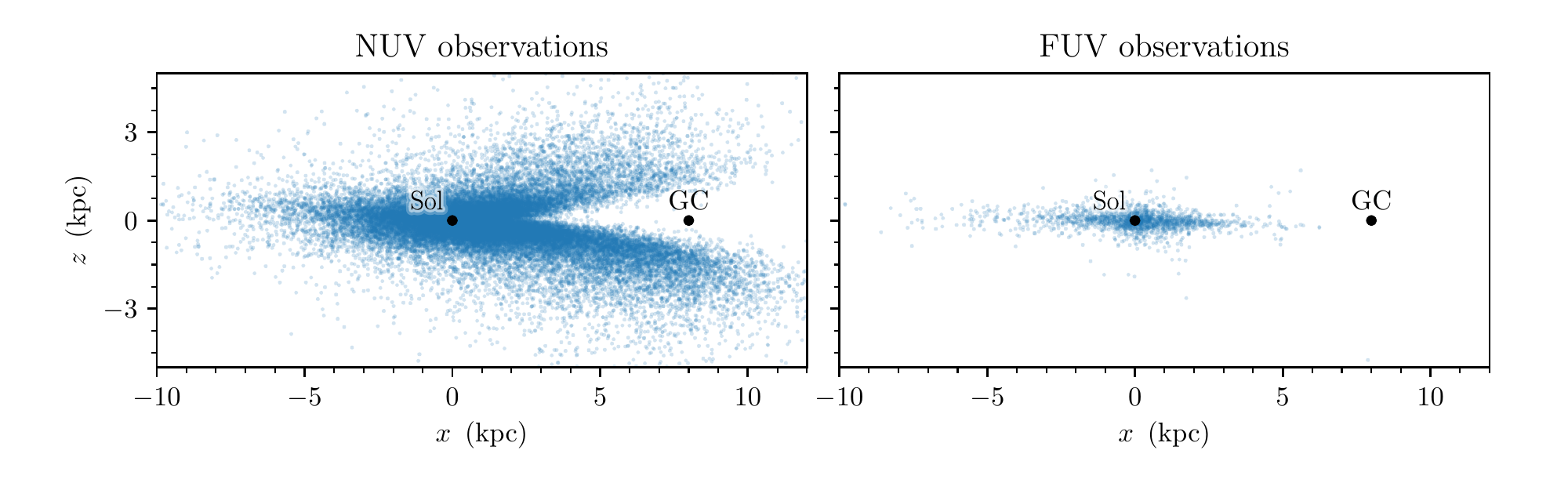}
    \caption{\small Expected spatial distribution of Milky Way stars
    observed by \uvex. The top panels show a bird's-eye view --
    centered on the Sun -- of stars within 100~pc of the mid-plane
    of the Galaxy that would be observed in the NUV (left) and FUV
    (right) bands. The bottom panels show the distribution of stars
    that would be observed in these two bands, projected onto the
    $\left(x,z\right)$-plane (in Cartesian Galactic coordinates,
    centered on the Sun). As can be seen from these panels, \uvex
    will observe stars through a large volume of the Milky Way,
    extending several kiloparsecs along sight-lines that do not pass
    through the inner Galaxy.}
 \label{fig:uvex_sources_cartesian}
\end{figure*}

\subsubsection{Metallicity Mapping in the Milky Way} \label{sec:metalmap}

The resolved stellar populations of the Milky Way encode its formation history. The stellar halo of the Milky Way, in particular, hosts a wide variety of substructure, such as streams, globular clusters, and disrupted dwarf galaxies, accumulated through a combination of \textit{in situ} star formation and accretion (e.g., \citealt{helmi1999, helmi2008, blanhawthron2016}). A key diagnostic for unraveling the formation history of the Milky Way is stellar metallicity since metallicities of individual stars can be used to identify substructures and infer chemical enrichment processes.  Metallicities are typically derived from iron lines in optical spectroscopy, and dedicated surveys (e.g., APOGEE, LAMOST, GALAH, 4MOST, \textit{Gaia} radial velocity survey; e.g., \citealt{deng2012}, \citealt{desilva2015}, \citealt{majewski2017}) will provide such measurements for $\sim30$~million stars -- a small fraction of the $>$1 billion Milky Way stars surveyed by \textit{Gaia} given the flux limits of  spectroscopic surveys.

Photometric metallicities offer access to a much larger population of stars. Measuring a photometric metallicity usually involves combining UV/$u$-band imaging with an optical band. UV wavelengths are particularly metallicity sensitive (capturing, e.g., the Balmer break and iron line blanketing in the mid-UV) when combined with optical imaging. For example, the UV color excess method (\citealt{1979ApJ...233..211C}) takes advantage of the metallicity dependency of $u-g$ color at constant optical color.  Using this method, SDSS mapped the metallicity of a volume-complete sample of two million F/G dwarfs in the Milky Way disk and halo, with typical uncertainties of 0.2~dex in $\left[\mathrm{Fe/H}\right]$ (\citealt{2008ApJ...684..287I}). These tomographic maps of Milky Way stellar metallicity were a boon to Galactic archaeology, identifying chemically distinct subcomponents within our Galaxy. 

Combined with Rubin optical photometry in the Southern Hemisphere and PS1 optical photometry in the Northern Hemisphere, \uvex\ will allow an even more sensitive determination of stellar metallicity. Although FUV-NUV color alone is not sufficient to determine metallicity, the combination of \uvex\ photometry and optical colors will allow a much more sensitive determination of stellar metallicity than possible with optical data alone. Specifically, NUV$-g$ color is far more sensitive to metalicity than $u-g$ color. Because the \uvex\ all-sky survey is slightly shallower than the Rubin $u$-band, it will observe fewer stars than Rubin. However, \uvex\ will obtain much more precise metallicities per star, particularly for hot stars (see Figure~\ref{fig:feh_cramer_rao}). \uvex\ will obtain NUV photometry of $\sim$300 million Milky Way stars, probing a significant fraction of the entire Galaxy (see Figure~\ref{fig:uvex_sources_cartesian}).

\subsection{Galaxy Formation}
\label{sec:GalaxyFormation}

\begin{figure}
 \centering\includegraphics[width=\columnwidth]{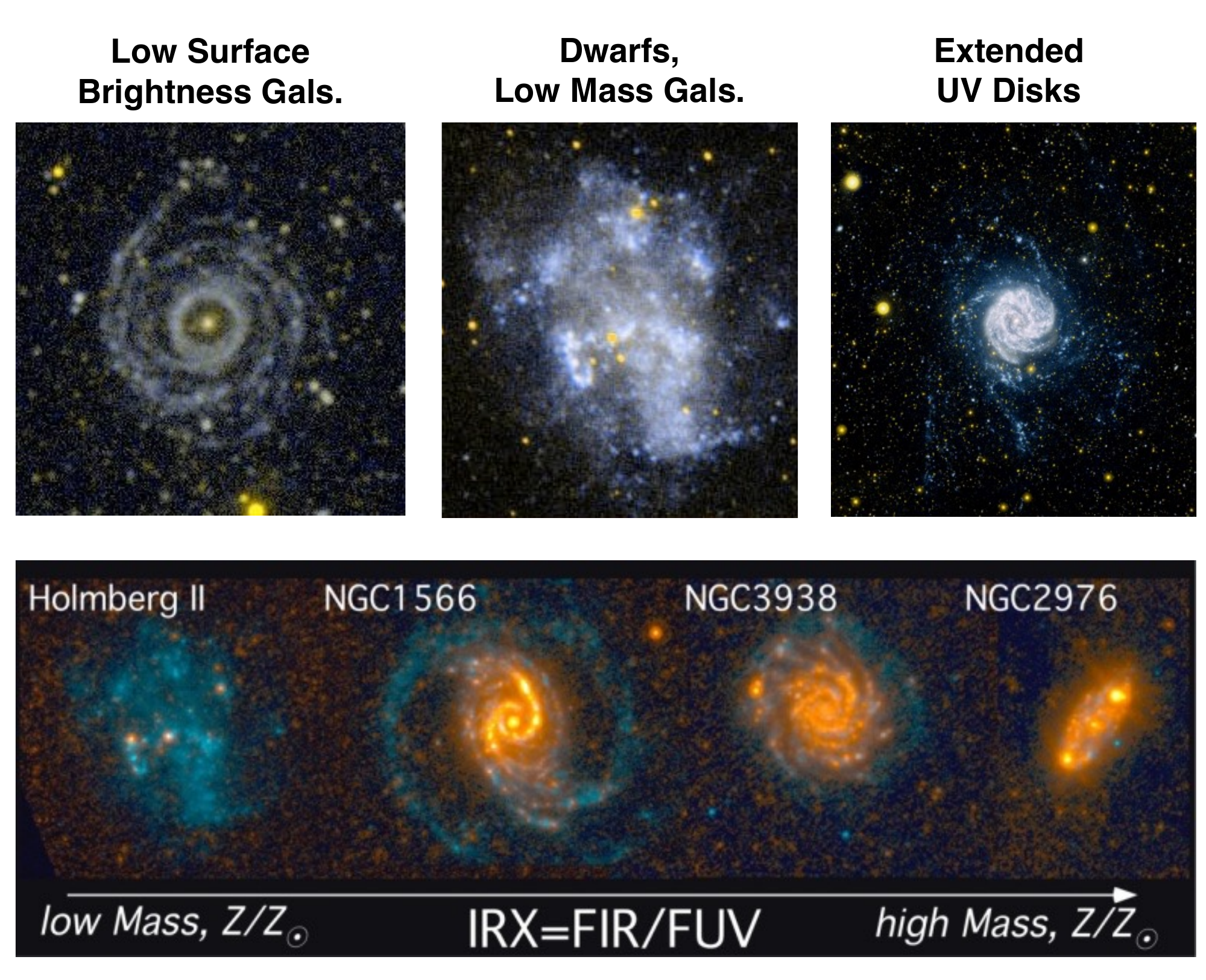}
  \caption{\small Low mass star forming galaxies are highly visible
  in the UV because their metallicity and extinction are low, and
  the UV sky is dark.}
 \label{fig:low_mass_uv}
\end{figure}

\begin{figure*}
 \centering
  \includegraphics[width=\textwidth]{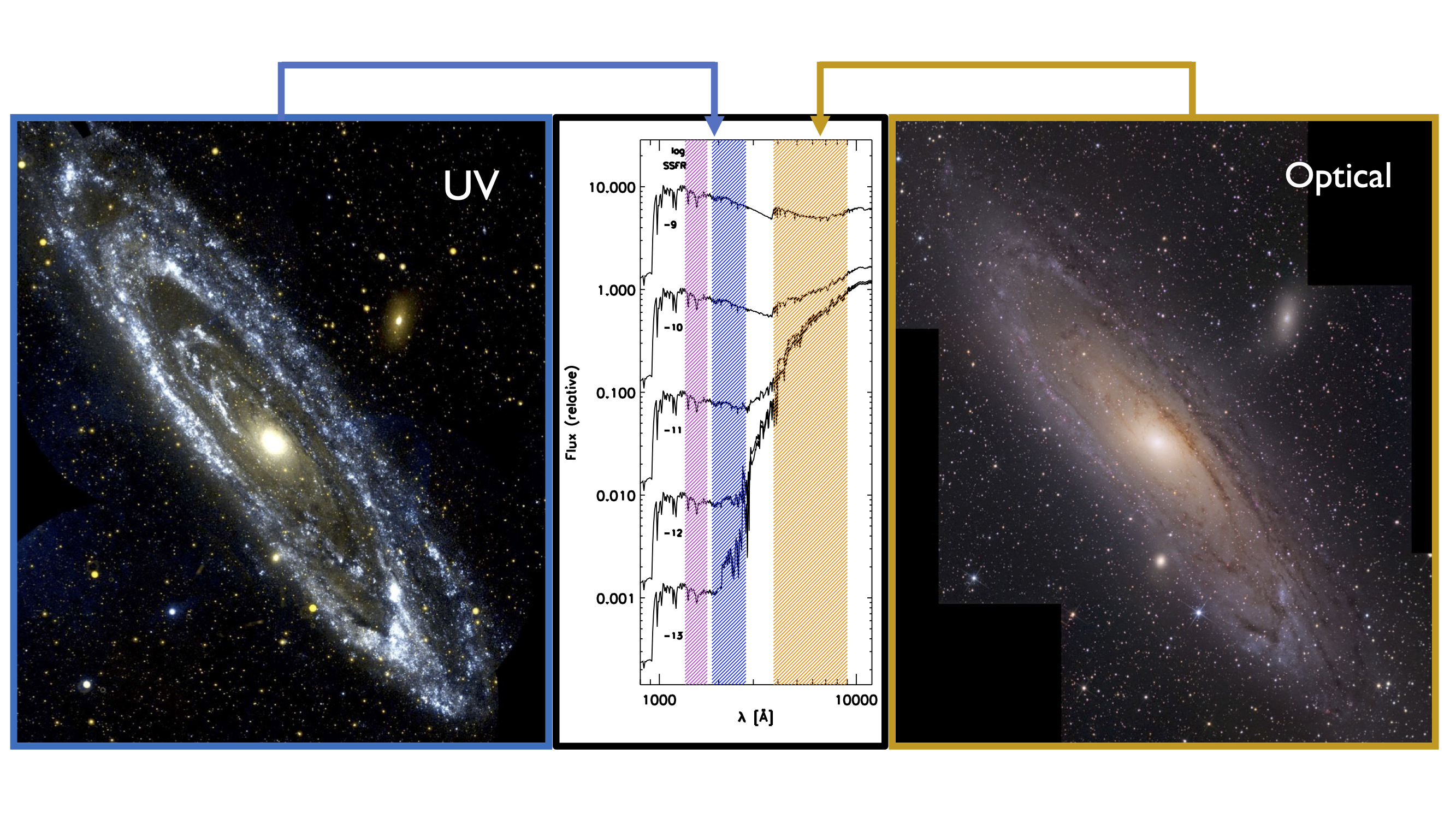}
   \caption{\small The optical band traces $\sim$ 1–5 Gyr of star formation
   history; UV traces 100–300 Myr, and can measure small amounts
   of residual star formation superimposed upon old stellar
   populations [image credits: M31 optical: Adam Block/NOAO/AURA/NSF;
   UV: \textit{GALEX}/JPL/NASA].}
 \label{fig:m31}
\end{figure*}

One of the primary motivators in extragalactic astronomy is to
understand how galaxies form and evolve over cosmic time. This is
an enormous enterprise that has spanned decades, engaged hundreds
if not thousands of astronomers world-wide, and occupied major
fractions of observing time on the ground and in space. While much
is known, next-generation surveys could revolutionize our
understanding of the field if and only if they exploit the full
power of the electromagnetic spectrum. UV imaging and spectroscopic
surveys are an essential component of this revolution. This is
because the UV measures SFR averaged over a
timescale which is matched to evolutionary changes in galaxies,
$\sim 100$ Myr. Optical probes such as H$\alpha$\ are noisy averages
of only $\sim 10$ Myr, while IR traces reprocessed UV radiation in
massive, metal-rich galaxies. UV is uniquely sensitive to the lowest
mass, lowest metallicity systems that are virtually dust-free
(\citealt{fisher14}) and represent analogues to the first galaxies
(Figure \ref{fig:low_mass_uv}). The UV also cleanly probes small amounts of
residual star formation in otherwise passively evolving galaxies,
essential for understanding how and why galaxies quench individually
and collectively (Figure \ref{fig:m31}). UV spectroscopy probes key
elements including carbon, whose abundance is otherwise poorly constrained.
UV spectroscopy also provides unique access to signatures of the circumgalactic medium (CGM) and feedback-driven outflows in nearby galaxies including through \lya{} and metal absorption lines, processes which lie at the heart of our modern understanding of galaxy evolution.
In this Section, we discuss the potential applications
of a UV imaging and spectroscopic survey mission on Galaxy Formation,
Galaxy Evolution, and Galaxy-Halo Co-evolution.

\subsubsection{The Galaxy ``HR-Diagram''}

\begin{figure*}
 \centering
  \includegraphics[width=\textwidth]{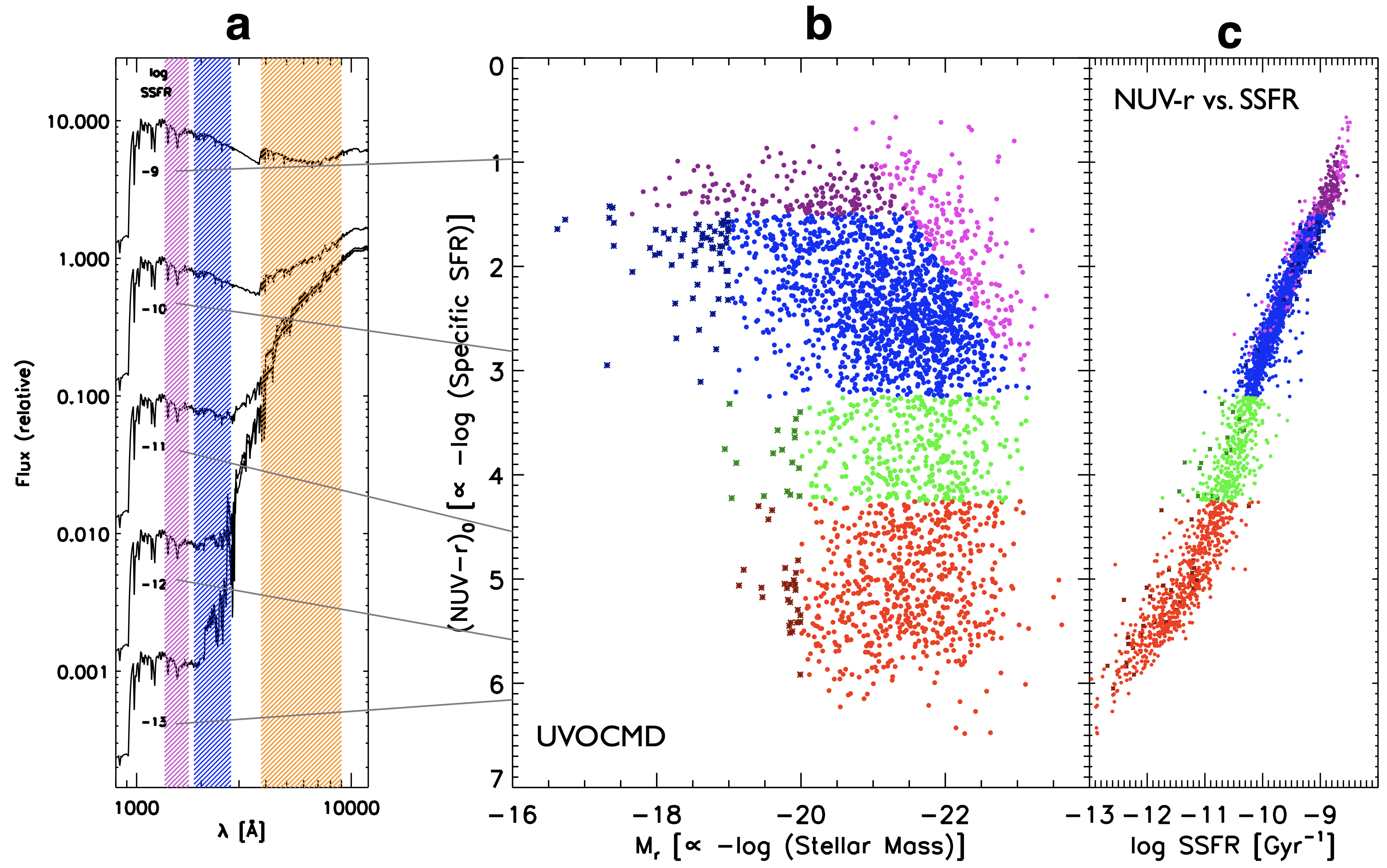}
   \caption{\small Galaxy HR Diagram constructed from mass-tracing
   optical/NIR color and FUV or NUV color. a) Old population with
   superimposed star-forming population with different specific
   star-formation rates (sSFR). b) Distribution of galaxies on the
   GHR showing the relation between star forming (main sequence)
   galaxies [blue], galaxies bursting above the SFMS [purple],
   transition galaxies [green], and quenched galaxies [red]. c)
   There is a tight correlation apparent between the extinction-corrected NUV-r
   color and sSFR.}
 \label{fig:ghr1}
\end{figure*}

A fundamental tool for probing galaxy evolution is the UV-optical
color-magnitude diagram (UVOCMD; Figure \ref{fig:ghr1}). The
extinction-corrected UVOCMD is often viewed analogously as a 
galaxy ``HR-diagram'' (GHR),
because the optical/NIR magnitude traces stellar mass ($\mstar$),
and the UVO color (e.g., NUV-$r$) traces specific star formation rate
(sSFR) to remarkably low levels. 
The long-term buildup of stars moves galaxies to the right.
New star formation bursts move galaxies up,
while quenching moves them down. 
Thus we can trace on this diagram the influence of the manifold processes driving galaxy evolution.

The \galex\ mission and NASA Great Observatories provided a
huge leap forward in our understanding of galaxy formation and
evolution. We now know that the galaxy distribution can be described
to first order as bimodal (i.e., either red or blue), with a population
of transitional galaxies in the so-called ``Green Valley'' between
the actively star-forming blue cloud galaxies and the passively
evolving red sequence systems
(\citealt{wyder07}; \citealt{schiminovich07};
\citealt{martin07a}; \citealt{martin07b}). The UVOCMD
that first identified this bimodality and the intermediate green
phase can also be extended beyond the low redshift galaxies observed
by \galex\ into higher redshift regimes to help explore evolution
over time. As large populations of galaxies at higher redshifts
have been added to the UVOCMD from more recent surveys (e.g.,
\citealt{2013Ilbert}), we can now state that the mass fraction of
galaxies in the red sequence vs. the blue sequence has grown by a
factor of at least $\sim$ 3 since $z\sim$1. This indicates that
there is a global evolution over the last $\sim 8$~Gyr of cosmic time that we are only just
beginning to understand and characterize. 
A new generation of galaxy studies anchored by the \uvex\ photometric and spectroscopic surveys would enable us to directly measure the evolution of galaxies across the GHR.
These data will directly address three key questions:
1) What equilibrium processes
create the star formation ``main-sequence''? 2) What processes drive
galaxies out of equilibrium leading to star formation ``quenching''?
and 3) How do galaxy halos and galaxies co-evolve, and how does
this co-evolution govern galaxy evolution?

\subsubsection{The Star Formation Main Sequence}

Star forming galaxies form a tight ``main sequence'' with SFR
proportional to stellar mass (\citealt{noeske07};
\citealt{wyder07}; \citealt{salim07}): the star formation main sequence (SFMS).
To understand why, we need to understand the processes
driving a return to equilibrium when galaxies are perturbed by
mergers/accretion/gas exhaustion. The responsible processes 
likely include galactic winds/fountains, accretion of new gas, 
and adjustments in the star formation efficiency in star-forming (SF) regions. 
Untangling these influences requires a large sample of SF galaxies that can be split into
subsamples based on e.g., mass, deviation from the main sequence, and environment. The combination
of FUV/NUV and optical/IR photometry and targeted spectroscopy
allows determination of recent SF history, which probes deviations
and returns to equilibrium.

Key to this investigation is the power of combining integral and
differential constraints on evolution, which can only be accomplished
with very large and homogeneous photometric samples spanning the
UV/O/IR. The growth of stellar mass in galaxies is an integral
constraint, while the SFR history is a differential constraint
(\citealt{madau14}). The UVOCMD vs. redshift is an integral constraint,
while a measure of the flux of galaxies across the UVOCMD provides
the next level of differential constraint
(\citealt{martin07a}; \citealt{martin17}; \citealt{darvish18}; \citealt{2021MNRAS.tmp.2945D}). By
combining GHR integral and differential constraints with galaxy-halo
connection methods (e.g., the Halo Occupation Distribution [HOD]), we
can take a step towards the ideal of ``watching'' galaxies evolve
by statistically weaving together the snapshots we observe. Because
stellar and dark halo mass can only increase with time, and because
the integral and differential constraints are linked by a continuity
equation, we can begin to assemble an ensemble of individual galaxy
star formation histories, and correlate these with halo mass and
environment. 
This approach will provide a powerful new constraint on galaxy
formation and evolution models and numerical simulations.

How does this work in the case of the SFMS? The best
way to diagnose equilibrium processes is to observe them out of
equilibrium, where they will be maximally subject to the
return-to-equilibrium processes. If we consider galaxies on, below,
and above the SF main sequence, the integral constraint comes from
how many galaxies are in each bin vs. redshift. The differential
constraint comes from how fast galaxies are moving between these
states. The distribution in bursting ``Star Formation Acceleration''
($SFA\equiv d(NUV-H)/dt \sim d(sSFR)/dt>0)$ and quenching ($SFA<0$)
gives the relative fraction vs. the speed of these processes. Fast
bursting could be produced by mergers, and slow bursting by changes in
gas accretion. Fast quenching could be produced by galactic winds
evacuating the gas, while slow quenching by starvation of gas
(accretion strangulation by hot gas or ram pressure/tidal stripping
of satellites). Therefore the spread in SFR and the distribution in
star formation acceleration give strong constraints on the equilibrium
processes on the SFMS.

We do not know whether the lowest mass galaxies exhibit a SFMS, and
what the dispersion around the relation is. It might be the case
that very low mass galaxies have long periods of quiescence punctuated
by SF bursts, as is suggested by resolved stellar photometry of
nearby dwarf galaxies (\citealt{2014ApJ...789..147W}; 
\citealt{2014ApJ...789..148W};
\citealt{2014ApJ...794L...3W}; \citealt{2015ApJ...804..136W}). \uvex will
supply the necessary sample to determine the star formation history
in the lowest-mass galaxies in the universe. UV observations provide
a direct measurement of SFR in these low mass galaxies because of
their low mass, metallicity, and dust extinction.

\subsubsection{Cosmic Quenching}

Cosmic quenching has shut down star formation by more than an order
of magnitude since $z\sim2$. Quenched galaxies exist at all masses
but dominate the high mass population.
Yet this fundamental evolutionary process is still poorly understood. Why
do most high mass galaxies quench? Why do some intermediate and low
mass galaxies quench, while others do not? What are the physical
processes responsible for quenching: merging, starvation/strangulation,
stripping, feedback? How does quenching depend on halo mass,
environment, position in the cosmic web, central/satellite
identification? What causes rejuvenation in quenched galaxies?

The approach discussed in the previous section
allows us to measure the flux of galaxies from the SFMS to
the red sequence (quenching through the green valley, $SFA>0$) and
the flux of galaxies experiencing periods of rejuvenation (bursting,
$SFA<0$). The speed of quenching and bursting is determined by the
physical processes at work, such as strangulation (slow) and
mergers/AGN winds (fast). Combining the integral and differential
constraints will test whether we have fully accounted for the
evolutionary tracks of galaxies across the GHR. The distribution
of star formation acceleration vs. position on the GHR, vs. 
environment (e.g., local density),
AGN presence, and central/satellite status will provide powerful
new constraints on models and simulations of galaxy quenching. The
FUV and NUV provide the maximum leverage to study quenching because
the UV traces SFR even for very low sSFR galaxies, where quenching
has already begun. In other words, they magnify the green valley,
where galaxies in transition reside.

\begin{figure*}
 \centering
  \includegraphics[width=\textwidth]{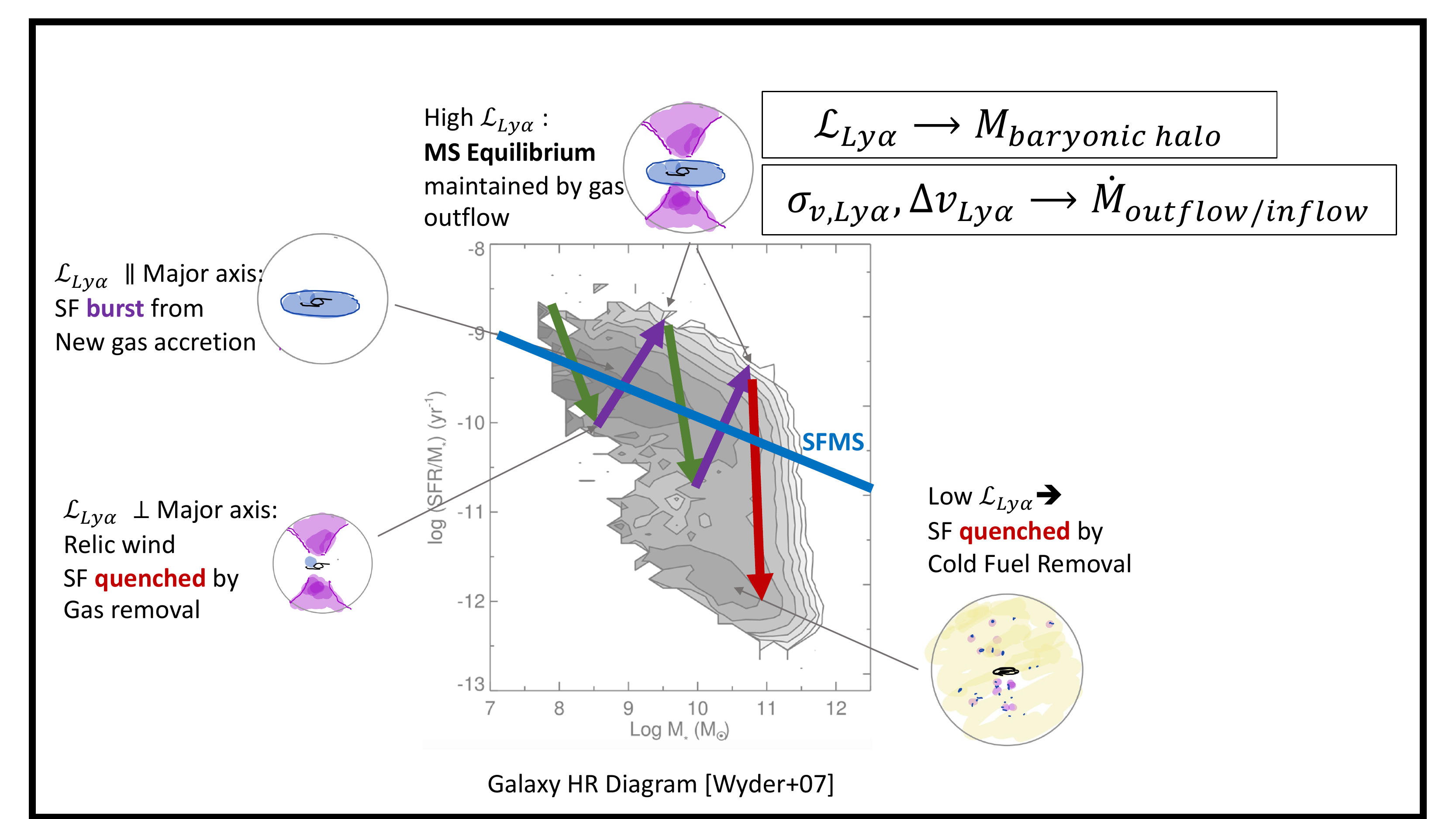}
   \caption{\small Galaxy HR diagram related to baryonic halo traced
   by \lya. Halo \lya\ luminosity gives baryonic halo mass, and
   line profile gives mass flux in outflows or inflows through the
   halo. \uvex-traced SFR and star formation acceleration give 
   recent bursting/quenching history.}
 \label{fig:ghr_halo}
\end{figure*}

\subsubsection{Galaxy-Halo Co-Evolution}

Galaxy formation and evolution is governed by the flow of gas,
metals, and energy into and out of the halo. Galaxy evolution is
really galaxy-halo co-evolution. Most of the baryons in the universe
are located in the halos and CGM of galaxies and
the intergalactic medium (IGM; \citealt{tumlinson17}). The baryons 
required for star
formation in galaxies are delivered through the halo and returned
there by galactic winds. Thus, the SFMS and the causes of cosmic quenching may
ultimately be tied to the flow of gas in the halo. A census of this
halo gas vs. galaxy mass, star formation history, and environment
is required to understand the co-evolution of galaxies and their
halos.

\lya is the most sensitive tracer of halo gas in emission. \lya
halos can be measured photometrically by stacking analysis of
galaxies with redshifts that place \lya\ at the peak of the FUV band 
(compared to neighboring redshifts). Because of its long slit and 
fast spectrograph design, and the fact that the spectrograph 
will be operating continuously during 
imaging exposures, \uvex\ will obtain serendipitous spectra of 
thousands of \lya galaxy halos, providing information about 
kinematics, inflows, and outflows.
\lya halo measurements can be stacked on galaxy properties
on the GHR diagram in order to determine global scaling laws between
halos and galaxies (Figure \ref{fig:ghr_halo}). The key observables
are the \lya luminosity ($L_{\alpha}$), line profile (width, mean),
and mass to light ratio ($M/L_{\alpha}$). For example, low mass
galaxies that are above and below the SFMS
can be compared in $M/L_{\alpha}$ to determine the role of halo gas
flows in maintaining main-sequence equilibrium. Quenching galaxies
may show lower $M/L_{\alpha}$ than galaxies of the same mass on the
SFMS. Metal lines (\ion{O}{4}, \ion{O}{6}, \ion{N}{5}, \ion{C}{4}, 
\ion{C}{3}], \ion{Si}{4}) may also be detected from warm baryonic halos.

\subsubsection{Tests of Fundamental Baryonic Structure Formation Processes}

In order to understand a fundamental astrophysical process such as
star formation, it is critical to vary key parameters to extremes
in order to explore how the process changes. In particular, the
lowest mass, lowest metallicity, and lowest gas density regimes of
star formation probe the impact of mass, metallicity, and density
on this process. UV is unique in its ability to trace star formation
in low mass, low metallicity, and low density regimes, providing
key tests of star formation scaling laws, the IMF, and the root
cause of low baryon efficiency and low metallicity in the lowest
mass galaxies.

\begin{figure*}
 \centering
  \includegraphics[width=\textwidth]{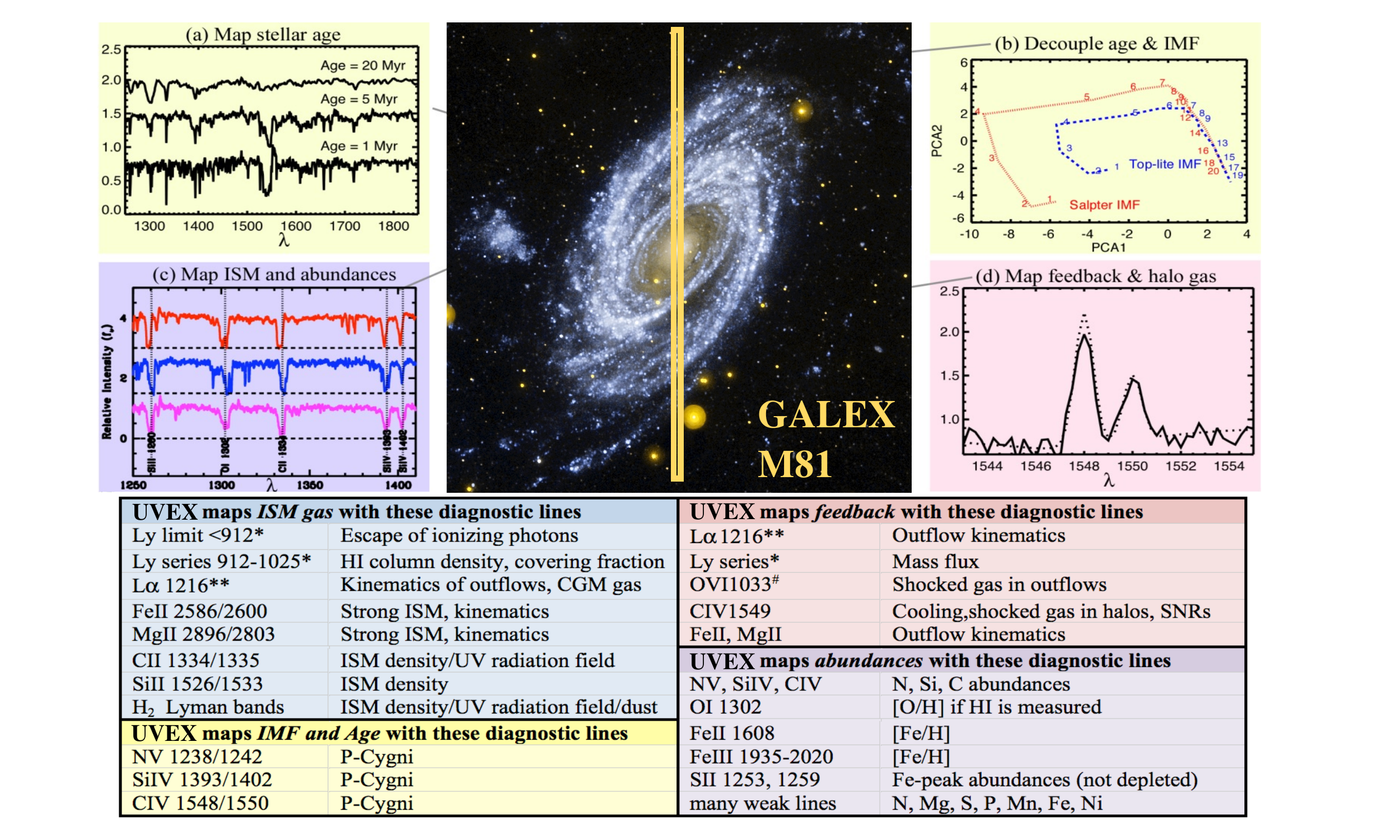}
   \caption{\small UV spectroscopy provides key physical diagnostics
   of galaxies, including metallicity (notably carbon), IMF, stellar
   age, and the presence of feedback. \uvex\ long-slit spectroscopy will
   provide exquisite spectra of nearby and distant galaxies in the
   FUV and NUV.}
 \label{fig:uv_spectra}
\end{figure*}

For example, \galex discovered that many galaxies have ``Extended
UV'' (XUV) disks (\citealt{thilker05}; \citealt{thilker07}; 
\citealt{lemonias11}),
extended regions of low-density gas and star formation unlike the
bulk of most disk galaxies. These extended regions may be produced
by on-going accretion of gas from the IGM, and if so could be used
to study the effects of IGM accretion on galaxy growth. XUV disks
and low surface brightness galaxies are laboratories for studying
the extreme low end of the Schmidt-Kennicutt star formation
scaling law (\citealt{1998Kennicutt}). \galex showed
that there is a sharp transition in the power-law Schmidt-Kennicutt 
dependence of SFR density with gas density (\citealt{wyder09};
\citealt{2008Bigiel}). Star
formation continues, but the dependence is much steeper, suggesting
a physical transition perhaps produced by the lack of dust and
resulting low H$_2$ formation rate and H$_2$/\ion{H}{1} fraction
(\citealt{krumholz13}). A deep \uvex survey would provide a definitive
sample of low-mass galaxies and XUV disks, allowing the study of
the most extreme examples. Follow-up of these with long-slit UV
spectroscopy would constrain stellar populations, star formation
rate and history, metallicity, and dust content. 

The IMF has been assumed to be universal
based on relatively limited data. The high-mass end of the IMF may
be particularly sensitive to key processes, since these low-probability
objects require the most massive molecular clouds to form without
the negative feedback of stellar winds. A powerful constraint on
the high-mass end of the IMF comes from comparing H$\alpha$ and UV
luminosity, since the former traces the highest mass stars and the
later traces high- to intermediate-mass stars. \galex provided strong
but not definitive evidence for a top-light IMF in low mass galaxies
(\citealt{uvex13}; \citealt{meurer09}). The challenge in interpretation is
that the burstiness of star formation history in low mass galaxies
produces a noisy H$\alpha$/UV ratio which could be biased
(\citealt{uvex14}). H$\alpha$ and UV respond differently to dust
extinction \citep{calzetti00}, which adds uncertainty and possible
bias (\citealt{2005Seibert}; \citealt{2007Johnsona}; \citealt{2007Johnsonb};
\citealt{salim20}).
However, with a large, homogeneous, multi-wavelength sample it should
be possible to simultaneously solve the star formation history (SFH), 
dust, and IMF problem
using the connect-the-dots approach for SFH described above 
and recent techniques to constrain extinction and extinction laws
(\citealt{salim20}).

The lowest mass galaxies are also the least successful at making
stars (lowest baryon-to-dark matter ratio) and at making (or
keeping) metals. Since galaxies are as a whole inefficient at making
stars, low mass systems are perfect laboratories for uncovering the
processes which prevent efficient star formation and metal retention,
determining the low-mass end of the baryon-to-dark matter efficiency
relation (\citealt{moster13}) and the mass-metallicity relation
(\citealt{tremonti04}). 
It should be possible to use a differential constraint on SFH, 
in combination with measurements of \lya\ emission in low mass galaxy halos,
to track the ebb and flow of star formation in low mass systems, and relate
this to the feedback-driven outflows which are believed to be the
root cause for the low baryon efficiency and metallicity of low
mass galaxies.

A UV spectroscopic subsample of low mass galaxies and low density star
formation regions would provide critical calibration of the photometric
sample. With fast \uvex\ long-slit spectroscopy, it will be possible
to measure UV spectra in low surface brightness conditions where
star formation is below the canonical threshold. Spectroscopy
provides critical diagnostics of stellar age, metallicity, feedback,
and the IMF (Figure \ref{fig:uv_spectra}).

\subsection{Cosmic Explosions}
 \label{sec:CosmicExplosions}
 
\begin{comment}
\begin{figure*}
 \centering
  \includegraphics[width=0.75\textwidth]{Cartoon2.png}
   \caption{\small Cartoon showing the mapping between the last nuclear
   burning stages before core-collapse, (lower left plot),  the
   ejection of shells of material by the progenitor as a result of instabilities developed in this process, and the observables (i.e., UV light-curves and spectra; plots on the right). Rapid
   turn around UV spectroscopy of infant explosions in the hours after core-collapse (Fig. \ref{Fig:IIPspecfig})  can uniquely reveal the composition, kinematics
   and mass distribution of shells of material ejected during the
   last moments of life of the progenitor star. In this way, UV
   observations can reveal details about core burning instabilities
   that would not be otherwise accessible. } 
 \label{Fig:Cartoon2}
\end{figure*}
\end{comment}
 
Supernovae (SNe; \citealt{Branch2017}) play a major role in the structure and evolution of galactic ecology and thus constitute a vibrant research area of modern astrophysics. Stellar explosions inject energy, momentum and newly synthesized metals into the interstellar medium. With the recent discovery of kilonovae resulting from double NS coalescence (e.g., \citealt{MarguttiChornock21}), astronomers now have a better understanding of how cosmic explosions contribute to the buildup of the periodic table. 

We identify three main frontier areas in the field of stellar explosions.  Firstly, massive stars that eventually undergo gravitational collapse (core collapse SNe) leave stellar residue in the form of NSs and BHs. The natal properties of neutron stars -- rotation and magnetic field strength -- appears to range over many orders of magnitude. Newly formed BHs can be born spinning slowly and dominated by fallback or spinning rapidly and generating tremendous amount of power via accretion of stellar debris. In either case, in some fraction of the events the central object injects power following the explosion (e.g., luminous magnetar or power generated by an accreting black hole), and the resulting supernova is distinct and bright. In fact, the spinning accreting black hole is an accepted model for GRBs (e.g., see  \citealt{Woosley2006,Hjorth12,Cano2017} for reviews of GRB-SNe) and the magnetar model is a popular explanation for super-luminous SNe (SLSNe; see below) and might be behind some observational manifestations of ``Fast and Blue Optical Transients'' as well (FBOTs; see below).

Next, the structure and chemical composition of stars at the time of explosion, and their very recent mass-loss history in the final $\sim0.1-100$  years before stellar death are among the least understood aspects of stellar evolution (e.g. \citealt{smith2014}) and have direct consequences on the explodability of a star (e.g., \citealt{Janka17} and references therein). Specifically, a major open question is whether the physical origin of instabilities that act in the final moments of stellar life ($\delta t<1$ yr) and that originate deep down in the stellar core (e.g., \citealt{Quataert16}, \citealt{Fuller18}, \citealt{Morozova20}, \citealt{Leung20,Wu21}) can trigger the most extreme episodes of mass loss in massive stars across the mass spectrum that we have just started to uncover with observations (for some examples see, e.g., \citealt{Pastorello13,Margutti14,smith2014}, \citealt{Margutti17}, \citealt{Bruch21}, \citealt{Strotjohann21, JacobsonGalan21}). Extreme mass loss timed with core-collapse might play an important role in defining the thermal UV-optical emission of the emerging, albeit motley, class of FBOTs (\citealt{Drout14}, \citealt{Arcavi16}, \citealt{Pursiainen18}, \citealt{Ho2021}). 
From a different perspective, the non-thermal emission of long GRBs and FBOTs represents one of the few real-time observational manifestations of the compact object formed in core-collapse: the properties of their relativistic or sub-relativistic jets directly link back to those of the newly-formed neutron star or black hole (\citealt{Margutti19},
\citealt{Ho19}, \citealt{Ho19b}, \citealt{Ho20}, \citealt{Coppejans20},
\citealt{Perley19}, \citealt{Perley21}). 

Last is our limited understanding of the progenitor star populations of several types of SNe. In particular, we lack a clear picture of the progenitors of Hydrogen-poor supernovae (which together comprise $>$50\% of SNe by volume, \citealt{Li11_LOSS}), including normal core-collapse explosions and SLSNe (e.g., \citealt{Quimby13}, \citealt{Chomiuk12}). Similarly, a complete picture of the origin of SNe of Type Ia, which have been widely employed as cosmic distance ladders to reveal the accelerating Universe \citep{Riess98}, continues to be shrouded (e.g., \citealt{Maoz14}).

In summary, the past decade has uncovered a dizzying range of new phenomena, but our overall understanding of them is poor. This lack of understanding is significant, as it further impacts the estimates of the initial stellar mass function in galaxies and star formation through cosmic time (e.g., \citealt{smith2014} and references therein). Stated succinctly: linking stellar progenitors to SN types and to their compact object remnants is a fundamental goal in modern astrophysics.

It is clear that further progress requires observational guidance, and indeed one of \uvex's main science objectives is to perform follow-up spectroscopy of core-collapse supernovae (see Section~\ref{sec:ccsne}). Below we outline further the ability of \uvex to advance our understanding of the phenomena and physics of the frontier areas discussed above. We first summarize the unique diagnostic value provided by UV spectroscopy (Section~\ref{sec:UVspectra}). We then discuss the key progress that will result from focused UV spectroscopic studies of the most common types of SNe (Section~\ref{sec:UVnormal}). 
We conclude with how the all-sky \uvex survey will naturally allow for exploration of exotica, in particular the most luminous and also the rarest events (Section~\ref{sec:exotica}).

\subsubsection{The Unique Role of UV Spectroscopic Observations of Cosmic Explosions} 
 \label{sec:UVspectra}

Following a stellar explosion, the debris is very hot. The peak emission naturally cascades from high energy to low energy as time goes by. 
The first radiation able to escape the explosion is the shock breakout (see \citealt{Waxman17} for a recent review), which peaks in the UV on timescales of $\approx$~hours for many extended stellar progenitors (e.g., \citealt{Campana06, Bersten18}). The next phase of UV emission can arise from two distinct phenomena: (i) the SN shock interaction with a companion star (e.g., \citealt{Kasen10}, \citealt{Liu15}); (ii) the SN shock interaction with the very nearby CSM, which was sculpted by the recent mass loss by the progenitor star before stellar death (e.g., \citealt{smith2014,Chevalier11,Chevalier17} and references therein). For massive stellar explosions, scenario (ii) applies.
The gas then rapidly expands and cools, shifting the peak of the emission towards increasingly longer wavelengths. This combines with the higher line blanketing at lower temperatures to quickly suppress the UV flux within a few days after the end of the interaction phase. This final phase of declining UV emission (and progressively emerging optical emission) from the shock heated ejecta can thus be studied by using ground-based observations (see the temporal evolution of the UV and $r$-band light-curves from a red supergiant star explosion in Figure \ref{Fig:PhaseSpace}).

Our focus here is on the second phase of UV emission, which probes the mass-loss history of the star.
This phase is rich with diagnostics (e.g., mass and composition of the ejecta; mass and radius of the pre-explosion ejected shell or shells; etc.). The short-lived nature ($\delta t\lesssim$ 48 hours) of the UV emission from the explosion's shock interaction either with the companion or with confined CSM shells deposited in the environment by the dying star in the years before explosion, coupled with the current complete lack of UV spectroscopic facilities with rapid repointing capabilities, makes this an assuredly fertile field for both detailed studies and exploration.

Obtaining rapid-response broadband UV spectroscopic sequences is beyond the capabilities of current missions (see Section~\ref{sec:ccsne}). 
As we detail in the next section, at early times optical spectra are relatively featureless (see the UV-optical spectrum displayed in Figure \ref{Fig:PhaseSpace}), and rapid optical spectroscopy does not access enough bright emission lines to constrain parameters (see, e.g., Section 4 of \citealt{Groh14}, for the case study of SN\,2013cu, a SN with arguably the best optical flash spectroscopy, and yet an unconstrained stellar progenitor type because of the lack of UV spectroscopic coverage).

\begin{figure*}
 \centering
  \includegraphics[width=0.75\textwidth]{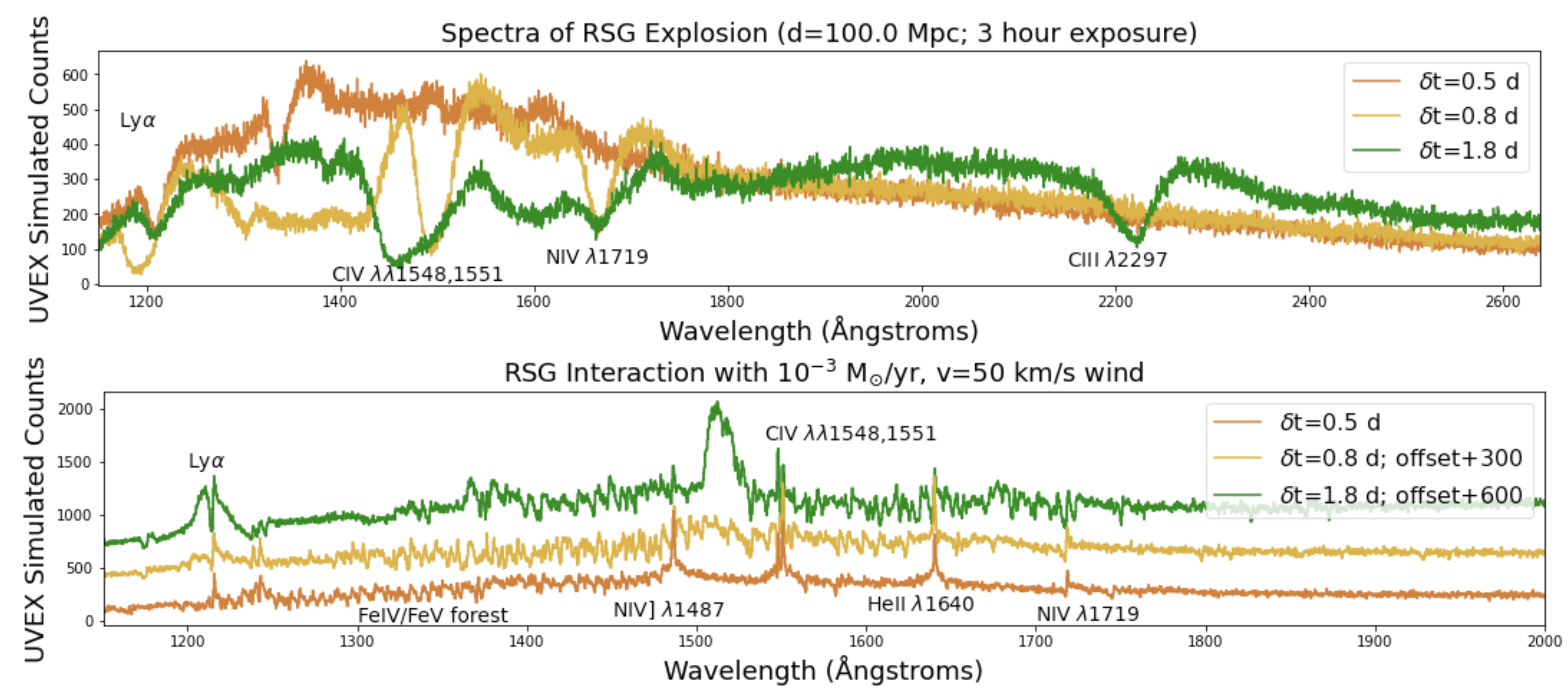}
  \caption{Simulated multi-epoch early \uvex spectra of the RSG explosions of Figure \ref{Fig:PhaseSpace}. We assume a representative distance of 100 Mpc and a \uvex exposure time of 3 hours. \emph{Upper panel}: RSG explosion without CSM. \emph{Lower panel}: RSG explosion embedded in thick CSM created by large pre-explosion mass-loss with rate of $10^{-3}\,\rm{M_{\sun}\,yr^{-1}}$ and ejected with velocity of $v_w=50\,\rm{km\,s^{-1}}$.
  The presence of dense CSM in the explosion's surroundings completely changes the spectroscopic appearances of the SN (and increases the fraction of flux in the UV, Fig. \ref{Fig:PhaseSpace} ). In both cases (with or without CSM), the UV spectrum undergoes  very rapid evolution during the first few days after the explosion, and allows us to constrain the kinematics, chemical composition and ionization stage of the emitting material. At these epochs the optical emission is significantly fainter. Simulated spectra taken from \citet{gezari08} and \citet{dessart17}.
  }
 \label{Fig:IIPspecfig}
\end{figure*}

\uvex, with its ToO capabilities and highly sensitive low-resolution spectrometer, is perfectly matched to realize the vision described above. \uvex spectroscopic observations of stellar explosions will be capable of mapping for the first time the chemical composition, kinematics, and location of the innermost layers of CSM of normal (Section~\ref{sec:UVnormal}) and exotic (Section~\ref{sec:exotica}) stellar explosions, providing information that would not be otherwise available. In the case of Type Ia SNe, \uvex observations have the potential to unveil the nature of the companion stars to exploding CO white dwarfs. Very early UV spectroscopy thus provides a direct probe of the immediate explosion's environment and progenitor system.

\subsubsection{UV Spectroscopic Studies of ``Ordinary'' SNe} \label{sec:UVnormal}

Recent observations of outbursting behavior in stars before core-collapse have shaken the traditional understanding of mass loss in evolved massive stars (e.g., \citealt{smith2014} for a recent review). A combination of pre-explosion optical imaging and post-explosion optical spectroscopy have demonstrated that a large fraction of massive stars, spanning all known classes (from ordinary Type IIP SNe to rare broad-lined Type Ic SNe), undergo major instabilities in the years preceding stellar death that spew dense shells of material into the surrounding environment (e.g., \citealt{Ofek10}, \citealt{Ofek14}, \citealt{Margutti14}, \citealt{Margutti17}, \citealt{Milisavljevic15}, \citealt{Ho2019gep}, \citealt{Bruch21}, \citealt{Tartaglia21}, \citealt{Strotjohann21}). In some cases the amount of expelled matter can reach $\gtrsim 1~\msun$, completely changing the observable properties of the resulting transient. This was not predicted on theoretical grounds and challenges physical mechanisms that drive mass loss in evolved massive stars \citep{smith2014}. One leading model suggests that these mass ejections can result from nuclear burning instabilities in the final stages of stellar evolution (\citealt{Quataert12,Fuller18,Wu21}). 

To advance our understanding of the physics behind these mass ejections it is necessary to know the chemical composition, ionization stage and kinematics of the ejected material.
UV spectroscopy is crucial because: (i) compared to the optical, it accesses a significantly larger number of spectral transitions, adding crucial constraints to an otherwise under-constrained problem (Figure \ref{Fig:PhaseSpace}, lower panel; Figure \ref{Fig:IIPspecfig}); (ii) virtually all young stellar explosions are bright UV emitters, with a spectral energy distribution that peaks in the UV, and a UV flux that is several times larger than the optical flux (Figure \ref{Fig:PhaseSpace}, upper-left panel); (iii)  UV probes resonance lines such as C IV $\lambda\lambda$1548,1551, He II $\lambda$1640, and N IV $\lambda$1719 which, given their high optical depths, are detectable at significantly lower wind densities compared to the optical and allow a much more precise determination of the wind velocity structure; and (iv) finally, UV also probes highly-ionized Fe lines at $\lambda\lambda$ 1200–1450, which can be used as direct probes of the close CSM metallicity and would otherwise be inaccessible (at these high temperatures, no Fe transition is available in the optical).

\uvex\ will perform dedicated follow-up of core-collapse SNe as part of its primary science objectives (see Section~\ref{sec:ccsne}), but will also have the capacity to investigate other ``ordinary'' types of SN. For example, in the specific context of Type Ia SNe, a novel way to gain insight into their progenitor systems is by studying the short-lived excess of UV emission that is expected to originate from the SN shock interaction with the companion star in the first $\sim48$ hrs after explosion (\citealt{Kasen10}, their Figure~3). 
Type Ia SNe are believed to originate from a binary system where the exploding star is a Carbon-Oxygen White Dwarf (C/O WD; e.g., \citealt{Maoz14}). The nature of the companion star is a matter of intense debate, as current observations point to WDs in some cases and non-degenerate companions (e.g. main sequence stars) in others. The UV burst properties (temperature, luminosity, duration) depend on the properties of the companion star and thus provide direct insight into its nature (e.g., \citealt{Brown12}). 

\uvex has the potential to acquire the first UV spectra of Type Ia SNe at $\delta t< 48$~hrs and constrain the nature of their progenitor systems.  To date, due to limitations of current observing facilities, the earliest UV spectroscopic observation of a Type Ia SN has been acquired at $\delta t \approx $+5 days with \hst\ (R.~J. Foley, private communication), well after the end of any emission from shock interaction with the companion.

To conclude, the acquisition of the first UV spectroscopic sequences of stellar explosions at very early times by \uvex will open an entirely new window of investigation on stellar death. 
\uvex will acquire rapid, high-cadence, UV spectroscopic sequences of the youngest stellar explosions identified by ground-based surveys (in 2028 we expect a large number of surveys to be operational, including ASAS-SN, BlackGEM, LS4, ATLAS, ZTF, PS-1 and 2, LAST, an upgraded EVRYSCOPE, and more) as well as SNe discovered by other spacecraft (e.g., \ultrasat). Additionally, \uvex is equipped to discover $\approx$6-10 SNe within hrs of explosion through its synoptic survey over the course of two years. 

\subsubsection{Probing the Exotica: the Rarest and Most UV-luminous Stellar Deaths}
 \label{sec:exotica}

\begin{figure}
 \centering
  \includegraphics[width=\columnwidth]{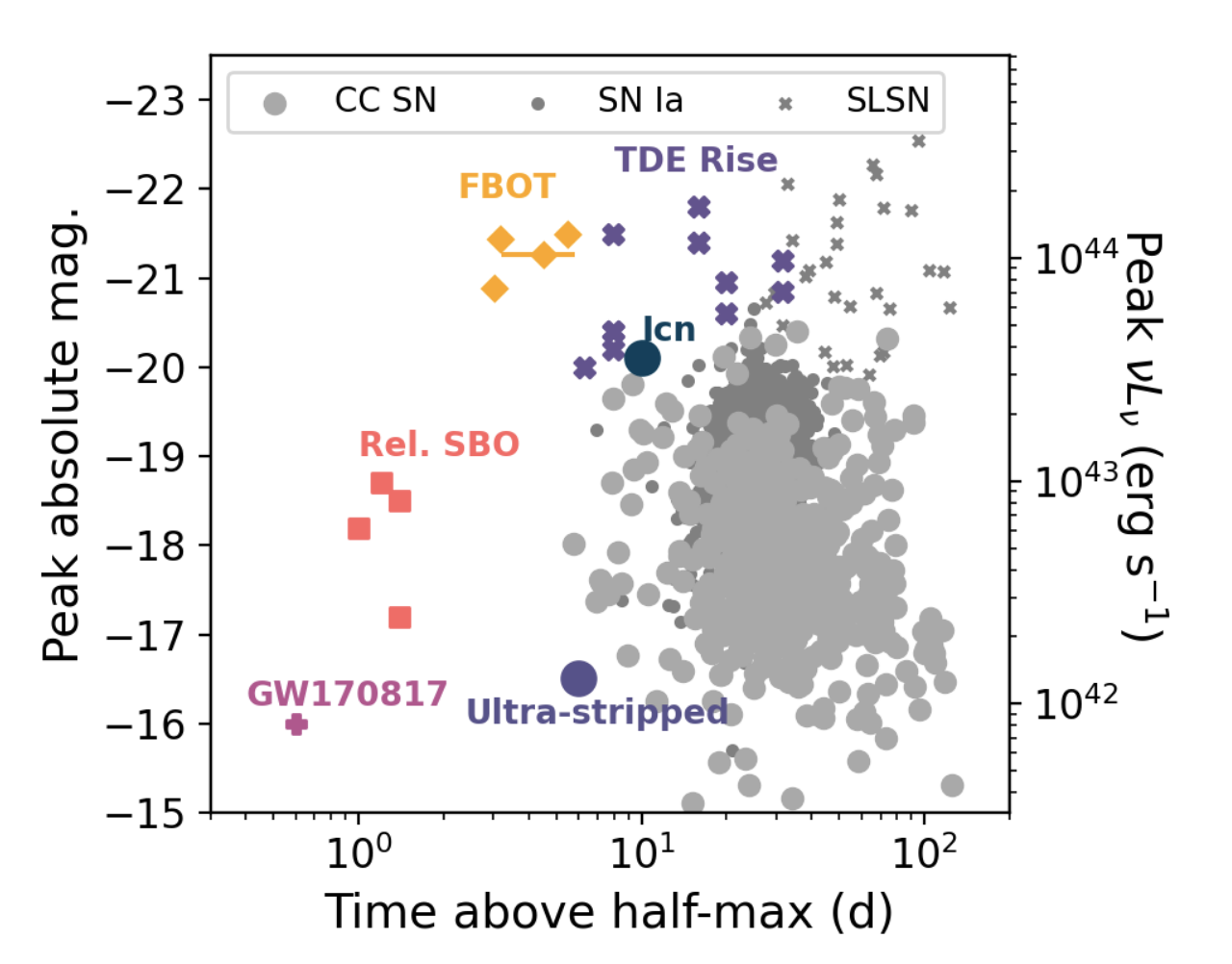}
   \caption{\small Luminosity vs. duration of optical transients, highlighting  classes of relativistic explosions that are prime targets for \uvex: Fast and Blue Optical Transients (``FBOTs''; yellow diamonds ), events powered by relativistic shock breakout (``Rel. SBO''; red squares), as well as ultra-stripped SNe (blue circle), counterparts to GW sources (here exemplified by GW\,170817, purple plus sign), the new class of Type Icn SNe and Tidal Disruption Events (``TDEs''; purple crosses). We place these transients in the context of Ia SNe, core collapse (CC) SNe, and SLSNe  from the ZTF Bright Transient Survey. References: \cite{Perley19,Fremling20,Margutti19,Coppejans20,Ho20,Perley21}.
     Modified from \citet{Ho2021}.
   }
 \label{fig:lum-duration-opt}
\end{figure}

In a typical core-collapse SN, the explosion is largely isotropic and the ejecta is accelerated to velocities up to about 10 percent of the speed of light -- the consequence of a neutrino-mediated spherical shock produced following core bounce \citep[e.g.,][]{Janka17}. However, since the late 1990s it has been realized that some massive stellar explosions are driven by a distinct mechanism involving the production of relativistic jets by a central engine: a rapidly-spinning neutron star or a black hole. The most extreme examples are long-duration gamma-ray bursts (GRBs; \citealt{Piran2004,Hjorth12,Cano2017}): extremely rare explosions involving ultra-relativistic (Lorentz factor $\Gamma > 100$) jets and almost exclusively discovered by high-energy satellites.  However, the discovery of ``low-luminosity'' GRBs (or ``X-ray flashes''; \citealt{Galama1998,Kulkarni98,Campana06,Soderberg06,Liang07}), and observations of relativistic explosions with no associated GRB detected by wide-field surveys at other wavelengths \citep{Soderberg2010,Cenko2013,Margutti14,Milisavljevic15,Ho2020blt}, suggest that GRBs are only the tip of the iceberg of a broader landscape of engine-driven phenomena spanning a wide variety of engine timescales, beaming angles, shock velocities, and CSM properties \citep{Lazzati2012,Margutti14,Milisavljevic15,Gottlieb2021}.  This suggests that the role of jets in end-of-life stellar explosions may be more significant than was previously appreciated by most of the astronomical community.

Of particular importance to this mission is the population of transients sometimes termed ``FBOTs'' (fast blue optical transients, Figure \ref{fig:lum-duration-opt}) and typified by the intensely-studied event AT\,2018cow (\citealt{Prentice18}, \citealt{Perley19}, \citealt{Kuin19}, \citealt{Margutti19}, \citealt{Ho19}), discovered in 2018.  These events rise and fade on timescales of just a few days (an order of magnitude faster than a typical SN; \citealt{Ho2021}), retain very high temperatures long after peak (\citealt{Perley19}, \citealt{Margutti19}), and have been shown in several cases to be accompanied by very luminous radio and sub-millimeter emission, indicating an energetic and mildly relativistic shock ($v\sim0.1$--0.6$c$) in a very dense CSM \citep{Margutti19,Ho19,Coppejans20,Ho20,Ho2021xnd,Bright2021}. Hydrogen and helium were detected in the late-time spectra of AT2018cow (\citealt{Perley19}, \citealt{Margutti19}), indicating an important distinction from the progenitors of GRBs, which are exclusively accompanied by H/He-poor SNe.

FBOTs are fundamentally UV phenomena -- the emission before, at, and (in many cases) after maximum light peaks in the UV \citep{Drout14,Pursiainen18}.  However, they have only so far been discovered via ground-based optical surveys, and the only constraints from spectroscopy have come from the optical. The result is that even in this era of wide-field optical surveys, the discovery rate is low ($\sim$1 per year), and spectroscopy has been minimally constraining. Almost all of our knowledge of this class of phenomena originates from AT2018cow itself, and there are few constraints on how these events are related to FBOTs and relativistic transients more broadly.  The next ten years are unlikely to change this paradigm, since even the most powerful new time-domain facilities (e.g., the LSST carried out by the Vera C. Rubin Observatory) will not be capable of recognizing similar transients at the critical, short-lived early phases of their evolution.

A dedicated UV facility would offer several key advantages in the study of this event class. FBOTs are more luminous at UV wavelengths, the background is greatly reduced compared with the optical, and UV spectral diagnostics (with the extensive set of strong resonance lines) will be far more powerful at examining the properties of the outflow and surrounding CSM compared to what can be done with optical observations alone.

By the launch of \uvex, the entire sky will be surveyed in the soft X-ray bands by facilities such as Einstein-Probe \citep{Liu2021}, transforming the study of relativistic explosions. Low-Luminosity GRBs (LLGRBs, i.e. GRBs with significantly lower-luminosity $\gamma$-ray prompt emission) may represent phenomena intermediate to classical GRBs and ordinary SNe (\citealt{Soderberg06b}, \citealt{Liang07}, \citealt{Nakar2015}, \citealt{Cano2017}). \uvex\ will obtain the first UV spectra of LLGRBs. The volumetric rate is uncertain, but could be between 0.1\% and 1\% of the core-collapse SN rate (\citealt{Soderberg06}, \citealt{Liang07}). Their luminous UV emission is accessible to \uvex for spectroscopy out to $z=0.06$, so we estimate between 6 and 60 candidates per year.

In addition, an entirely new class of strongly interacting SNe has been identified in the last few months: Type Icn SNe, i.e. SNe that show clear spectroscopic signatures of shock interaction with a He- and H-poor medium (\citealt{Fraser21,GalYam2021,Perley21csp}). Type Icn SNe join the groups of Type Ibn and Type IIn SNe, which show interaction with He-rich and H-rich CSM, respectively. Type Icn SNe descend from stellar progenitors that shed their envelopes at significantly earlier times before collapse, compared to their Ibn and IIn cousins. The physical nature of Type Icn SNe is unknown. These SNe are rare but UV-luminous (Figure \ref{fig:lum-duration-opt}), and thus detectable out to large volumes. \uvex has the potential to acquire the first UV spectra of a Type Icn SN and provide key information on its origin.

In summary, \uvex\ will shed exciting new light on the rarest and most mysterious cosmic explosions, transforming our understanding of the many manifestations of stellar death. 

\subsection{Active Galactic Nuclei}
 \label{sec:AGN}
 
Active galactic nuclei (AGN), corresponding to the phases in a galaxy's
life when its central supermassive black hole (SMBH) is actively accreting
material, are intrinsically UV phenomena.  Gravitational potential
energy from in-falling material becomes kinetic energy, and is then
released as thermal energy from the hot accretion disk that naturally
forms. Accretion disk temperature is inversely proportional to the
central mass, so while emission from the accretion disks of stellar
mass compact objects in the Galaxy, i.e., Galactic binaries, peak
in the soft X-ray regime, emission from the accretion disks of
SMBHs peak in the UV.  Indeed, the first quasars
were identified as unusually blue, quasi-stellar counterparts to
radio sources \citep{Schmidt1963}, and the so-called ``Big Blue
Bump'' which dominates quasar SEDs
in the spectral range from $\sim 100$~\AA\ to 3000~\AA\ is dominated
by 10,000-100,000~K thermal emission from the accretion disk
\citep{Sanders1989}.

Thermal UV emission from the accretion disk provides the source
photons for two other distinguishing features in AGN SEDs. UV
emission from the accretion disk is reradiated in the IR by
dust, often assumed to be in form of a torus of material that
obscures the higher energy emission along certain lines of sight
(e.g., \citealt{Stern2005}). And UV emission is Compton up-scattered
into the X-ray range by the AGN corona, creating the characteristic
power-law X-ray spectrum of actively accreting, unobscured AGN.

Thus, it should come as no surprise to find that UV observations play 
an outsized role for AGN studies.  In
the following section, we consider the scientific potential of \uvex\ for
AGN studies, with an emphasis on the discovery space enabled by
sensitive, synoptic UV imaging, UV spectroscopy, and the $\geq 50-100\times$
increase in two-band UV imaging sensitivity enabled by \uvex\
relative to \galex.  The related phenomenon of tidal disruption events
(TDEs) are discussed in the next section (Section~\ref{sec:TidalDisruptionEvents}). However, we do 
note that TDEs are also expected to happen in active galaxies.
\citet{Ricci2021} present one dramatic candidate event, while
\citet{Frederick2021} discuss optical transients in narrow-line
Seyfert~1 galaxies, some of which they associate with likely TDEs.  
In addition, \citet{Stein2021NatAs} discuss the likely coincidence of a PeV neutrino with a TDE in an active galaxy. UV observations are a key aspect of their analysis, which shows that the EM observations can be explained with a multi-zone model: a UV-bright photosphere powering an extended synchrotron-emitting outflow in which high-energy neutrinos are produced.  Their model suggests that TDEs with mildly-relativistic outflows are likely important contributors to the cosmic neutrino flux, particularly at high energies, and also shows the key role that UV observations play in probing extreme AGN events, touching on AGN (this Section), TDEs (Section~\ref{sec:TidalDisruptionEvents}), and multimessenger astrophysics (Section~\ref{sec:MMA}).  Finally, from the theoretical side, \citet{McKernan2021} and McKernan
et al. (in prep.) discuss how stars embedded in AGN accretion disks
can lead to TDEs.

\subsubsection{Quasar Variability}

Time domain surveys have long been recognized as important tools
for studying AGN. Indeed, optical continuum variability was recognized
as a common feature of quasars within a year of their initial
discovery \citep{Matthews1963}, and was quickly exploited as a means
of identifying quasars, particularly those that might be missed by
the UV-excess technique due to their higher redshift (e.g.,
\citealt{vandenBergh1973}). Since then, several groups have used
optical synoptic studies to construct quasar samples based on their
unique optical variability, thereby avoiding the inherent biases
of color selection. A non-exhaustive list of such efforts include
studies of SDSS Stripe~82 \citep{MacLeod2011}, MACHO \citep{Pichara2012},
OGLE \citep{Kozlowski2013}, COSMOS \citep{DeCicco2019}, and GOODS-S
\citep{Pouliasis2019}. \citet{Graham2014} show how combining optical
variability from CRTS with mid-IR colors from \wise\ (e.g.,
\citealt{Stern2012}, \citealt{Assef2013}) improves completeness and
reliability of quasar selection relative to variability or color
selection alone, and optical variability selection of quasars has
long been heralded as one of the many promising scientific results
to come from Rubin (e.g., \citealt{Ivezic2019}).

\uvex\ will provide a powerful probe of quasar UV variability, a
wavelength where quasars are $\sim 5 \times$ more variable than at
optical wavelengths \citep{Gezari2013}. This will be useful for a
wide range of quasar studies, from simply identifying quasars, to
correlating their UV light-curves with variability at other wavelengths,
to mapping out the central engine and determining black hole masses, to
exploring the new territory of extreme quasar variability recently
opened from optical time-domain surveys. In particular, since quasar
UV emission comes predominantly from the hot accretion disk, UV
studies are uniquely sensitive probes of extreme events and activity
close to the supermassive black hole.

By studying quasars at a wavelength where they are more variable,
\uvex\ has significant promise for identifying AGN based on their
variability, particularly in the low-redshift Universe where the
rest-frame UV photons are detected in the observed UV bands. Of
particular interest will be identifying AGN in low-mass, low-redshift
galaxies, which are predominantly star-forming \citep{Geha2012}.
UV color excess techniques will therefore be somewhat compromised
(e.g., \citealt{Latimer2019}), making UV variability a
powerful tool for identifying low-luminosity AGN in dwarf galaxies.
Obtaining a comprehensive census of such systems is particularly
exciting as a tool to understand early BH growth and
potentially answering the key open question regarding SMBH
``seeds'' (e.g., \citealt{Reines2013}): how was the
universe able to create billion-solar-mass BHs in less than
a Gyr (e.g., \citealt{Banados2018})? Different models for these
high-redshift seeds have different predictions for the massive BH
occupancy fraction in low-mass galaxies in the local universe
\citep{Reines2016}, which \uvex\ will be well-placed to test.

\subsubsection{Reverberation Mapping}

Although the spatial structure of the AGN central engine cannot be
directly resolved, the time-variable nature of AGN emission makes
it possible to resolve the inner structure by correlating variability
from different emitting regions and associating time delays with
the light travel time (\citealt{Blandford1982}, \citealt{Cackett2021}).
This method, known as reverberation mapping, uses synoptic observations
to determine the geometry of the central region; spectroscopic
monitoring opens such studies up to dynamical studies as well (e.g.,
\citealt{Pancoast2014}). Reverberation mapping provides our largest
sample of robust SMBH mass measurements,
supplemented with a few nearby sources, such as Sgr~A*, where we
can kinematically resolve the central regions (e.g., \citealt{Ghez2008},
\citealt{Cohn2021}). \citet{Panda2019} discuss photometric reverberation
mapping with Rubin, essentially modeling the huge sample of quasars
with six-band synoptic photometry as a substitute for spectroscopic
monitoring of a smaller sample, as is the current standard for
reverberation campaigns (e.g., \citealt{Barth2015}). Providing two
additional photometric bands at the wavelengths where quasars are most
variable, \uvex\ will be an important supplemental dataset for
photometric reverberation programs.

\begin{figure}[htbp]
 \centering
  \includegraphics[width=\columnwidth]{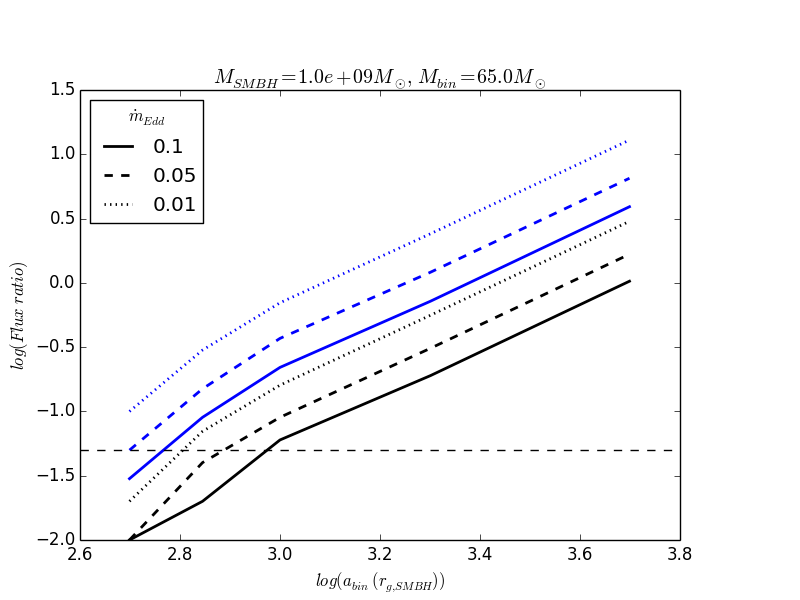}
   \caption{\small The merger of stellar mass BHs in an AGN accretion
   disk causes a bright UV flare. Plotted is the ratio of \galex\
   NUV flux (blue; similar to \uvex\ NUV) and ZTF $g$-band flux (black)
   from such an event relative to an unperturbed AGN disk. These
   curves are for a $10^9 \msun$ SMBH in an AGN accreting at a
   range of Eddington ratios: $\dot{m}_{\rm Edd} = 0.1$ (solid),
   $0.05$ (dashed), and $0.01$ (dotted), and assume that the merging
   BHs have a total initial mass of $65 \msun$, 5\% of
   the mass is lost to GWs from the merger, and a
   kick velocity of $100\, {\rm km}\, {\rm s}^{-1}$ for the resultant,
   merged BH. Ratios are plotted for a range of binary
   distance $a_{\rm bin}$ from the SMBH, in units of the SMBH
   gravitational radius, $r_{\rm g,SMBH} \equiv G\, M_{\rm SMBH}/c^2$.
   The horizontal dashed line corresponds to a flux increase of
   5\%.  The flux increase is larger at shorter wavelength. From
   \citet{McKernan2019}.
 \label{fig:AGN1}}
\end{figure}

\subsubsection{Stellar Mass Black Hole Mergers}

One exciting, though controversial, possibility is that a significant
fraction of BH-BH merger GW events occur within
AGN accretion disks and are detectable in EM (e.g., \citealt{Graham2020b}).
Stellar mass BHs are expected to be common in galactic
nuclei due to mass segregation, and accretion disk gas will dissipate
angular momentum, causing more massive embedded objects to in-spiral
more rapidly than less massive ones. This provides a natural
explanation for asymmetric mass mergers \citep{McKernan2020} such
as GW190814, which consisted of a $23 \msun$ BH coalescing
with a $2.6 \msun$ compact object \citep{Abbott2020a}, as well
as BHs more massive than 35-70 $\msun$, the maximum BH
mass expected from a supernova \citep{Woosley2017}, such as
GW190521 which consisted of an $86+66 \msun$ BH merger
\citep{Abbott2020b}.  As shown by Figure~\ref{fig:AGN1}, from
\citet{McKernan2019}, the disk gas provides baryons that are expected
to produce a UV flare due to the merger, assuming either a short
diffusion time or a thin AGN accretion disk. Jetted emission from a rapidly
spinning, merged BH can also yield UV photons, implying
that \uvex\ will be a powerful tool for studying counterparts not
only to NS merger events (discussed in Section~\ref{sec:emgw}), 
but also to BH merger events.

\subsubsection{Supermassive Black Hole Binaries}

Time-domain surveys have recently identified a population of quasars
with apparently sinusoidal, periodic light-curves (\citealt{Graham2015a},
\citealt{Graham2015b}, \citealt{Charisi2016}). With the important
caveat that many candidate periodic sources are claimed on the basis
of problematic statistical analyses (e.g., see discussion in
\citealt{Vaughan2016} and \citealt{Barth2018}), actual sustained
periodic or quasi-periodic variability is likely a signature of a
binary supermassive black hole (SMBH) system with a sub-parsec separation.
For example, \citet{DOrazio2015} showed that the periodicity of 
PG~1302-102 can be explained by relativistic Doppler boosting and 
beaming of emission from the mini-accretion disk around a secondary 
SMBH as it orbits a more massive primary at velocities of a few 
tenths the speed of light with a separation of $\sim 2000$~AU 
(i.e., $\sim 0.01$~pc). This model predicts a strong inverse 
correlation between variability amplitude and wavelength 
\citep{Xin2020}, which \uvex will test.

\subsubsection{Flaring AGN}

\uvex will also be important for studying AGN with unusual, extreme
optical light-curves, such as flaring AGN. For example, \citet{Graham2017}
reported on a systematic search for major flares in AGN in the
Catalina Real-time Transient Survey as part of a broader study into
extreme quasar variability. Requiring flares that are quantitatively
stronger than normal, stochastic quasar variability, \citet{Graham2017}
identified 51 extreme events from a sample of $> 900,000$ confirmed
and high-probability quasar candidates. The events typically lasted
900 days, with a median peak brightening of $\Delta m = 1.25$~mag.
The sample shows a range of flare morphologies, with some being more
symmetric, while others evolve with a fast rise followed by a slower,
exponential decay. While a subset of the sources appear consistent
with microlensing (e.g., \citealt{Lawrence2016}) or self-lensing
(e.g., \citealt{DOrazio2018}), \citet{Graham2017} attribute the
majority of the events to explosive stellar-related activity in the
accretion disk, such as super-luminous SNe, TDEs, and mergers of 
stellar mass BHs. Given the range
of potential phenomena, and that the flares are likely associated
with events close to the central engine in the UV-luminous accretion
disk, \uvex observations will be critical for refining our understanding
of these events and ultimately using them to improve our understanding
of the extreme and poorly understood physics of AGN accretion disks.

\begin{figure}[htbp]
 \centering
  \includegraphics[width=\columnwidth]{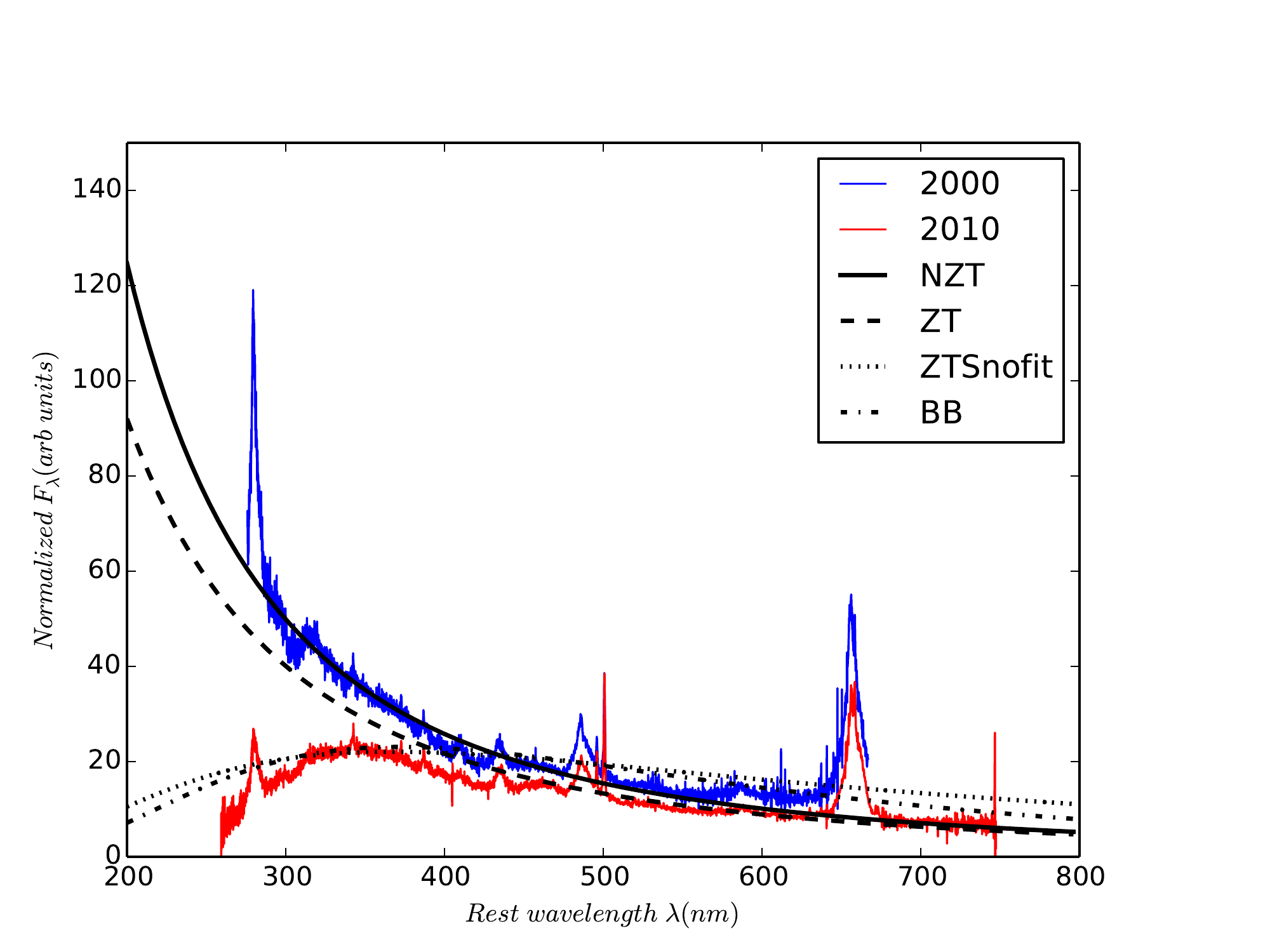}
   \caption{\small Example of a changing-look AGN, 
   SDSS J110057.70-005304.5,
   in which the UV emission from the AGN collapsed \citep{Ross2018}.
   The authors explore models where the source has non-zero torque
   (NZT model; solid line) at the inner, stable circular orbit
   (ISCO; \citealt{Afshordi2003}), and suggest that the spectrum
   is best explained by shutting down emission from the inner $\sim
   200 r_g$ of the accretion disk (ZTSnofit model; dotted line).
   See \citet{Ross2018} for details.
 \label{fig:AGN2}}
\end{figure}

\subsubsection{Changing-Look Quasars}

\uvex will also be important for studying ``changing-look quasars'',
quasars which rapidly rise or drop in optical brightness 
(e.g., \citealt{LaMassa2015}, \citealt{Ross2018}, \citealt{Stern2018}, 
\citealt{Graham2020a}).
Though a range of selection criteria and wavelength ranges have been
used to identify such sources, changes in the innermost accretion
disk occurring on the thermal or cooling/heating front timescale
seem to be the most plausible explanation for the year-scale
variability typically observed in changing-look quasars. For example,
\citet{Ross2018} present one example of a changing-look AGN where
the UV emission has collapsed (Figure~\ref{fig:AGN2}). The authors
argue that the most likely explanation is that some triggering event
caused the inner accretion disk to cool, and thus dim, though other
models have been favored for other events (e.g., \citealt{Guo2016}).
With few such events studied, particularly at the key UV wavelengths
which likely dominate the state changes, \uvex will have an important
role to play in determining the timescales, characteristics, and
spectral changes associated with changing-look quasars.

\subsection{Tidal Disruption Events}
 \label{sec:TidalDisruptionEvents}

It is now well established that SMBHs are a ubiquitous presence in the nuclei of almost all galaxies \citep{Magorrian1998, Kormendy1995, Ho2008}.  In fact, the remarkably tight correlation between the masses of central SMBHs and the mass, luminosity, and structure of their host galaxies \citep{Ferrarese2000, Gebhardt2000, Graham2001, Marconi2003} suggests coeval formation and growth over cosmic time. Despite these advancements, the formation mechanism of the primordial seeds from which SMBHs grow through accretion and mergers is yet unknown. The demographics of SMBHs in low-mass galaxies may place the most promising constraints \citep{Volonteri2009}. Unfortunately, the low-mass end of the SMBH mass function is difficult to detect, due to the smaller gravitational sphere of influence and lower Eddington luminosity of the black hole, which both scale linearly with SMBH mass. Tidal disruption events (TDEs) provide the most promising method to detect and weigh black holes below $10^{8} \msun$, where scaling relations between black hole mass and galaxy mass are poorly constrained.

A TDE will occur when an unlucky star's orbit passes close enough to a central SMBH to be tidally ripped apart, and the resulting stellar debris is slowly consumed by the black hole, producing a luminous accretion flare.
The physics of TDEs depends on the interplay of the stellar radius and the black hole tidal radius: black holes more massive than $\sim 10^8
\msun$ disrupt main sequence stars inside the event horizon and
thus do not produce a flare, while black holes less massive than
$\sim 10^{5.4} \msun$ do not have a tidal field sufficient
to disrupt main sequence stars. Compact stars, such as white dwarfs,
can be disrupted by intermediate mass black holes (IMBHs; $\sim 10^{4} \msun$), while red giants can be disrupted by the most massive black holes known.

TDEs are not only probes of SMBH demographics. They are also an ideal laboratory to study the physics of accretion. One of the primary parameters in accretion physics is the accretion rate. The mass accretion rate in a TDE undergoes a large variation over month- to year-long timescales, starting as super-Eddington and gradually decreasing as a power-law with time.
When a TDE occurs around an inactive black hole, it provides a unique opportunity to study the formation and physics of an accretion disk and its associated structure. There is increasing suspicion that TDEs are sources of very high-energy neutrinos (e.g. \citealt{vanVelzen2021_neutrino}; see also Section~\ref{sec:AGN}). Finally, TDEs can be used to probe the occupation fraction of massive black holes in the nuclei of low-mass galaxies, which allows for the evaluation of competing models proposed to explain the surprising presence of SMBHs with masses $>10^9 \msun$ in the first Gyr after the Big Bang \citep{Banados2018}.  

The census of TDE candidates has been steadily growing thanks to
dedicated searches for nuclear transients from quiescent galaxies
across the electromagnetic spectrum (see \citealt{Gezari2021} for a recent
review).  

\subsubsection{UV Properties of TDEs \& Uniqueness of \uvex}

\begin{figure*}[htbp!]
 \centering
  \includegraphics[width=0.85\textwidth]{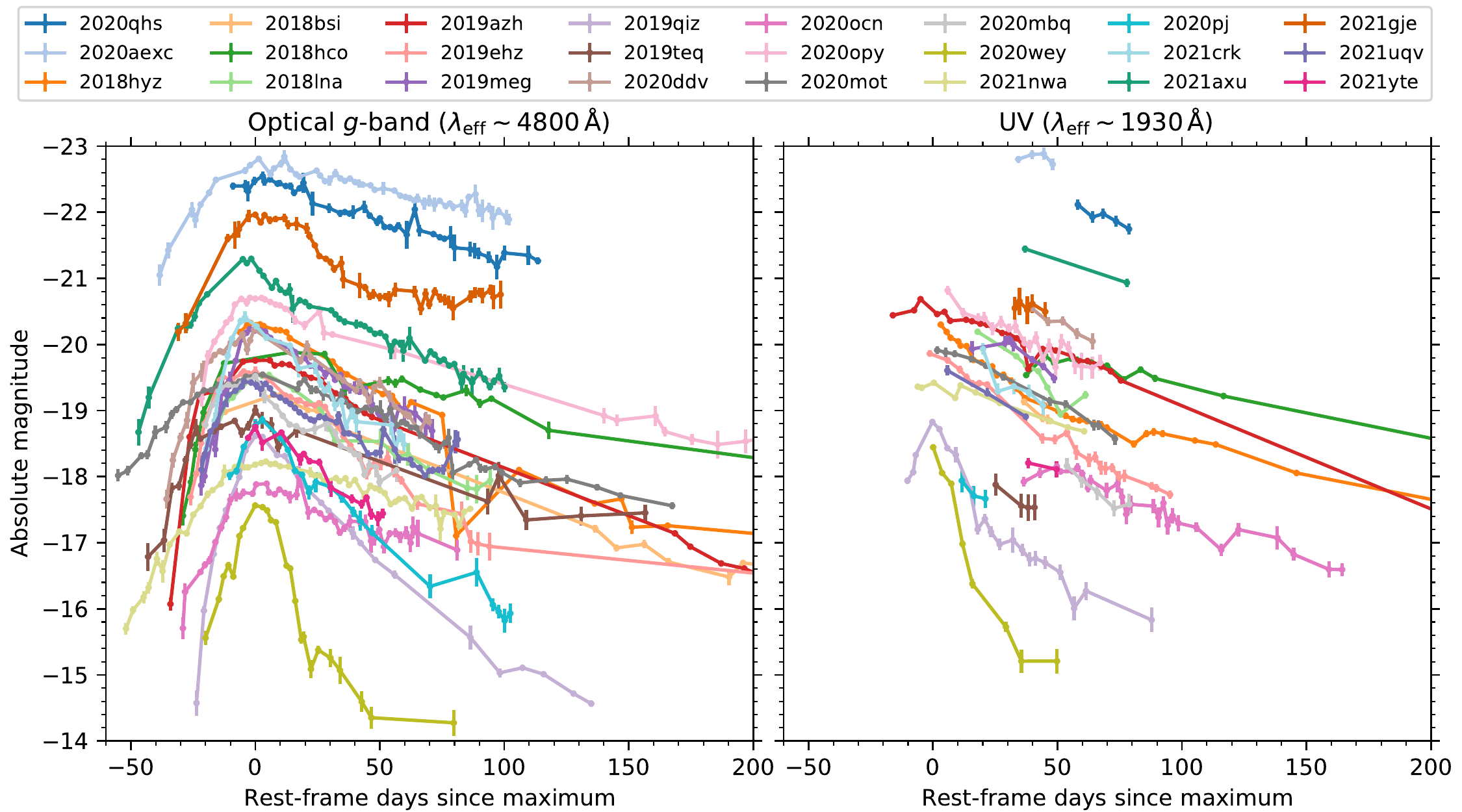}
   \caption{\small Optical (left panel) and UV (right panel) light curves of
    TDEs selected from ZTF. Since UV observations were carried out
    after the optical identification, only three TDEs had UV data
    before the peak of the light curves. 
 \label{fig:lc_library}}
\end{figure*}

TDE candidates were first identified from the \rosat X-ray All Sky Survey \citep{Bade1996, Komossa1999}. The modest number of epochs were sufficient to establish the phenomenon from extremely soft, luminous X-ray outbursts from otherwise quiescent galaxies.
The second era of TDE searches began with \galex, together with joint optical observations from the ground, which led to a class of UV-selected TDEs \citep{Gezari2008, Gezari2009, Gezari2012}. These TDEs are characterized by thermal emission with temperatures $T_{\rm bb}\sim (3-5) \times 10^4\rm\, K$ (and radius $R_{\rm bb}\sim 10^{14}\mbox{--}10^{15}\rm\,
cm$) and emission peaking in the UV -- much cooler than the original TDE candidates discovered in the soft X-rays ($T_{\rm
bb} \sim 10^{6}$ K and $R_{\rm bb} \sim 10^{12}\rm\,
cm$). Since those studies with \galex, the majority of TDE discoveries have been made in the optical.  ZTF is now routinely discovering 15 TDEs per year \citep{vanVelzen2021}.
Follow up with the \swift satellite has shown
that they are also very bright in the UV, with typical colors of $NUV -
r < -1$ mag, and peak luminosities of $-18 < NUV <-22$ mag
(\citealt{Gezari2021}; see Figure~\ref{fig:lc_library}).  In fact, TDEs
are some of the most luminous and long-lived UV transients in the Universe (see Figure \ref{fig:lum-duration-opt}).

The origin of the luminous UV-bright thermal emission from TDEs is still debated (see Section~\ref{sec:ConfrontingModels}). Some UV and optically selected TDEs are also bright in the soft X-ray band, indicating that these are two emission components that likely co-exist in the same TDE system.

Owing to the high luminosity and long lifetime of TDEs in the UV, 
\uvex will routinely discover TDEs. 
However, by 2027 the TDE landscape will be quite sophisticated
and the focus will be on {\it large and well-defined} samples of TDEs.
Therefore, to illustrate \uvex's capability in this field, we consider a hypothetical \uvex\ survey to discover and characterize over 1000 TDEs.
The framework to compute the discovery rate of TDEs is given
in Appendix~\ref{sec:TDE_rate}. 
This survey would be accompanied by a spectroscopic component to systematically probe the kinematics and structure of outflows launched by TDEs, and test competing models for the origin of the UV emission (Section~\ref{sec:ConfrontingModels}). 

The resulting data from such a survey will enable us to: (1) detect  the  lower energy tail of thermal emission from X-ray loud TDEs (Section~\ref{subsec:Xray_tail}); (2) address the 
``missing energy problem'' for optically loud TDEs (Section~\ref{subsec:missing_energy}); (3) identify the mechanisms of the UV/optical thermal emission (Section~\ref{sec:ConfrontingModels}); and (4) probe the BH occupation fraction of low mass galaxies, and even the spin distribution of high mass BHs (Section~\ref{subsec:BHdemographics}).

\begin{figure*}[htbp!]
 \centering
  \includegraphics[width=0.75\textwidth]{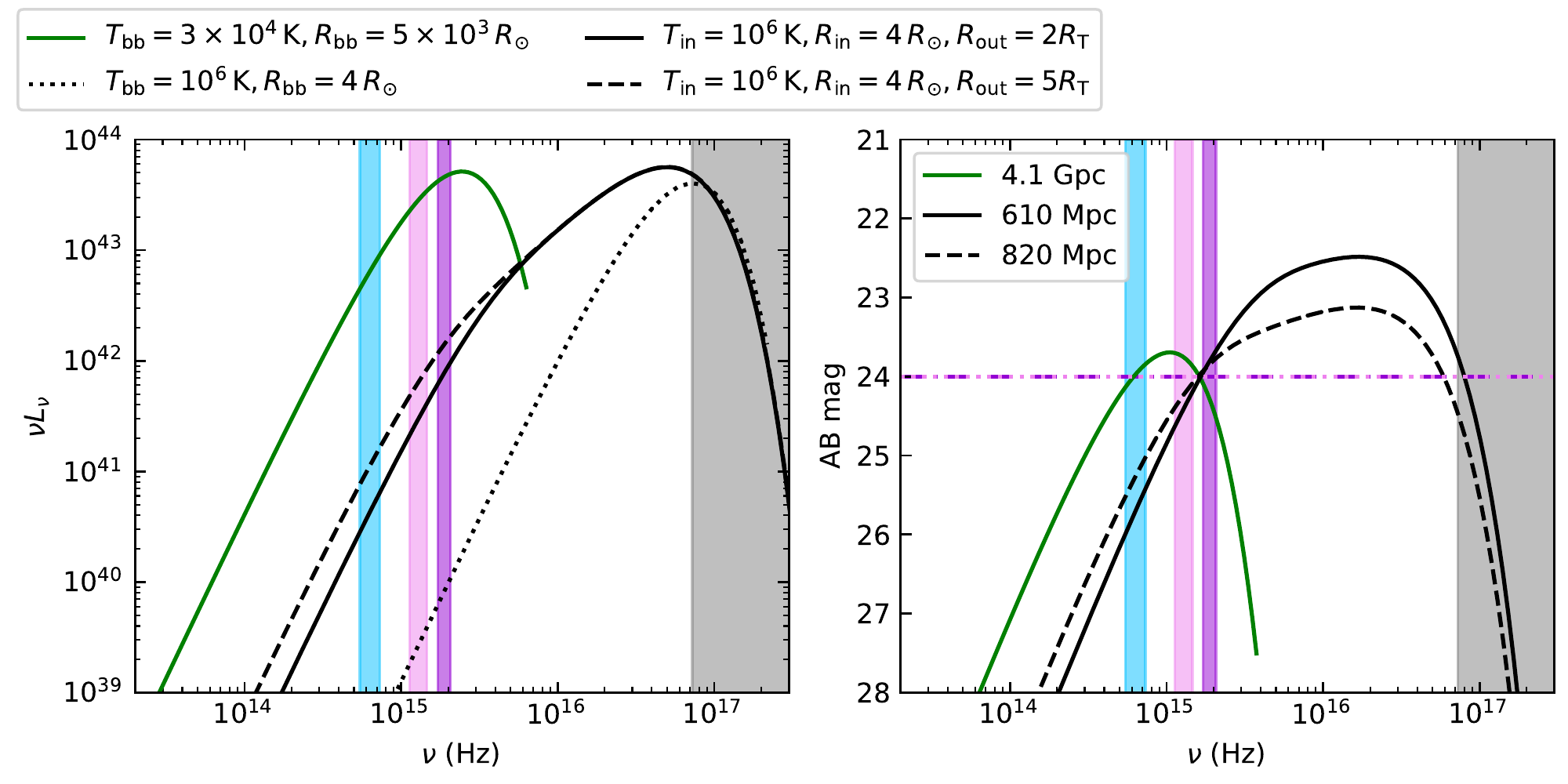}
   \caption{\small (Left) SEDs of typical optically-discovered (green) and
   X-ray-discovered (black) TDEs. Photometric bands are marked by the
   vertical lines in blue ($g$-band), light purple (NUV), deep
   purple (FUV), and gray (0.3--10\,keV). (Right) The same SEDs are now
   shifted to the largest distance that still meets the \uvex
   sensitivity threshold (conservatively assumed to be 24\,mag), including
   $K$-correction.
 \label{fig:OptXrayCompare}}
\end{figure*}

\subsubsection{UV Detections of X-ray Loud TDEs} \label{subsec:Xray_tail}
The Russian-German \textit{Spektr-RG} (\textit{SRG}) mission is now finding
TDEs at a higher rate relative to those found from optical surveys \citep{Sazonov2021}.
X-ray discovered TDEs \citep{Saxton2020} display temperatures of
$T_{\rm bb}\sim 10^6$\,K and $R_{\rm bb}\sim 10^{12}\rm\, cm$, as
expected from emission near the innermost stable circular orbit
(ISCO) of a black hole (BH) of $M\sim 10^6 \msun$. The black dotted
line in Figure~\ref{fig:OptXrayCompare} (left panel) illustrates a
typical blackbody inferred from soft X-ray observations. However,
from angular momentum conservation, we expect the tidal debris to
circularize at a few times the tidal radius $R_{\rm T} =
R_{\odot}(M/\msun)^{1/3}$ (for a solar-like star), which means
that the radius of the outer disk is of the order $R_{\rm out}\sim
10^{13}\rm\, cm$. One of the exciting capabilities of the FUV
sensitivity of \uvex will be to detect the lower energy tail of the
thermal emission of X-ray loud TDEs, even for those without the
additional UV/optical component.

Multi-color disk SEDs for two different choices of $R_{\rm out} =
2R_{\rm T}$ and $5R_{\rm T}$ are shown in solid and dashed black
lines in Figure~\ref{fig:OptXrayCompare}. For a detection threshold
of $24$\,mag, \uvex will be able to detect the emission from the
outer accretion disk out to at least 610\,Mpc.

\subsubsection{Addressing the "Missing Energy" Problem} \label{subsec:missing_energy}

The total radiated energies in optically-discovered TDEs are in the
range $10^{50}$--$10^{51}\rm \, erg$. This is at least one order
of magnitude below the theoretically expected energy release from
$0.1 \msun$ of accreted mass, even considering that the accretion
disk may be radiatively inefficient in the super-Eddington regime
\citep{Lu18}. Therefore, $>90\%$ of the expected energy released by 
TDEs has been missed by current observations. This is one of the 
major puzzles in TDEs.

A possible solution is that the majority of the disk mass is only
slowly accreted onto the BH on a timescale much longer than a few
years, provided that the accretion disk is very thin with a long
viscous timescale. Without optically thick reprocessing gas at large
distances, the disk emission is expected to be primarily in the UV
and soft X-ray bands, as shown by the multi-color blackbody SED in
Figure \ref{fig:OptXrayCompare}. This is supported by late-time
($\sim$5 yr) FUV/NUV observations of a handful of TDEs by \hst after
the optical emission had completely faded away
\citep{van_velzen19_late_time_HST}. The observed FUV fluxes are in
the range 23--25 mag, which are detectable either by a single \uvex observation or by stacking of multiple exposures. Thus, the \uvex
all-sky survey will provide late-time FUV measurements or stringent
upper limits for all $N\sim 10^3$ TDEs that will have been discovered
by optical and X-ray surveys before 2028 -- not feasible with individual 
\hst observations for such a large number of TDEs.

\subsubsection{Confronting Observations with TDE Models}
 \label{sec:ConfrontingModels}
 
There are two competing models for how the UV/optical emission seen
in some TDEs is generated.
 \begin{enumerate}

  \item \textbf{Reprocessing of Disk Emission.} In this scenario,
  an optically thick gas layer (most likely in the form of an
  outflow) at a distance of $10^{14}$--$10^{15}\rm\, cm$ absorbs
  the soft X-ray emission from the disk and re-emits in the UV-optical
  band (\citealt{metzger16_reprocessing}, \citealt{roth16_reprocessing},
  \citealt{Lu2020}).  As the fallback rate declines with time, the
  density of the ``reprocessing layer'' drops and there is less and
  less absorption.  Thus, the bulk of the emission is expected to
  shift to higher frequencies and correspondingly, $T_{\rm bb}$
  {\bf increases} and $R_{\rm bb}$ drops with time.

 \item \textbf{Stream-stream Collisions.} An alternate scenario
 is that the bound stellar debris stream intersects itself, 
 producing a self-crossing shock, and the kinetic energy 
 dissipated by the shock powers the
 optical emission (\citealt{piran15_shock_model},
 \citealt{jiang16_self_crossing_shock}). In this case, $R_{\rm
 bb}$ is of the order of the self-crossing radius, which stays roughly
 unchanged over time. As the fallback rate declines with time, the
 shock power drops (roughly as $t^{-5/3}$) and hence the blackbody
 temperature $T_{\rm bb}$ is expected to {\bf decrease} with time.

\end{enumerate}

While temperature evolution is a compelling discriminator between
these models, the optical band is on the Rayleigh-Jeans tail
(insensitive to $T_{\rm bb}$) of the TDE thermal continuum, and UV
photometry is required for a reliable measurement of the temperature
evolution. However, as can be seen from Figure~\ref{fig:lc_library},
early UV photometry is only available for a few TDEs. \uvex will for
the first time, provide UV photometry before and after the peak,
and map out the temperature evolution on timescales relevant for
the circularization of the debris streams.

\begin{figure}[htbp]
 \centering
  \includegraphics[width=0.85\columnwidth]{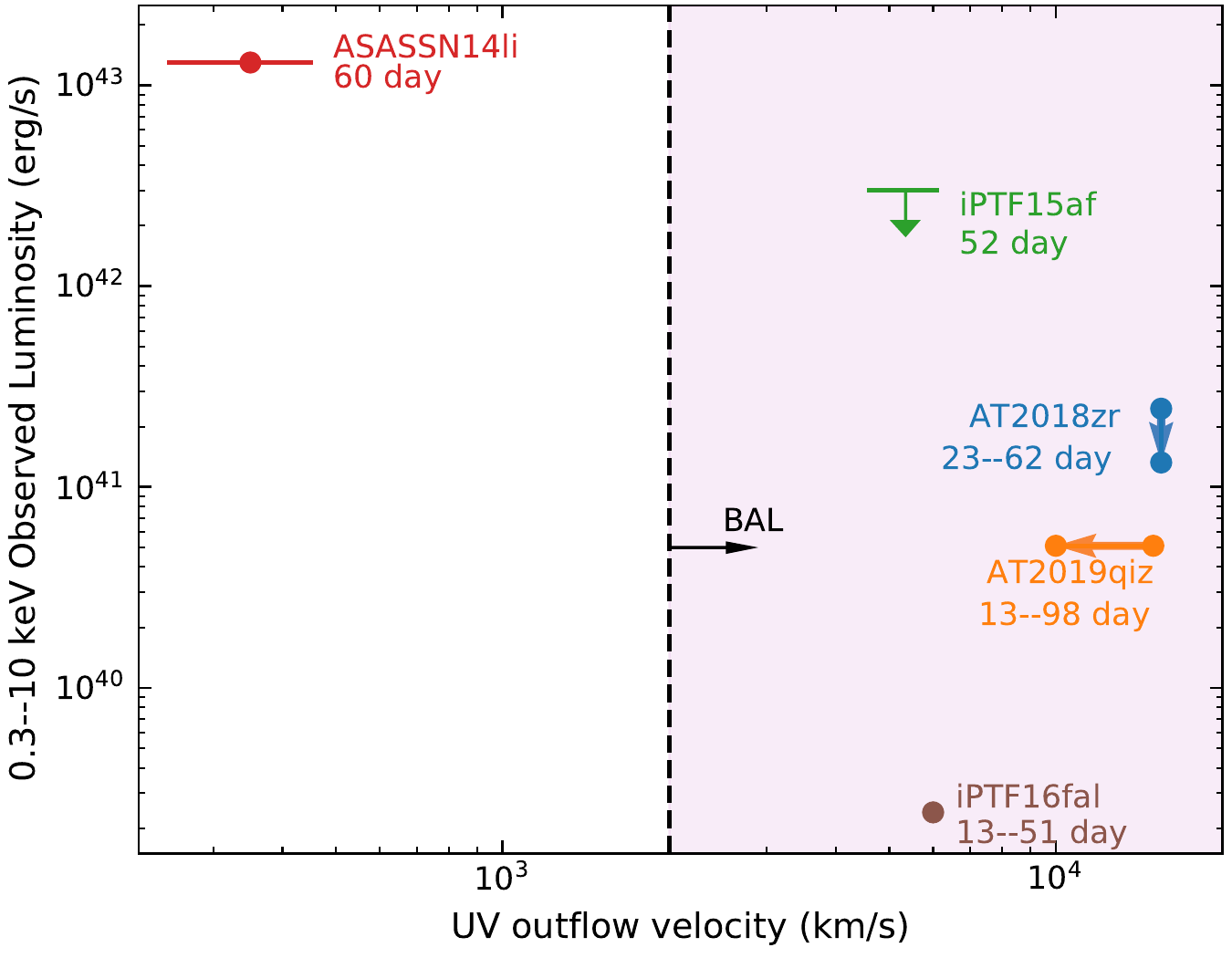}
   \caption{\small X-ray luminosity versus outflow velocity measured from UV spectroscopy. Among the five TDEs with published UV spectra, four exhibit broad absorption line systems (BALs; defined by $v>2000\,{\rm km\,s^{-1}}$).
 \label{fig:uvspec}}
\end{figure}

The typical ionization parameter near the optical photosphere of
TDEs is $\xi \gtrsim 10^2$. This means that most atoms lose their
outer-shell electrons, which are responsible for transitions in the
optical band, whereas the transitions of inner-shell (K and L)
electrons produce lines in the UV band. Thus, UV spectroscopy provides
a unique probe of the density and velocity structures of the gas
under high UV and soft X-ray fluxes.

As of late 2021 there are only five TDEs with UV spectra, ASASSN-14li
\citep{Cenko2016}, iPTF16fnl \citep{Brown2018}, iPTF15af
\citep{Blagorodnova2019}, AT2018zr \citep{Hung2019}, and AT2019qiz
\citep{Hung2021}. Broad absorption line (BAL) systems are observed in 
all sources, except for ASASSN-14li.
The broad absorption/emission features are thought to arise from
an outflow driven by the disk accretion or stream collision.

A direct prediction of the reprocessing scenario is that
if we observe into the funnel of the disk-driven wind, more X-ray flux
and less prominent UV absorption features are visible 
because most atoms are completely ionized by the high X-ray flux.
On the other hand, if we observe from higher angles where the X-rays
are obscured by the outflow, the UV absorption features should be
strong. An anti-correlation between BAL presence/strength and X-ray
luminosity will be a smoking gun for the reprocessing picture. The current sample of five TDEs with UV spectra so far supports this picture (see
Figure~\ref{fig:uvspec}). \uvex will $\sim$triple the sample of TDEs with UV spectra, enabling significantly more robust conclusions about such an anti-correlation.

The outflow velocity of the gas in the line formation region can
be measured by the P-Cygni profiles. If the outflow
is driven by disk accretion, the outflow speed should
increase as the accretion rate drops at later times, because
the disk-wind launching radius gets closer and closer to the 
innermost stable circular orbit (ISCO).
In the alternative scenario, if the outflow is driven by stream
collision, the outflow speed should stay roughly constant 
because the radius of stream self-crossing does not evolve 
with time \citep{Lu2020}.

In the past, the study of line-width evolution in TDEs has been stymied
by the relatively late optical discovery and the latency of \hst\ UV
ToO triggers. Up until now, the earliest UV spectrum of a TDE only
started at $\Delta t=+13$\,day (relative to the optical peak). 
An example \uvex\ campaign, tracking the velocity evolution of 
$\approx 20$ TDEs with five spectroscopic observations each 
(from $\Delta t \sim -4$\,days to $\sim+56$\,days), 
will provide a unique probe of the outflow's origin.
The \uvex spectral resolution of $R\geq1000$ is more than adequate to measure the outflow velocity (typically $>10^3\,{\rm km\,s^{-1}}$). 

\subsubsection{Probing Black Hole Demographics} 
 \label{subsec:BHdemographics}
 
\begin{figure}[htbp!]
 \centering
  \includegraphics[width=\columnwidth]{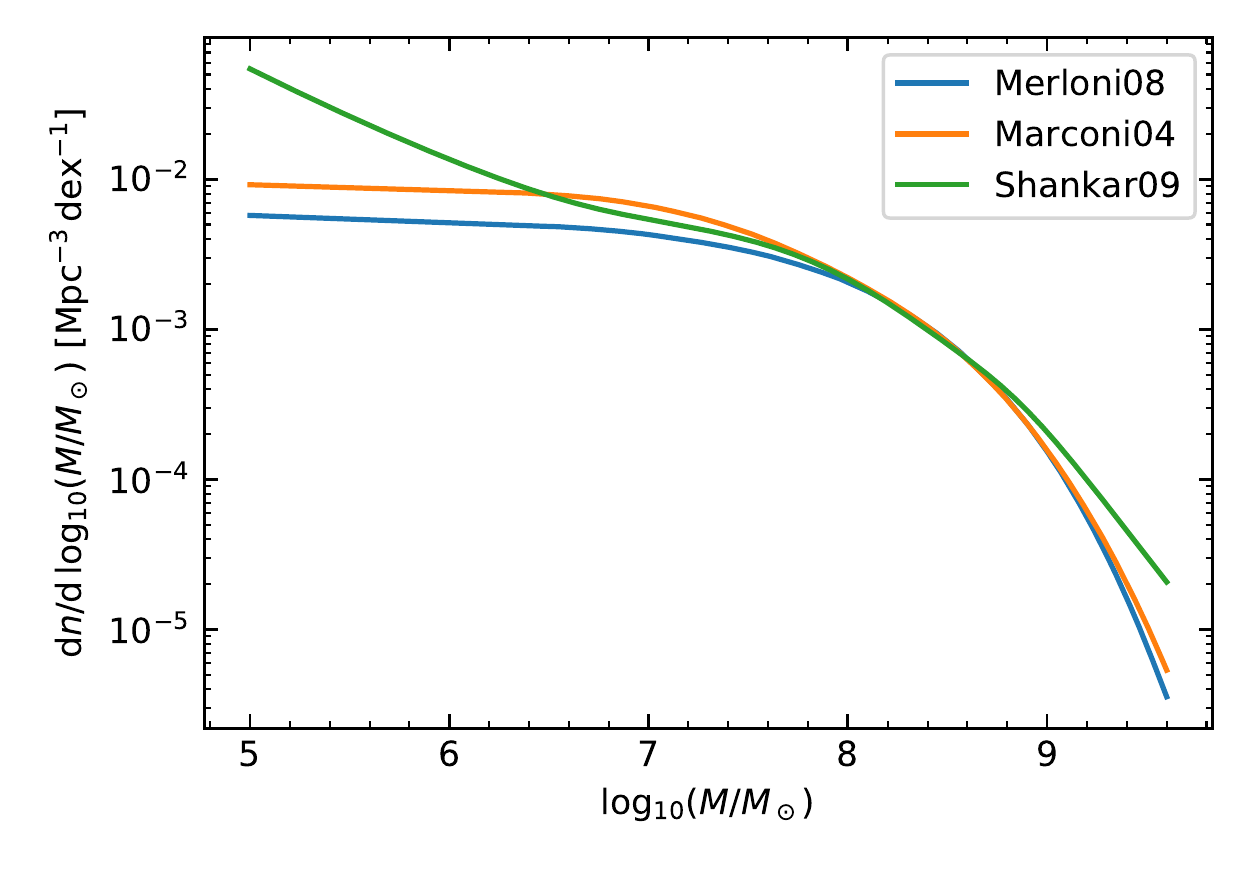}
   \caption{\small Three different BHMFs given in the literature
   (\citealt{marconi04_BHMF}, \citealt{merloni08_BHMF},
   \citealt{shankar09_BHMF}).
 \label{fig:bhmf}}
\end{figure}

The BH mass function (BHMF) below $\sim 3\times 10^6 \msun$ and
above $\sim 5\times 10^8 \msun$ is poorly known (see
Figure~\ref{fig:bhmf}). Our goal is to use TDEs as tracers of BH
demographics. Specifically, we would like to measure the BHMF at the
low mass end from $10^5 \msun$ to $10^6 \msun$, and to
investigate the upper bound of $M_{\rm BH}$ that can disrupt a star.
The TDE rate at the low mass end is sensitive to the BH occupation
fraction \citep{Stone2016}.

\begin{figure}[htbp!]
 \centering
  \includegraphics[width=\columnwidth]{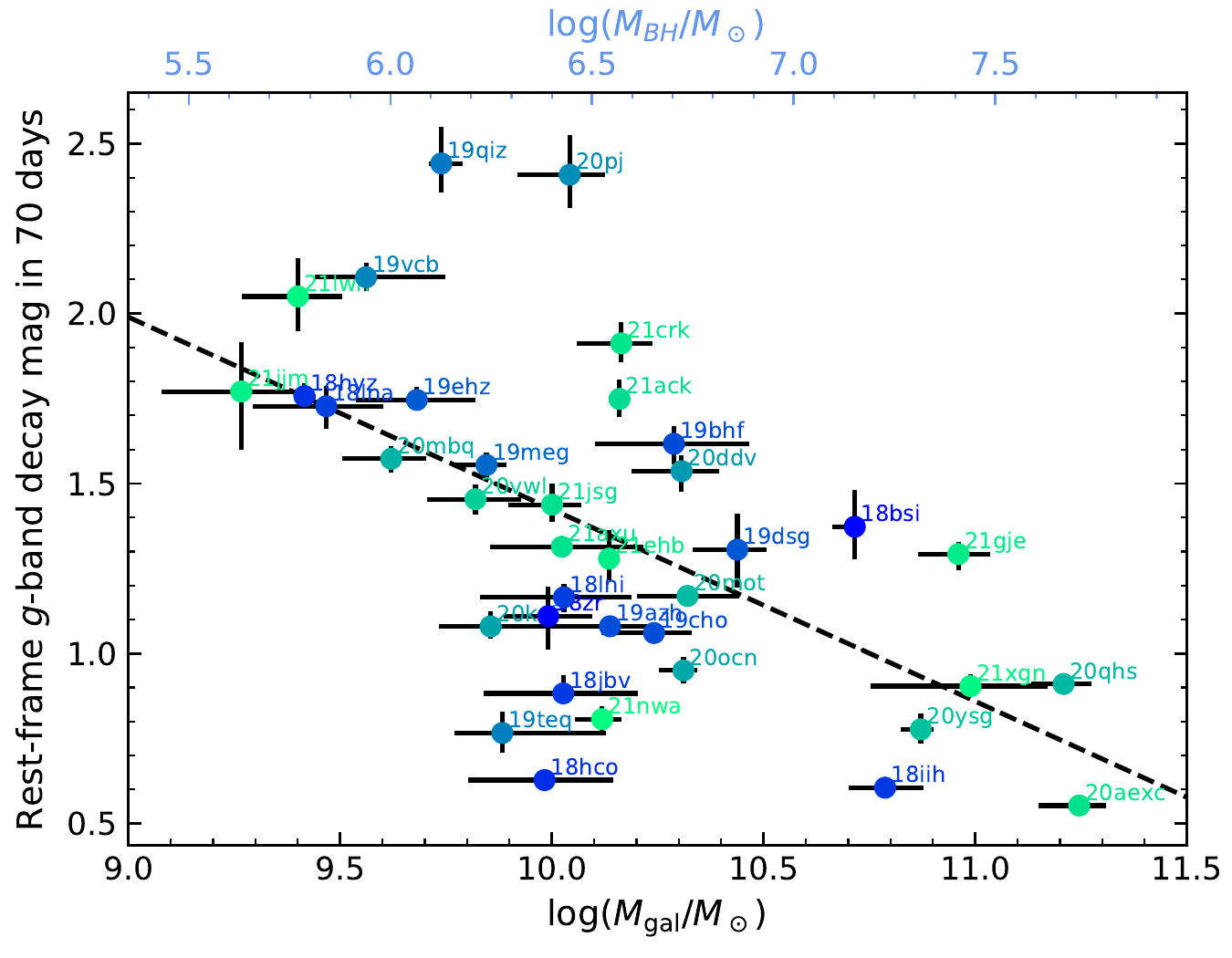}
   \caption{\small TDE optical decay rate as a function of host
   galaxy stellar mass $M_{\rm gal}$. The decay is quantified by
   the total magnitude the light curve fades from peak to 70\,days
   post-peak.  $M_{\rm gal}$ is measured by modeling the host galaxy
   UV--MIR SED following the procedures laid out in \citet{vanVelzen2021}.
   The upper x-axis marks rough estimates of $M_{\rm BH}$ using the
   $M_{\rm gal}$--$M_{\rm BH}$ relation from \citet{Reines2015}. A
   statistically significant correlation was found ($p=0.0006$ for
   a Kendall's Tau test), suggesting that TDEs disrupted by higher
   mass BHs fade slower.
 \label{fig:Mbh_tfb}}
\end{figure}

\begin{figure}[hbpt]
 \centering
  \includegraphics[width = 0.9\columnwidth]{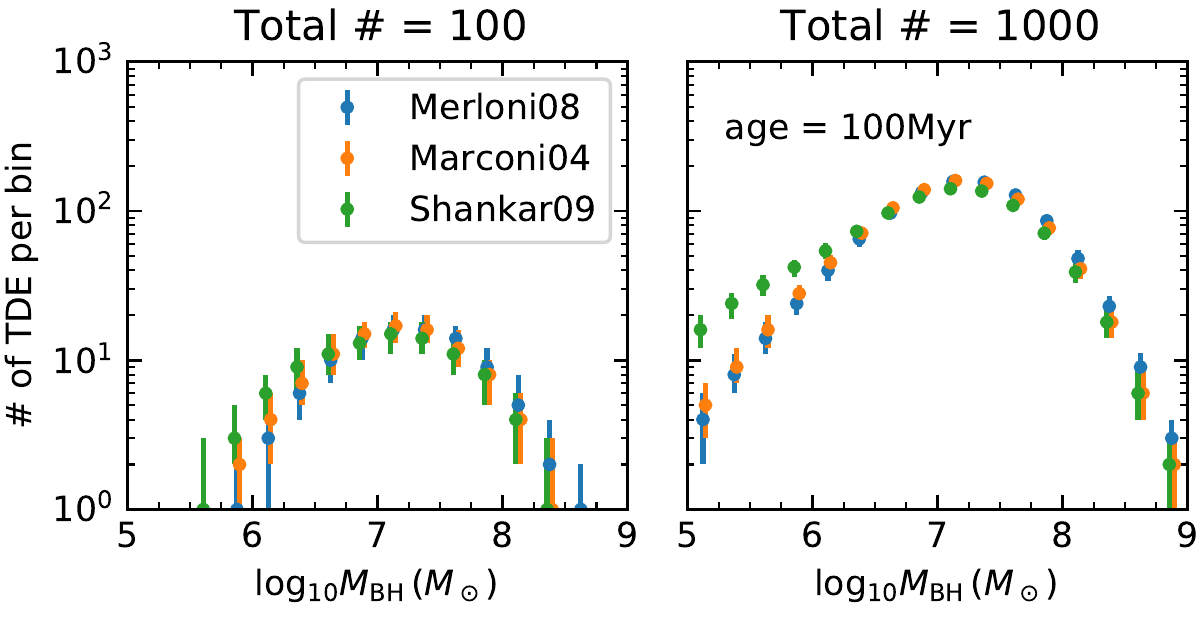}\\
   \includegraphics[width = 0.9\columnwidth]{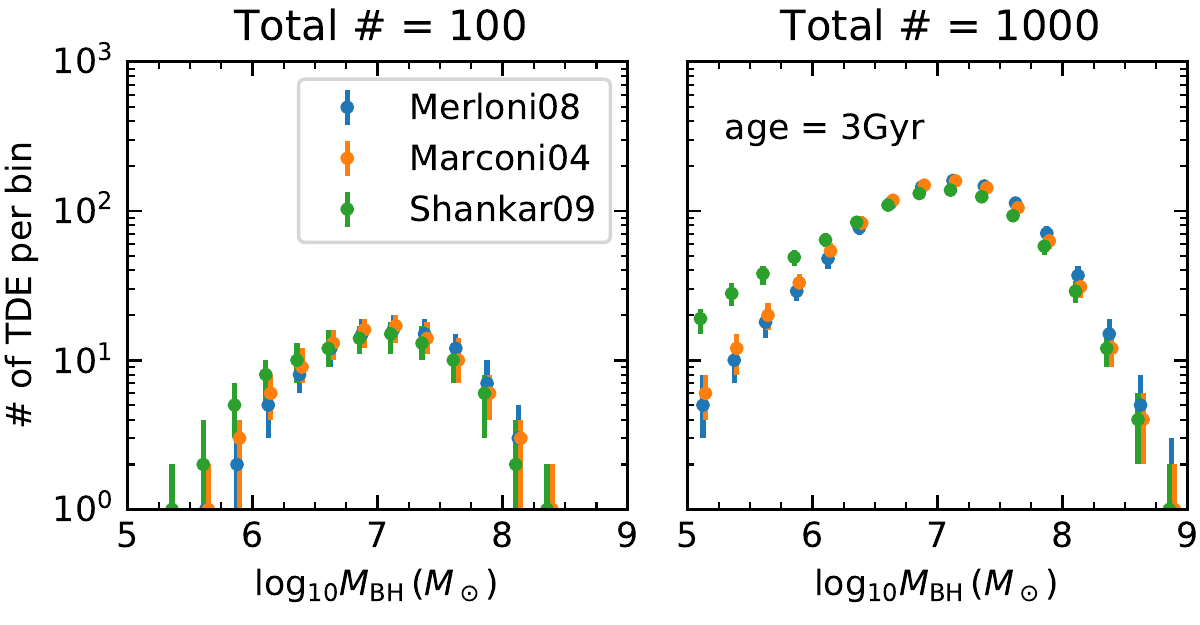}
    \caption{\small Observed number of TDEs as a function of $M_{\rm
     BH}$, under different assumptions of the total number of
     detected TDEs, the BHMF (Figure~\ref{fig:bhmf}), and age of the
     stellar population (upper panels: $100$\,Myr; lower panels:
     $3$\,Gyr).
 \label{fig:Ndet}}
\end{figure}

Given the relatively strong correlation between $M_{\rm BH}$ (inferred
from the host galaxy central velocity dispersion) and the 
fallback time $t_{\rm fb}$
(derived from fitting the decay timescale of the light curve;
\citealt{vanVelzen2020,Gezari2021}; see
Figure~\ref{fig:Mbh_tfb}), one should be able to use well-sampled
light curves to infer the BH mass independently of host galaxy
properties. This is particularly important for IMBHs, where the 
scaling relations between host galaxy mass
and central BH mass are poorly constrained \citep{Greene2020}.
A survey cadence of $\lesssim$25 days would be needed to track the
decay of IMBH TDEs (see AT2020wey in Figure~\ref{fig:lc_library}).
Figure~\ref{fig:Ndet} shows that a sample of $\approx 1000$ TDEs is
needed to constrain the BHMF between $10^5 \msun$ and $10^6 \msun$
and to measure the shape of TDE rate suppression due to the BH event
horizon.

The all-sky cadenced synoptic surveys carried out by \uvex\ will be 
able to build up the large sample of TDEs required to differentiate
between different BHMFs, as well as to probe other questions about
BH demographics that have not been possible before, such as the spin
distribution of the most massive non-active BHs
\citep{Kesden12_maximum_Mbh}. 

\subsection{Multi-Messenger Astronomy}
 \label{sec:MMA}

On August 17, 2017, the groundbreaking discovery of both gravitational
waves and electromagnetic radiation from a neutron star merger
(e.g., \citealt{2017ApJ...848L..12A}) marked a new era in multi-messenger
astrophysics (MMA). The discovery of a flaring blazar
\citep{2018Sci...361.1378I} and three tidal disruption events
\citep{Stein2021NatAs, Reusch2021, vanVelzen2021} associated with high energy neutrinos opened up yet another facet of MMA.

With planned advances in sensitivity of GW
interferometers, neutrino detectors and EM
surveyors, we could expect several MMA events to be discovered in
the years leading to the launch of \uvex. However, capabilities for
opening the UV window into MMA will be limited because
of the combination of wide FOV, sensitivity, and
rapid response time that is required.  While \hst
has the sensitivity, it is neither wide-field nor does it
respond fast enough.  \ultrasat  (launch date, 2025) will have a
larger FOV and rapid response, but will be much less sensitive
than \uvex (and lacks an FUV channel), and thus will only be able to pick the 
lowest-hanging fruit
(e.g., exceptionally nearby or bright sources). Fully realizing the
potential for population studies enabled by the A+ sensitivity
upgrades (e.g., $H_{0}$ constraints: \citealt{Chen+2018}; connection
to gamma-ray bursts: \citealt{Abbott+2017GRB}) requires the higher
sensitivity that  will be provided by \uvex.

Opening a new region of the wavelength-depth-FOV phase space,
particularly FUV imaging, \uvex
will significantly impact the study of neutron star mergers, black
hole mergers and SMBH flares. \uvex will tackle
open questions on where the heaviest elements are synthesized and
how relativistic jets are formed by identifying and characterizing
the UV emission from GW and high-energy neutrino events.  Neutron star mergers were discussed in Section~\ref{sec:emgw}, while
SMBH flares and TDEs within active galaxies as potential
sources of high energy neutrinos were discussed in Section~\ref{sec:AGN}.  Appendix~\ref{sec:GW_appendix} presents the expected GW event rates and light curve models for \uvex.

\subsection{Exoplanets}
 \label{sec:Exoplanets}

With thousands of discoveries at hand, the field of exoplanets is
rapidly pivoting from detection to characterization. Limited thus
far by the necessary advances in technology, we have nevertheless
been able to explore a small number of planets in unprecedented
detail. In particular, our ability to observe and understand the
atmospheres of exoplanets has increased significantly over the last
decade. Understanding exoplanet atmospheres, and refining how we
analyze and model them, is a crucial stepping stone towards the
detection and understanding of biosignatures, which are likely to
be first found in exoplanet atmosphere analyses.

Transmission spectroscopy -- measuring the absorbing cross section
of the planet as a function of wavelength during transit -- has been
one of the most productive paths of atmosphere investigation. One
finding is that clouds and hazes are prevalent across all types of
exoplanet atmospheres (\citealt{Wakeford2019}, \citealt{Fu2017},
\citealt{Iyer2016}, \citealt{Crossfield2017}, \citealt{Morley2015}).
Clouds and hazes can reduce the amplitude of spectral features, a
reduction that can also independently be produced by
high mean molecular weight atmospheres \citep{Madhu2015}. The
amplitudes of important molecular features, such as water, vary
significantly between planets in the same class -- hot Jupiters
\citep{Sing2016}, warm Neptunes \citep{Crossfield2017}, and
super-Earths \citep{Southworth2017} -- and understanding the origin
of these variations is crucial for unlocking the physical processes
that dominate these atmospheres. The degeneracy between the presence
of clouds and hazes and the impact of a higher mean molecular weight
atmosphere has hindered further understanding of these variations.

With the launch of \jwst, transmission
spectroscopy will be substantially boosted at IR wavelengths, an
important window into understanding the composition and structure
of exoplanet atmospheres. However, at IR wavelengths the
aforementioned aerosol/mean molecular weight degeneracy is severe. 
In addition, in order to holistically understand atmospheres and the
physical processes that sculpt them, IR spectroscopy on
its own is insufficient. We need to characterize exoplanet atmospheres
over a much broader wavelength range to capture physical processes
that manifest at other wavelengths and to understand the entire
energy budget of the planet. We also need to characterize exoplanet
atmospheres over a much larger set of exoplanet parameters than has
been previously accessible, such as planet size and insolation. The
upcoming ESA \textit{Ariel} mission with the NASA/CASE contribution will
fill in important gaps at optical and near-IR wavelengths for
a very large number of planets, partially satisfying these needs,
but there remain important atmospheric processes that are only
captured in the UV. Without complementary UV observations,
our ability to both plan and interpret 
\jwst\ and \textit{Ariel} observations
will be significantly hindered. There is a broad range of atmospheric
physics that can be studied with UV observations. These include
Rayleigh and Mie scattering properties of cloud and haze particles
(Section \ref{sec:hazes}); heavy metal condensation and disequilibrium
processes (Section \ref{sec:rainout}); and observations of atmospheric
escape in Lyman-$\alpha$ (e.g., \citealt{Zhang2021}). 
The former in
particular presents the opportunity to identify cloudy or hazy
atmospheres and break the aerosol/mean molecular weight degeneracy.
Looking forward to \jwst\ and \textit{Ariel}, being able to use UV
observations to predict which planets are most likely to have
measurable spectral features, and to interpret the very high quality
optical and IR transmission spectra that will be obtained,
will be crucial for maximizing the scientific return of these
missions.

In addition to atmospheric science, understanding the UV
environment of exoplanets is also crucial to understanding their
habitability. Much attention has turned to rocky planets orbiting
in the habitable zones of M dwarfs, as they are more readily
detectable and characterizable than their analogs around FGK stars.
Transit and radial velocity surveys, for instance, have optimized
their filter bandpasses for M dwarf spectral energy distributions
(e.g., \citealt{Ricker2015}, \citealt{Addison2019}). However, the
UV radiation from M dwarfs is much more stochastic than
that from FGK stars, with both more energetic flares and higher
flare rates in general (see, e.g., \citealt{Medina2020}). This could
have significant impact on the viability of these planets for hosting
life (\citealt{Estrela2018}, \citealt{France2020}). Understanding the
UV radiation history and current insolation of rocky planets
orbiting M dwarfs is another important question addressed by
UV observations.

While the \hst\ UV capabilities are
operational, we can continue to construct and exploit multi-wavelength
transmission spectroscopy and investigate the UV
environment of exoplanets. However, when those capabilities are
gone, we will have a large gap in our ability to constrain these
phenomena. In the following sections we outline two important science
cases that could be performed with an exoplanet transmission survey
undertaken with a UV telescope with the capabilities of \uvex.

\subsubsection{Constraining Hazes in Cool Exoplanet Atmospheres}
 \label{sec:hazes}

Whereas clouds are considered to be `grey', scattering stellar
insolation roughly equally with wavelength, Rayleigh scattering of
small haze particles in the upper atmospheres of exoplanets has a
very strong wavelength dependence ($\propto\lambda^{-4}$). This
leads to a much higher cross-section of a hazy planetary atmosphere
at bluer wavelengths, and correspondingly deeper transit depths.
The slope in the transmission spectrum as a function of wavelength
can thus be used to measure the particle size of hazes forming high
up in a planet's atmosphere. Hazes may become more significant for
planets cooler than 850~K 
(\citealt{Morley2015}, \citealt{Crossfield2017}), which
are of particular interest to the community as they approach the
conditions of habitability. Significant effort and observing time
will be expended by the next generation of optical and IR
missions to characterize the atmospheres of cool exoplanets, and
large remaining uncertainties remain about the fraction of their
atmospheres that are cloudy or hazy, and the characteristics of the
extant clouds and hazes. See, e.g., Figure~2 of \citet{Sing2016} for
a sample of hot Jupiters with significantly varying transmission
spectra between UV and IR wavelengths. Understanding
how hazes manifest across a large range of atmospheric compositions,
from hydrogen- and helium-dominated atmospheres to higher mean
molecular atmospheres dominated by, for instance, methane or carbon
dioxide, over a range of exoplanet sizes and temperatures is crucial
for breaking the aerosol/mean molecular weight degeneracy that mutes
the spectral features. By measuring a large sample of UV-IR
transmission spectral slopes, we can constrain trends with planet
properties, which can then be compared to theoretical and experimental
studies of haze production rates and compositions (e.g. \citealt{gao2017},
\citealt{horst2018}, \citealt{he2020haze}).

\subsubsection{Probing Metals, Clouds, and Rainout in Hot Exoplanets}
 \label{sec:rainout}

A few low-resolution observations of ultra-hot Jupiters ($>$2300~K)
show strong absorption at UV wavelengths which far exceeds that
expected from Rayleigh scattering (e.g. \citealt{2013MNRAS.436.2956S},
\citealt{2018AJ....156..283E}, \citealt{vonessen2019}, \citealt{fu2021}).
The source of the absorption is not clear, and several suggestions
(photochemistry, mass loss, disequilibrium chemistry) remain under
consideration. Theoretical transmission spectra of hot ($>$1000~K)
gas giant atmospheres from \citet{Lothringer2020} show strong
absorption lines from heavy metal atoms and ions at NUV
wavelengths, including Fe I, Fe II, Ti I, Ni I, Ca I, and Ca II.
These species had not often been included as opacity sources in
transmission spectra models because of their low abundances. However,
the authors show that since they have such strong absorption lines
in the UV, they can significantly increase the broadband
transit depths measured at these wavelengths. Indeed, evidence for
heavy metal absorption has been found for some ultra-hot exoplanets
at higher resolution, e.g. \cite{fossati2010}, \cite{gibson2020},
\cite{ehrenreich2020}.

\citet{Lothringer2020} demonstrate the use of a UV-optical spectral
index to test whether heavy metals are indeed the mystery UV
opacity source for hot exoplanets. One such index is defined in
their Eq. 2: 
$\Delta R_{\rm p,NUV-Red}= (R_{\rm p,0.2-0.3\mu m}-R_{\rm p,0.6-0.7\mu
m})/H_{eq}$, where $R_{\rm p,0.2-0.3\mu m}$ is the radius of the planet
as measured between 0.2--0.3\,$\mu m$, $R_{\rm p,0.6-0.7\mu m}$ is the
radius of the planet as measured between 0.6--0.7\,$\rm \mu m$, and
$H_{\rm eq}$ is the atmospheric scale height at the equilibrium
temperature. If the hot atmospheres are in equilibrium, then $\Delta
R_{\rm p,NUV-Red}$ should increase rapidly from 3 to 9 between 1000 and
2500\,K (see their Fig.~4). Between 2500\,K and 4000\,K, $\Delta
R_{\rm p,NUV-Red}$ will slowly decrease to 6. This predicted characteristic
shape to the spectral index could be tested with a large sample of
$\Delta R_{\rm p,NUV-Red}$ measurements across the 1000--4000\,K temperature
range measured with \uvex. Further, we can test if the metals are
raining out after forming clouds. If this is the case, then certain
metal species should have depleted abundances in the atmospheric
regions probed by transmission spectroscopy. This will lead to a
shallower slope in $\Delta R_{\rm p,NUV-Red}$ between 1000 and 2500\,K.

\subsubsection{A Systematic NUV/FUV Survey with \uvex}

\begin{figure*}
 \centering
  \includegraphics[width=0.65\textwidth]{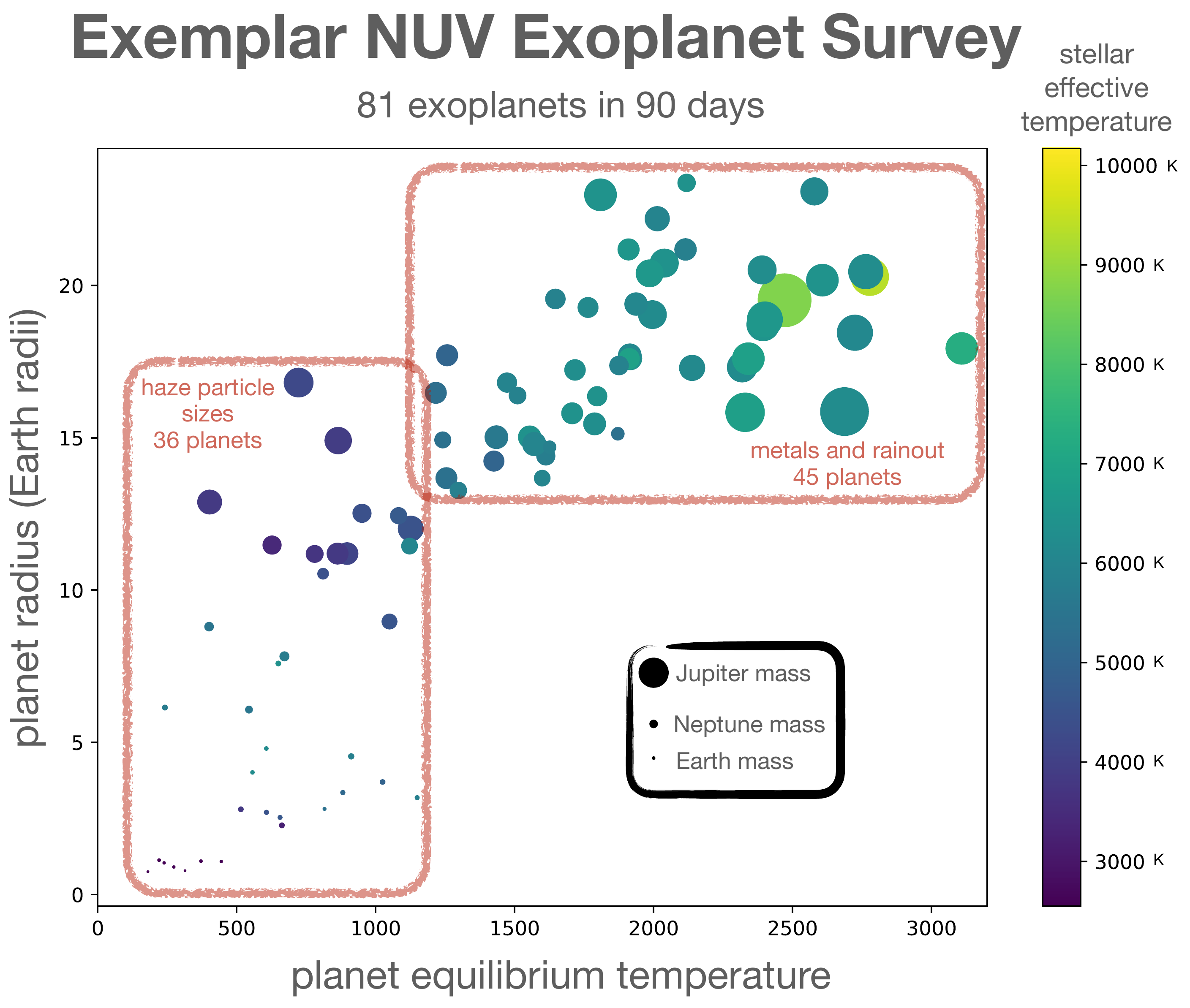}
   \caption{The planet radii and equilibrium temperatures of a sample of 81 planets that would be used to answer two critical exoplanet atmosphere questions: how haze properties of cooler exoplanets depend on planet size, equilibrium temperature, and host star properties; and the extent to which metals in ultra-hot atmospheres are raining out. The full sample would take $\sim$3 months to observe, and leave a legacy of FUV/NUV spectra of exoplanets that will be observed by \jwst\ and \textit{Ariel}/CASE.}
 \label{fig:exosurvey}
\end{figure*}

\begin{figure*}
 \centering
   \includegraphics[width=0.65\textwidth]{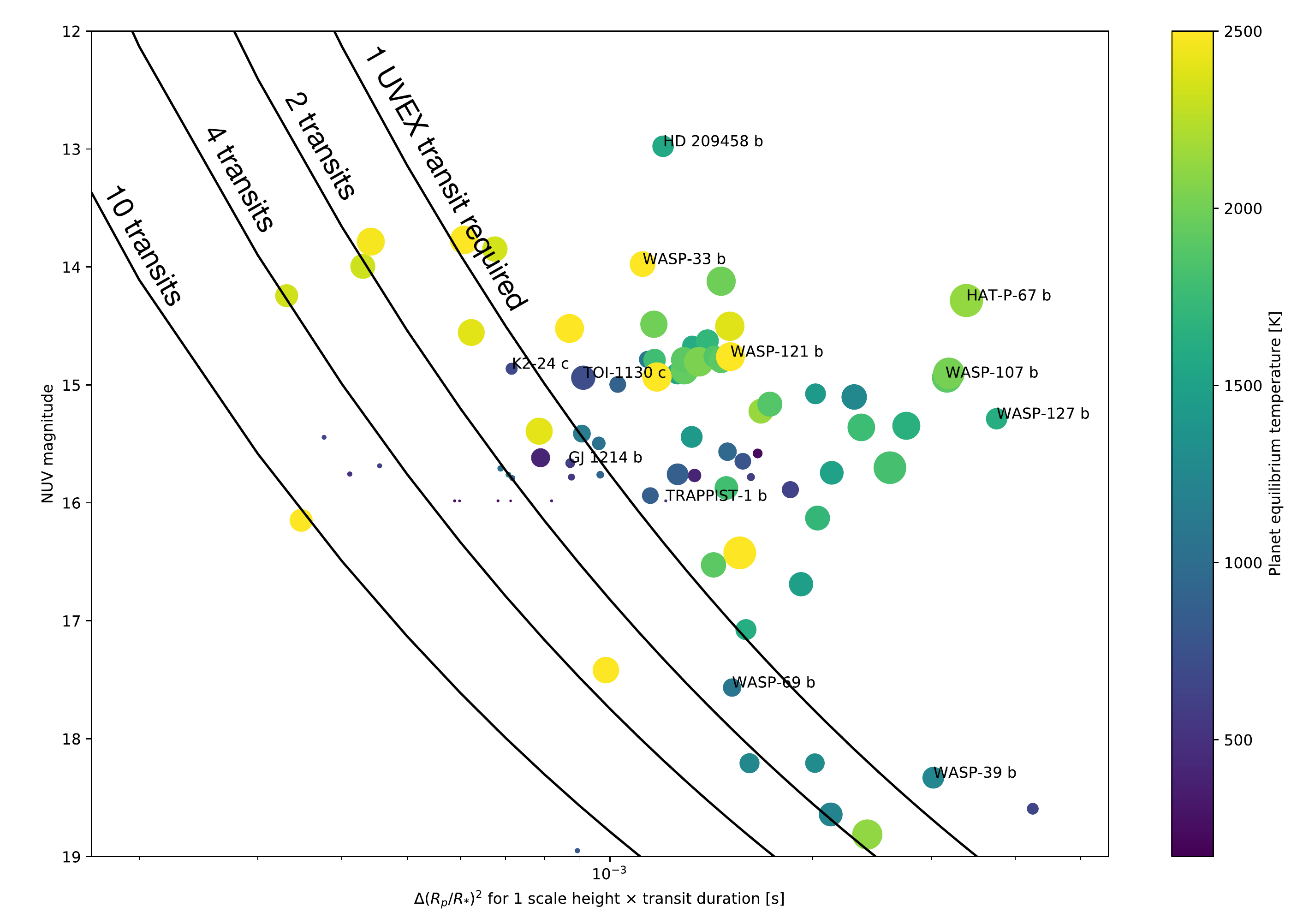}
    \caption{The observable atmospheric signal for each planet in the exemplar survey (defined as the fractional transit depth for one atmospheric scale height scaled by the transit duration), as a function of the NUV brightness of the host star. Solid lines show the number of \uvex\ transits (1, 2, 4, and 10) required to measure the transit depth to a precision of one atmospheric scale height.}
 \label{fig:exosnr}
\end{figure*}

We can address the two aforementioned
questions and provide a legacy survey of UV exoplanet measurements
with a \uvex survey spanning a broad range of both planet equilibrium
temperature and radius. To construct a sample survey, we searched
the NASA Exoplanet Archive for all transiting exoplanets with mass
measurements, and calculated the change in transit depth of one
atmospheric scale height, assuming zero albedo and a heat re-circulation
factor of 0.36. We calculated the number of transit observations
with \uvex required to reach a precision of one atmospheric scale
height for each planet. This was motivated by the typical size of
atmospheric features in the NUV, which are around 3–5 atmospheric
scale heights (e.g. \citealt{2018AJ....156..283E}, 
\citealt{wakeford2020}, \citealt{lothringer2022})
 -- thus allowing transit depth measurements
with a typical precision of $\sim$3$\sigma$. To do this, we estimated
the stellar flux in the \uvex passbands by first searching for the
exoplanet host stars in the \galex\ database (in the \galex\ NUV band).
For systems without \galex\ data, we estimated the NUV flux by scaling
the closest stellar model from the PHOENIX grid, and summing the
flux in the \galex\ NUV passband using \texttt{Pyphot}. We then 
designed the sample survey of the planets most amenable to atmospheric
characterization with \uvex (those requiring fewer transit observations
to reach the required precision in transit depth). To achieve a
sufficient spread in equilibrium temperature, we selected the 15
planets with the highest signal-to-noise in five temperature bins:
0--800~K, 800--1300~K, 1300--1800~K, 1800--2300~K, and $>$2300~K. To 
provide a sufficient spread in planetary radii, we additionally selected
the 15 planets with the highest SNR which were smaller than Neptune
(not counting duplicates from the previous step), since larger planets
are over-represented when ranking by SNR. This resulted in a total
sample of 81 planets ranging from $V$=7--17~mag, shown in Figures~\ref{fig:exosurvey} and \ref{fig:exosnr}. For each transit visit of each target we estimate the total observing time as equal to the duration of two full transits (to acquire sufficient out-of-transit baseline) plus one hour of overheads. In total the proposed survey of 81 planets would take 2,230 hours to complete---approximately three months. Of these planets, 36 can be used to investigate how haze properties of exoplanet atmospheres change with size (from $\sim$1.0--15~$R_{\oplus}$) and temperature (from $\sim$200--1250~K), and 45 can be used to investigate metal rain-out in ultra-hot giant planets from $\sim$1300--3100~K. All of our targets will have red-optical transit depths measured by the \textit{TESS} all-sky survey, so their $\Delta R_{\rm p,NUV-Red}$ index can be measured. Besides its use in identifying haze-formation and rainout trends, we can use this index to flag planets that exhibit large offsets, and effectively select the most favorable targets for more detailed atmosphere characterization with \jwst\ and \textit{Ariel}.

There are many critical exoplanet atmosphere processes that can 
only be observed in the UV. For example: the
presence and composition of clouds and hazes in cool exoplanets;
the rain-out of metals in hot exoplanets; atmospheric escape and
mass-loss processes on highly irradiated planets; and the
integrated UV radiation experienced by potentially
habitable planets. The three-month survey of 81
exoplanets outlined here would directly address the first two processes. 
It would also provide a legacy database of spectra for additional study,
and, crucially, would enable interpretation of expensive optical and IR
spectra obtained by upcoming NASA investments.

\section{Conclusion}
The \uvex\ mission opens a powerful new window to simultaneously explore both the static and dynamic UV sky. With a planned launch date in the late 2020's, \uvex\ fills a critical capability gap -- wide-field UV imaging -- in the era when Rubin, \textit{Roman}, and \textit{Euclid} will be transforming our understanding of the cosmos. Here we have described just a sampling of the anticipated science yield offered by \uvex, which covers the gamut of modern astrophysics, including all three science goals of NASA's Astrophysics division and the Astro2020 decadal survey. As evidenced by the success of projects like the Sloan Digital Sky Survey, by providing timely and high-quality data products to the entire community, the resulting \uvex\ discoveries will be limited only by the creativity of astronomers worldwide.

\appendix

\section{\uvex\ C/O Line Ratio Calculations}
\label{sec:co_ratios}

A primary goal of the \uvex\ metal-poor dwarf-galaxy science is to obtain new gas-phase UV observations of 
O\iii] \W\W1661,1666 and C\iii] \W\W1907,1909 in the first sample of metal-poor dwarf galaxies 
with young stellar populations.
Using the C\iii] \W\W1907,1909/O\iii] \W1666 line ratio method benefits from the fact that C/O 
exhibits minimal uncertainty due to reddening,
as the interstellar extinction curve is nearly flat over the wavelength range of interest ($1600-2000$ \AA)
and the O\iii] and C\iii] lines have similar excitation and ionization
potentials such that their ratio has little dependence on the physical conditions of 
the gas (i.e., nebular $T_e$ and ionization structure).

In order to fill in the C/O relationship with O/H in the sparsely measured 
metal-poor, young-age regime, we selected objects from the Sloan Digital Sky Survey (SDSS) Data Release
15 (DR15) with large equivalent widths (EWs) of emission lines (e.g., EW([O\iii] \W5007)$>750$ \AA)
and low metallicity (12 + log(O/H) $\leq 7.45$ or $Z\leq 0.05 Z_\odot$). 
High-ionization H\ii\ regions are needed given the energies required to ionize
C$^{+}$ and O$^{+}$ are 24.8 eV and 35.1 eV respectively.
High nebular electron temperatures ($T_e$) in low-metallicity 
environments allow the collisionally excited C and O transitions of interest to be observed
despite their large excitation energies (6--8 eV).
Below we describe the calculations that led to this selection criteria.

\subsection{Nebular Emission Line Measurements} \label{sec:lines}

Each of the targets in our initial UVEX sample has been previously observed as part of the SDSS DR15.
We used the publicly available SDSS data \citep{york00}, 
which have been reduced with the SDSS pipeline \citep{bolton12}.
Emission line fluxes and uncertainties were determined by Jarle Brinchmann using an upgraded pipeline 
similar to that of the MPA-JHU catalog, where
groups of nearby lines are fit simultaneously, constrained by a single Gaussian 
FWHM and a single line center offset from the vacuum wavelengths (i.e., redshift).

The line measurements were corrected for Galactic plus dust extinction 
(a reasonable method at low redshift) using the $E(B-V)$ color excess determined from
the observed-to-theoretical H$\alpha$/H$\beta$ ratios with the
\citet{Cardelli1989} reddening law.
All of our targets have low extinction with $E(B-V)$ values $\leq0.1$.

\subsection{Chemical Abundances} \label{sec:abundances}

\begin{table*}[]
    \centering
    \begin{tabular}{rll}
        && \\ \hline\hline 
         Ion		& Radiative Transition Probabilities & Collision Strengths      \\ \hline
        {C$^{+2}$}	& {\citet{wiese96}}		        	& {\citet{berrington85}}    \\
        {C$^{+3}$}	& {\citet{wiese96}}			        & {\citet{aggarwal04}}	    \\
        {O$^{+}$}	& {\citet{fft04}$^{\star\dagger}$}	& {\citet{kisielius09}}		\\
        {O$^{+2}$}	& {\citet{fft04}$^{\star\dagger}$}	& {\citet{aggarwal99}}      \\ \hline
    \end{tabular}
    \caption{The atomic data used with the P{\sc y}N{\sc eb} package to calculate ionic abundances
    and determine predicted emission line fluxes. 
    Note that the O$^{+2}$ collision strengths from \citet{aggarwal99} are calculated from a 6-level atom 
    approximation, which is required to predict the O\iii] \W\W1661,1666 fluxes.  \\
    $^{\star}$ Agrees with updated values from \citet{tayal11}. \\
    $^{\dagger}$ Equivalent to \citet{tachiev02}, as recommended by \citet{stasinska12}.}
    \label{tab:co}
\end{table*}

For all remaining calculations, we used the P{\sc y}N{\sc eb} package in {\sc python}
\citep{luridiana12, luridiana15}, assuming a five-level atom model \citep{derobertis87} and default atomic data
in most cases.
However, our O\iii] \W\W1661,1666 predictions require the use of a 6-level atom for which we 
adopt \citet{aggarwal99} who calculated the necessary collision strengths.
The adopted atomic data is listed in Table \ref{tab:co}.

\subsubsection{Temperature and Density} \label{sec:temden}

We required all of the galaxies in our initial UVEX sample to have SDSS spectra with significant 
[O\iii] \W4363 detections that allowed us to directly calculate the electron temperature, $T_e$,
of the high-ionization gas via the O\iii] I(\W\W4959,5007)/I(\W4363) ratio.
The low ionization zone temperatures were then determined using the 
photoionization model relationship from \citet{garnett92}:
\begin{equation}
    T_e{\rm [O II]} =  0.70\times T_e{\rm [O III]} + 3000 {\rm\ K,}
\end{equation}
The [S\ii] \W\W6717,6731 ratio was used to determine the electron densities.

\subsubsection{Ionic And Total Abundances} \label{sec:ionic}

Ionic abundances relative to hydrogen are calculated using:
\begin{equation}
	{\frac{N(X^{i})}{N(H^{+})}\ } = {\frac{I_{\lambda(i)}}{I_{H\beta}}\ } {\frac{j_{H\beta}}{j_{\lambda(i)}}\ },
	\label{eq:Nfrac}
\end{equation}
where the emissivity coefficients, $j_{\lambda(i)}$, are functions of both 
temperature and density.
\citet{berg21} demonstrated that O$^{+}$ and O$^{+2}$ are the dominant O ions
in nebular gas, while contributions from O$^{+}$ and O$^{+3}$ (requiring an ionization potential of 54.9 eV) 
are negligible, even in very high-ionization galaxies.
Therefore, we calculated total oxygen abundances (O/H) as
\begin{equation}
    Z = 12+\log({\rm O/H}) = 12+\log(\frac{{\rm O}^+}{{\rm H}^+} + \frac{{\rm O}^{+2}}{{\rm H}^+}).
\end{equation}

\subsection{UV Flux Requirements} \label{sec:co}

While observed C/O abundances vary significantly amongst dwarf galaxies, 
the emissivity of the O\iii] \W1666 line is roughly $6\times$ weaker than the emissivity
of C\iii] \W\W1907,1909, making the O\iii] \W1666 line the limiting emission line of the two. 
We, therefore, determined the minimum flux requirement to achieve our science goals based on the 
predicted O\iii] \W1666 fluxes of our initial metal-poor, young dwarf galaxy sample.
We determined the O\iii] \W1666 fluxes from the observed SDSS [O\iii] \W5007 fluxes using 
emissivity ratios from P{\sc y}N{\sc eb}, assuming the high-ionization gas electron temperatures 
determined above. 
The predicted O\iii] \W1666 fluxes were then reddened using the \citet{calzetti00} reddening
law and the $E(B-V)$ color excesses determined from the Balmer decrement. 
This provides the minimum required observed emission line fluxes.

\section{\uvex\ Magellanic Cloud Survey in Context of other Observations  }
 \label{appendix:MC_context}
 
The Magellanic Clouds have long been targets of deep, wide-area, and cadenced imaging surveys in the optical and near-IR.  Concerted optical and near-IR efforts have provided a detailed view of the stellar contents of the LMC and SMC (e.g., \citealt{udalski2000}, \citealt{zaritsky2002}, \citealt{zaritsky2004}, \citealt{cioni2011}, \citealt{nidever2017}), while observations in the mid- and far-IR reveal much about the nature of evolved stars, dust production, and the ISM at low-metallicities (e.g., \citealt{meixner2006}, \citealt{meixner2013}).   Upcoming surveys (e.g., Rubin, {\it Roman}) will provide an unprecedented deep and cadenced inventory of LMC and SMC stars in the optical and near-IR (i.e., with sub-arcsecond angular resolution and depths $m\gtrsim25$).  

Missing among these exquisite datasets is a matching, modern UV imaging survey of the LMC and SMC.  Though several UV facilities have imaged the LMC and SMC, none are well-matched to properties of modern optical and near-IR datasets.  \galex\ surveyed the LMC and SMC only in the NUV band (\citealt{simons2014}).  Its angular resolution (5\arcsec\ PSF) and limiting depth ($m_{\rm NUV} \sim21$) is far shallower and coarser than the optical and near-IR imaging.  \swift/UVOT ($\sim2.5$\arcsec) is limited to $m_{\rm UV} \sim$ 19 for point sources and only available in NUV bands (\citealt{hagen2017}).   \hst\ has the sensitivity and angular resolution in the UV to match (or exceed) modern optical and near-IR LMC and SMC imaging (e.g., \citealt{sabbi2013}), but its small FOV makes it unrealistic to survey the entirety of the LMC and SMC. Moreover, \hst\ is also limited to NUV wavelengths.  Other UV missions, such as \ultrasat\ (13\arcsec\ PSF), do not have characteristics that are suitable for resolved stellar populations in the LMC and SMC (e.g., Figure~\ref{fig:uvex_lmc_image}).

Spectroscopic surveys of massive stars in the LMC and SMC are far less complete and more heterogeneous than imaging surveys.  The largest systematic spectroscopic survey of the LMC is the VLT/Tarantula survey which obtained multi-epoch modest resolution optical spectroscopy of $\sim$800 OB type stars in the Tarantula star-forming region of the LMC (e.g., \citealt{evans2011}).  Despite including only a single star-forming region and being over a decade old, this survey accounts for the majority of our knowledge of massive star properties (e.g., spectral type, binarity) in the LMC.  In the SMC, combined efforts with the VLT and 2DF spectrograph have obtained optical spectra for $\sim300$ massive stars (e.g., \citealt{evans2004}, \citealt{evans2006}).  Other optical spectroscopic efforts are generally smaller and/or focus on specific sub-classes of massive stars.  

UV spectroscopy of massive stars in the LMC and SMC is essential for understanding stellar winds of massive stars at sub-Solar metallicity (see Section~\ref{sec:massivestars}). However, compared to optical data, UV spectroscopy in the LMC and SMC is sparse. The challenge and expense of obtaining UV spectra at the distances of the LMC and SMC have limited most studies to handfuls of stars per publication. The two most prominent systematic surveys are of R136, the cluster within 30~Doradus in the LMC.  \citet{massey1998} acquired UV spectra of 65 stars with \hst, while \citet{crowther2016} obtained spectra of 57 OB stars.  There are fewer UV spectra of massive stars in the SMC (e.g., \citealt{walborn2000}).

Though it has taken tremendous effort to assemble the above, and other, spectroscopic datasets in the LMC and SMC, they only represent a small fraction of the entire massive star content.  Estimates place the number of OB type stars in the LMC and SMC at $\sim30,000$ (based on recent star formation rates, counting resolved stars; e.g., \citealt{harris2004}, \citealt{harris2009}).  Of course, progress in understanding metal-poor massive star physics does not require spectra of every possible massive star. But far more are needed than are currently available.

Two ongoing and upcoming efforts stand to, at least partially, remedy the paucity of massive star spectra in the LMC and SMC. The first is from the 4MOST consortium. The 1001MC survey will target several thousand massive stars in the LMC and SMC with low-resolution optical spectroscopy (\citealt{cioni2019}). Another 4MOST effort proposes to monitor $\sim20,000$ massive stars in the LMC and SMC over the course of 5 years with medium-resolution spectroscopy, with the goal of determining binarity through radial velocity variations (\citealt{sana_shenar_2019}). 4MOST will be on sky by the mid-2020s. Multi-epoch spectral constraints on binary star characteristics from 4MOST are likely to be too late to guide substantial new \hst\ UV spectra in the LMC and SMC. However, 4MOST data will be well-timed relative to \uvex, which will uniquely provide the stellar wind measurements.

The second effort is the {\it Hubble} UV Legacy Library of Young Stars as Essential Standards\footnote{\url{https://ullyses.stsci.edu}} (ULLYSES; \citealt{roman-duval_2020}).  This program will combine new and archival \hst\ UV spectra in the LMC and SMC to form the first spectral atlas of metal-poor massive stars. This large effort is being accompanied by a VLT survey that will provide high resolution optical spectroscopy for all ULLYSES targets. ULLYSES is the first systematic program targeting UV spectra in the Magellanic Clouds. However, as we discuss in Section~\ref{sec:massivestars}, it should be viewed as an important first step of an effort that needs more data.

In an effort to increase number statistics and homogeneity, ULLYSES is an \hst\ director's discretionary program that combines new and archival STIS and COS UV spectra of $\sim300$ massive stars in the LMC and SMC (\citealt{roman-duval_2020}). It will create the first large UV spectral library of sub-Solar-metallicity massive stars. ULLYSES provides an important platform for improving our understanding of winds and mass loss in low-metallicity massive stars. Complementary ground-based optical spectra with the VLT will ensure an exquisite multi-wavelength dataset.

However, even with $\sim300$ spectra, ULLYSES will only sparsely sample the upper HR diagram. For single stars, the mass, metallicity, rotation rate, and phase of evolution are considered primary determinants of massive star winds. ULLYSES will sample approximately ten stars for each reasonable permutation of these parameters. The dimensionality and complexity of the problem grows when binarity is included. At least 50\% of massive stars are in binaries and the mass ratios and separation, in tandem with single star parameters listed above, can drive winds through mass transfer, Roche lobe overflow, and/or common envelope ejection (e.g., \citealt{uvex15}). By design, ULLYSES primarily targets single stars in order to provide benchmarks in absence of additional complexity due to binarity (\citealt{roman-duval_2020}).  

Given the (i) large variations in wind parameters at fixed stellar property (Section~\ref{sec:massivestars}; e.g., \citealt{crowther2016}); (ii) importance of binarity to so many stellar end states (e.g., Figure~\ref{fig:uvex_smc}); and (iii) the sparse (but substantially improved) sampling provided by ULLYSES, it is imperative to improve the state of UV spectra for low-metallicity massive single and binary stars.

\uvex\ will substantially expand on the foundation established by ULLYSES.  During its two year prime mission, \uvex\ will obtain UV spectroscopy of 1000 OB stars in the LMC and SMC. The high throughput, modest spectral resolution, and broad wavelength coverage of \uvex\ are ideal for filling out areas of the HR diagram in which ULLYSES is known to be sparse, such as binary stars.  \uvex\ will provide over three times as many UV spectra as ULLYSES, ensuring the robust statistics needed to quantify variations in stellar wind properties across the HR diagram and at fixed stellar parameters (e.g., luminosity, temperature, metallicity).  Crucially, it will also provide the UV spectra necessary to establish the link between binary configuration and stellar winds, a connection that is known to be important, but is essentially unconstrained empirically (e.g., \citealt{crowther2016}).   By the time \uvex\ is on sky, 4MOST will have acquired multi-epoch optical spectral for thousands of massive stars in the LMC and SMC (e.g., \citealt{cioni2019}, \citealt{sana_shenar_2019}). Spectroscopic targets for \uvex\ will be selected based on a wealth of optical spectra and time series optical and \uvex\ imaging, ensuring \uvex\ spectra covers all relevant stages of single and binary massive star evolution.

\section{Exposure Time Estimates for Hot Stars in the Magellanic Clouds} \label{sec:lmc_exp_time}

Here we detail calculations for \uvex\ exposure times needed for measuring wind velocities of stripped and hot, massive stars in the Magellanic Clouds.  Though we have simulated observations for a wide variety of stellar types, for brevity, we provide illustrative examples.   For these examples, as assume an LMC distance of 50~kpc, the \citet{Gordon2003-Ext-Curves} extinction curve, an extinction of $E(B-V)=0.1$~mag (typical of the LMC; \citealt{zaritsky2004}), and version \texttt{0.2.dev0} of the \uvex\ exposure time calculator (ETC).

\subsection{Stripped Stars}

In order to measure the wind velocity of a stripped star in the LMC, we require SNR $\ge10$ for $F_{\lambda} = 6\times10^{-15}$ erg s$^{-1}$ cm$^{-2}$ \AA$^{-1}$  in 3600s at 1240 \AA.  This estimate is based on a stripped star with a current mass of 2.7 $\msun$ ($M_{\rm initial} = 9 \msun$ ) in the LMC, which corresponds to $m_v = 18.5$~mag.  The model spectrum used is from the suite of stripped star models published in \citet{2018A&A...615A..78G}.  

\subsection{Hot Massive Stars}

We compute the \uvex\ time needed to measure the terminal wind velocity of an 18 $M_{\odot}$ O-type main sequence star ($T_{\rm eff} = 3 \times 10^4$\,K, $\log(g)=4.2$~dex, $m_v = 15$~mag) with a modest mass loss-rate of $\dot{M} = 1\times10^{-8} \msun$ yr$^{-1}$ (or $v_{\infty} = 2900$ km s$^{-1}$) and the CMFGEN stellar atmospheres. To accurately recover the terminal velocity, our simulations require a SNR $\ge10$ at 1550 \AA, which translates to $F_{\lambda} = 1.1\times10^{-13}$ erg s$^{-1}$ cm$^{-2}$ \AA$^{-1}$.  According to the ETC, this requires $\sim50$\,s of exposure time with \uvex.

\section{ \uvex TDE Discovery Rate }
\label{sec:TDE_rate}

Here we calculate the \uvex TDE discovery rate. At a typical
temperature of $3 \times 10^4$\,K, we have $L_{\rm UV} / L_{g} =
6.0$ in the FUV band which we focus on in the following calculations.
Based on the \textit{g}-band luminosity function measured 
by \citet{vanVelzen2018}, we infer the volumetric rate of TDE as a
function of UV luminosity $L$:
 \begin{align}
    \frac{\mathrm{d}\dot N}{\mathrm{d}L} = \frac{\dot N_0}{L_0}
    \left(\frac{L}{L_0}\right)^{\alpha}
 \end{align}
where $L_0 = 10^{43}\,{\rm erg\,s^{-1}}$, $\alpha \approx - 2.5$,
and $ \dot N_0 = \frac{\dot N_{0, g} }{\mathrm{ln}(10)} \times
6.0^{-(\alpha +1)} \approx  1.2\times 10^{-6}\,{\rm Mpc^{-3}\,yr^{-1}}$.
The number of detected events per year will be \begin{subequations}
 \begin{align}
    \R & = \int_{0}^{D_{\rm max}} \Omega D^2 \mathrm{d} D  \int_{L_{\rm
    min}}^{L_{\rm max}}\frac{\mathrm{d}N}{\mathrm{d}\luv} \mathrm{d}
    \luv\\ & = \frac{\Omega}{3} \frac{\dot N_0}{L_0} \int_{L_{\rm
    min}}^{L_{\rm max}} D_{\rm max}^3 \left(\frac{L}{L_0}\right)^{\alpha}
    \mathrm{d} L
 \end{align} \end{subequations}
where $\Omega$ is the solid angle of the surveyed area.

We note that the sample of luminous TDEs is currently poorly
characterized -- among the 13 TDEs used by \citet{vanVelzen2018},
only ASASSN-15lh peaked at $>10^{44}\,{\rm erg\,s^{-1}}$, and it
has distinct spectral properties compared with other events. The
shape of the luminosity function at the high end is poorly explored.
Therefore, to be conservative, hereafter we only consider the rate
of TDEs peaking below $10^{44}\,{\rm erg\,s^{-1}}$.

For each $L$, there is a maximum distance out to which the UV rise
of a TDE can be characterized (not just simply detected). 
We have $4 \pi D_{\rm
max}^2 f_{\rm thre} =  L$. Note that the threshold flux, $f_{\rm
thre}$, does not correspond to $f_0 \equiv 0.56\times
10^{-14}\,{\rm erg\,s^{-1}\,cm^{-2}}$ at 25\,mag (limit magnitude
with a 900\,s dwell). Instead, we require $f_{\rm thre} = 10 f_0$
to ensure that the TDEs peak at least $2.5$\,mag above the survey
threshold, such that the light curve UV rise and decay can be well
measured. Hence, we have,
 \begin{align}
   \R & = \frac{\Omega}{3} \frac{\dot N_0}{L_0}\int_{L_{\rm
   min}}^{10^{44}\,{\rm erg\,s^{-1}}} \left(\frac{L}{4 \pi f_0
   \times 10} \right)^{3/2} \left( \frac{L}{L_0}\right)^{-2.5}{\rm
   d}L
 \end{align}

We define a new parameter $x \equiv L / L_0$.
\begin{subequations}\label{eq:R1} \begin{align}
     \R & =\frac{\Omega}{3} \dot N_0   \int_{x_{\rm min}}^{10}
     \left( \frac{L_0}{4\pi f_0 \times 10}\right)^{3/2} x^{-1}{\rm
     d}x\\ & = \frac{\Omega}{3} \dot N_0  (1216\,{\rm Mpc})^{3}\times
     {\rm ln} \left( \frac{10}{x_{\rm min}}\right) \\ & = 2531
     \Omega
\end{align} \end{subequations} where we have assumed $x_{\rm min}
= 10^{42.5}\,{\rm erg\,s^{-1}}$ (iPTF16fnl is the faintest and
fastest TDE ever discovered, see \citealt{Blagorodnova2017}). Note
that $\mathcal{R}$ does not strongly depend upon $x_{\rm min}$.

We assume a mission lifetime of 2\,years and require that TDEs peak
after the first 2\,months and before the last 2\,months (such that
both the rise and decay can be characterized). Therefore, the
effective survey period is 1.67\,yr. In order to characterize 1000
TDEs (see requirement justified in Section~\ref{subsec:BHdemographics}),
using Eq.~(\ref{eq:R1}), we have
 \begin{align}
    2531 \times \Omega\, {\rm yr^{-1}} \times (1.67)\,{\rm yr} & =
    1000 \\ \Omega = 0.24 {\rm ster}
 \end{align}

The observed fields should be selected to avoid low Galactic latitude,
low declination, and regions of high Galactic extinction. The total
number of visits per band per field will be $365\times 2 \times 0.7
/ 25 = 20.5$. The proposed exposure time per dwell is 900\,s
(simultaneously in FUV and NUV). We assume that a
given field is visible to \uvex for 70\% of the survey time (due
to Sun constraint, etc). Thus, the surveyed solid angle should be
$0.24 / 0.7 = 0.34$ ster, corresponding to 1200\,deg$^2$ ($\approx 110$
\uvex\ fields). Taken together, such an imaging survey would cost 3--4\%
of the total \uvex baseline mission time.

\section{ \uvex GW Event Rates and Light Curve Models}
\label{sec:GW_appendix}

\subsection{Event Rates}

Currently, the LIGO-Virgo-KAGRA GW interferometers are fully funded for a fifth observing run (O5) in 2025--2026 \citep{Abbott2020LRRprospects}. The sixth observing run (O6) is expected to be 18--24 months in duration in the years 2028--2029, and will overlap with \uvex. In addition to LIGO Hanford, LIGO Livingston, Virgo and KAGRA, the LIGO India interferometer is also expected to join the O6 run \citep{Abbott2020LRRprospects}.

We undertake a detailed end-to-end simulation of the GW network performance to quantify \uvex measurement requirements. Our simulation assumes that all five interferometers are active in O6, but only at the funded A+ sensitivity, each with a duty cycle of 70\%.  
The methodology is similar to that described in \citealt{Petrov+2021}, updated to reflect population modeling results from the most recent GW transient catalog through the end of O3, GWTC-3. Fig. 7 of the GWTC-3 paper \citep{GWTC3-rates-and-populations} shows that the data currently support a fairly broad and flat NS mass distribution. Therefore, we adopted a distribution that is uniform from 1 to 2 $\msun$. 
We use the GWTC-2 rate estimate of $R_{\text{BNS}} = 320^{+490}_{-240}$\,Gpc$^{-3}$yr$^{-1}$ \citep{LIGO_GWTC2_pop_2020} as it is quoted for a uniform mass distribution consistent with what we learned from GWTC-3.

From this sample of simulated BNS mergers in O6, we select only those targets that are localized to better than 100\,deg$^2$. We then determine the necessary exposure times and tiling scheme to map 90\% of the enclosed probability while meeting our astrophysical requirements. Our goal is to achieve the depth necessary to detect and characterize a kilonova regardless of which model is dominant (see Section~\ref{subsec:lc_models} below). Thus we choose a conservative peak FUV absolute magnitude of  --12.1\,mag, which corresponds to an apparent magnitude of 24.4\,mag at 200\,Mpc (Figure \ref{fig:lightcurves}). We use the \uvex ETC to estimate the required exposure time for each event depending on the distance to the source and the UV background at that simulated location. For the selected events, the estimated required exposure time varied between 500\,s and 5,250\,s (with a median of 1080\,s). 

\begin{figure}
 \centering
  \includegraphics[width=\columnwidth]{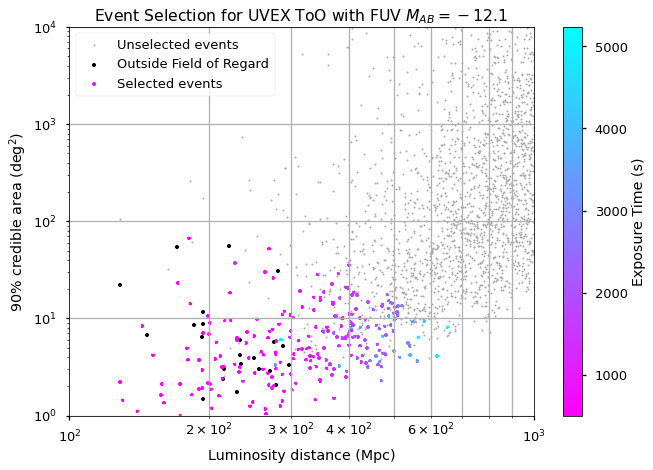}
   \caption{\small Simulation of GW triggers in the LIGO-Virgo-KAGRA sixth observing run (O6). The y-axis shows the 90\% credible area and the x-axis shows the luminosity distance. Only those events with a credible area $<$100\,deg$^2$ and distance such that a depth of $-$12.1 mag can be achieved in each pointing will be selected; selected events are color-coded by the optimal exposure time per the \uvex ETC. Events that fall outside the \uvex field of regard are excluded and denoted as black points. (Note that the number of points in the simulation shown above is proportional to the number of predicted events in O6 but this scaling is not 1:1.)}
 \label{fig:simulation}
\end{figure}

GW events are selected as observable by \uvex\ if their entire 90\% localization region can be observed to the required depth within 10 ks (Figure \ref{fig:simulation}). As a result of this analysis, $\sim 2.8\%$ of the events satisfy the localization, tiling, and exposure time criteria. Of these, 53\% are fully within the \uvex field-of-regard (accounting for sun exclusion). We conclude that 35 ToO triggers are expected to pass our selection criterion during the 18 months of the GW O6 run (20 ToO triggers would pass our selection criterion using the O5 configuration).


\subsection{Light Curve Models}
\label{subsec:lc_models}

Early kilonova data do not yet exist in the UV, hence we calculate
light curves for a GW170817-like event for three possible models
powering the UV transient at early times.

\begin{enumerate}

\item The first is a semi-analytical, nucleosynthesis-powered model
(\citealt{hotokezaka2020radioactive}; also, \citealt{Li1998},
\citealt{Metzger10}), where the radiation is purely generated by
radioactive decay of $r$-process nuclei. 
The model is described by 7 parameters: the mass of the ejecta (M$_{ej}$), the minimum and maximum velocity of the ejected material (v$_{min}$, v$_{max}$), the transition velocity between low and high opacity $\kappa$ (v$_{\kappa}$), the effective grey opacity for v $\leq$ v$_{\kappa}$ and v $>$ v$_{\kappa}$ ($\kappa_{low}$, $\kappa_{high}$), and the power law index of the velocity distribution across the mass space (n).

\item The second model is a shock-powered analytical prescription
\citep{piro2018evidence}, where the kilonova is powered through
shock-cooling of material surrounding the merger remnant, that has
been heated by a jet depositing energy into the material. It is
described by four parameters, the mass of the shock-heated material
($M_{sh}$), the minimum velocity of the material ($v_{sh}$), the
initial radius of the material ($R_{0}$) and the opacity of the
material ($\kappa_{sh}$).

\item For the third model we use predictions that, on top of the
nucleosynthesis-powered model, there is additional radiation coming
from the $\beta$-decay of free neutrons that have not been captured
by nuclei through r-process \citep{Metzger15}, with a total mass
of $M_{\mathrm{fn}} = 10^{-4} \msun$. This model is only
considered in combination with the nucleosynthesis-powered model,
and only affects the very early behavior of the light curve (t
$\leq$ 6h).

\end{enumerate}

\begin{figure}
 \centering
  \includegraphics[width=\columnwidth]{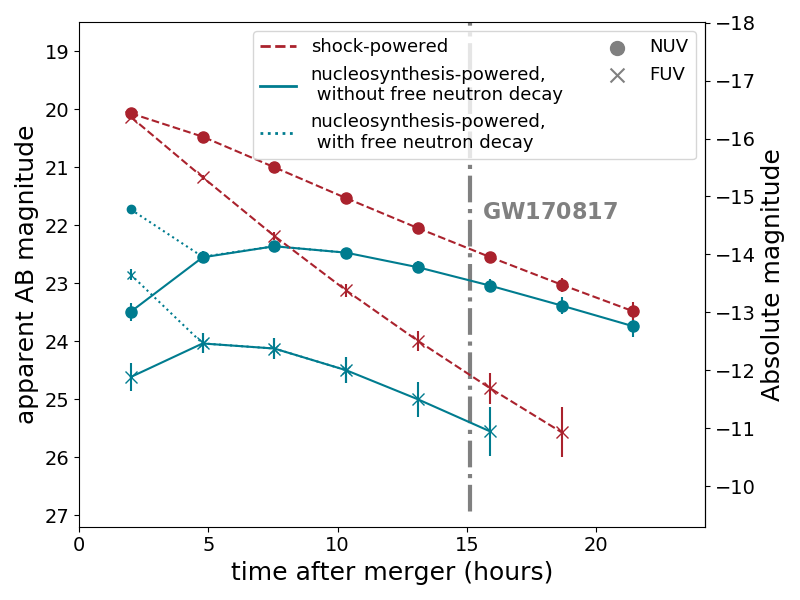}
   \caption{\small UV light curve predictions for a shock-powered
   model (dashed), and a nucleosynthesis-powered model with (dotted) and
   without (solid) a free neutron decay component. This light curve
   assumes a  distance of 200\,Mpc, exposure time of 1000~s and
   cadence of 10,000~s (due to ten tiles needed to map the localization
   area).  Note that 24.4 mag at 200 Mpc corresponds to an absolute
   magnitude of $-12.1$\,mag, which means that the targeted \uvex depth
   of $-12.1$\,mag is sensitive to all models in both filters.}
 \label{fig:lightcurves}
\end{figure}

In Table \ref{tab:parameters} we summarize the parameters for the
models, and indicate the ranges within which these parameters can vary,
noting that the free neutron model is fixed. Within these ranges,
90\% of the peak absolute magnitude lies between $[-15.6, -12.4]$
($[-14.5, -10.2]$) for the nucleosynthesis powered model and $[-17.8,
-15.3]$ ($[-17.9, -15.0]$) for the shock-powered model in the NUV
(FUV) band. In Figure~\ref{fig:lightcurves} we show the apparent
AB magnitude of all three models in the two \uvex filter bands, for
a GW170817-like event at a distance of 200\,Mpc, calculated with
the fiducial model parameters from Table \ref{tab:parameters}. The
cadence, 10,000 seconds, is determined by the amount of tiles needed
to map the localization area as estimated from the GW analysis, for
which we take a fiducial value of 100\,deg$^2$, and a 1,000s exposure time.

Based on the model light curves, especially the fast fading shock-model
and neutron-precursor model, our goal is for \uvex to respond within
3 hours on average for BNS/NS--BH events. Once \uvex is on target, it will map the localization area and repeat the sequence for 24 hours. The median number of tiles per event will be 5. 
Thus, a median of 17 epochs in the \uvex light curve is expected.

These results demonstrate that \uvex will generate high-SNR, well-sampled
light curves in both FUV and NUV  bands for all selected events in the O6 simulation, even in the more pessimistic case where the early UV
emission is powered entirely by heavy-element nucleosynthesis. 

\begin{table*}[ht]
\caption{The parameters describing the nucleosynthesis-powered and
shock-powered models in neutron star mergers, with the allowed ranges within
each parameter can vary. The fiducial values are the values used
to generate the data in Figure~\ref{fig:lightcurves}. }
\centering
\resizebox{\textwidth}{!}{%
\begin{tabular}{llll}
\hline
\textbf{Parameter (Unit)}               & \textbf{Description}                                  & \textbf{Range}                & \textbf{Fiducial value} \\ \hline
\multicolumn{4}{c}{\textit{Nucleosynthesis Powered Model} }                                                    \\ \hline
$M_{\mathrm{ej}}$ ($M_\odot$)     & Ejecta mass                                           & (0.01, 0.1)                          & 0.05                    \\
$v_{\mathrm{min}}$ ($c$)            & Minimum ejecta velocity                               & (0.05, 0.2)                          & 0.1                     \\
$v_{\mathrm{max}}$ ($c$)            & Maximum ejecta velocity                               & (0.3, 0.8)            & 0.4               \\
$n_{\mathrm{ej}}$                & Power law index of ejecta density distribution        & (3.5, 5)                             & 4.5                     \\
$v_{\mathrm{\kappa}}$ ($c$)         & Transition velocity between high and low $\kappa$     & ($v_\mathrm{min}$, $v_\mathrm{max}$) & 0.2               \\
$\kappa_\mathrm{high}$ (cm$^2$~g$^{-1}$)  & Effective grey opacity for $v \leq v_\mathrm{\kappa}$ &(1, 10)                              & 3                       \\
$\kappa_\mathrm{low}$ (cm$^2$~g$^{-1}$) & Effective grey opacity for $v \geq v_\mathrm{\kappa}$ & (0.1, 1)             & 0.5     \\ \hline
$M_{\mathrm{fn}}$ ($M_\odot$) & Free neutron mass & - & $10^{-4}$ \\ \hline
\multicolumn{4}{c}{\textit{Shock Interaction Powered Model}}                                                                                                        \\ \hline
$M_{\mathrm{sh}}$ ($M_\odot$)     & Shocked ejecta mass                                   & (0.005, 0.05)                        & 0.01                    \\
$v_{\mathrm{sh}}$($c$)             & Shocked ejecta velocity                               & (0.1, 0.3)                           & 0.2                     \\
$R_0$ (10$^{10}$ cm)              & Initial shock radius                                  & (1, 10)                              & 5                       \\
$\kappa_\mathrm{sh}$ (cm$^2$~g$^{-1}$)   & Effective grey opacity of shocked ejecta              & (0.1, 1)                             & 0.5                     \\ \hline
\end{tabular}%
}
\label{tab:parameters}
\end{table*}

\end{document}